\documentclass[structabstract]{aa}

\usepackage{graphicx,units}
\usepackage{color}
\usepackage{natbib}
\usepackage[pdftex]{hyperref}
\usepackage{txfonts}

\newcommand{\ebv}{E(B-V) }
\newcommand{\photoz}{photometric redshift }
\newcommand{\pdf}{probability density function }
\newcommand{\pdfs}{probability density functions }

\begin{document}

\title{A new method to improve photometric redshift reconstruction}
\subtitle{Applications to the Large Synoptic Survey Telescope}

\author{Alexia Gorecki$^1$, Alexandra Abate$^{2,3}$, R\'eza
Ansari$^2$, Aur\'elien Barrau$^1$, Sylvain Baumont$^1$, Marc Moniez$^2$ and Jean-Stephane Ricol$^1$}

\institute{Laboratoire de Physique Subatomique et de Cosmologie, UJF/INP/CNRS/IN2P3, 53 avenue des
Martyrs, 38026 Grenoble cedex, France
\and
Laboratoire de l'Acc\'el\'erateur Lin\'eaire, Univ. Paris Sud/CNRS/IN2P3, Bat. 200, 91898
Orsay cedex, France
\and
Physics Department, University of Arizona, 1118 East 4th Street, Tucson, AZ 85721, USA\\ \email{abate@email.arizona.edu}
}
\date{Received ???; accepted ???}

  \abstract
   {In the next decade, the Large Synoptic Survey Telescope (LSST) will become a major facility for the astronomical community. However accurately determining the redshifts of the observed
   galaxies without using spectroscopy is a major challenge.}
   {
   Reconstruction of the redshifts with high resolution and well-understood uncertainties is mandatory
   for many science goals, including the study of baryonic acoustic oscillations (BAO).
   We investigate different approaches to establish the accuracy that can be
   reached by the LSST six-band photometry.}
   {We construct a realistic mock galaxy catalog, based on the Great
     Observatories Origins Deep Survey (GOODS)
   luminosity function, by simulating the expected apparent magnitude distribution for the
   LSST.  To reconstruct the photometric redshifts (photo-z's),
we consider a template-fitting method and a neural network method. 
   The photo-z reconstruction from both of these techniques
   is tested on real Canada-France-Hawaii Telescope Legacy Survey
(CFHTLS) data and also on simulated
   catalogs. We describe a new method to improve photometric redshift reconstruction that
   efficiently removes catastrophic outliers via a likelihood ratio statistical test. This test uses the posterior probability functions of the fit parameters and the colors.}
   {We show that the photometric redshift accuracy will meet the stringent LSST
   requirements up to redshift $\sim2.5$ after a selection that is based on the likelihood ratio
   test or on the apparent magnitude for galaxies with signal-to-noise
   ratio $S/N>5$ in at least 5 bands.
   The former selection has the advantage of retaining roughly 35\% more galaxies for a similar photo-z performance compared to the latter. Photo-z reconstruction using a neural network algorithm is also described. 
   In addition, we utilize the CFHTLS spectro-photometric catalog to outline the possibility of combining the neural network and template-fitting methods.}
   {We demonstrate that the photometric redshifts will be accurately estimated with the LSST if a Bayesian prior probability and a calibration sample are used.}
   

	\authorrunning{A. Gorecki \textit{et al.}}
	\titlerunning{Photometric redshift reconstruction techniques and methods to reject catastrophic
	outliers.}
   \maketitle


\section{Introduction}
The Large Synoptic Survey Telescope (LSST) has an optimal design for investigating the mysterious dark energy. With its large field of view
and high transmission bandpasses, the LSST will be able to observe a tremendous amount of galaxies, out to high redshift, over the visible sky from Cerro Pach\'on over ten years. This will lead to an unprecedented study of dark energy, among other science programs such as the study of the Milky Way and our Solar System \citep{ScienceBook}. 

One of the main systematic uncertainties in the cosmological analysis will be tightly 
related to errors in the photometric redshift (photo-z) estimation. Estimating the redshift from 
the photometry alone \citep{Baum:1962} is indeed much less reliable than using 
spectroscopy, although it does allow measurements to be obtained for
vastly more galaxies, especially for those that are very faint and distant.  Photo-z estimates are mainly sensitive to characteristic changes in the galaxy's
spectral energy distribution (SED), such as the Lyman and the Balmer breaks at 
$100 \unit{nm}$ and $400 \unit{nm}$, respectively. Incorrect identifications between 
these two main features greatly impact the photometric redshifts and are an example of how catastrophic photo-z outliers can arise.
Mischaracterizing the proportion of these outliers will strongly impact the level of 
systematic uncertainties.

There are basically two different techniques to compute the photo-z. 
On the one hand, \textit{template-fitting methods} \citep[e.g.][]{Puschell:1982,Bolzonella:2000js} fit a model galaxy SED to the photometric data and aim to identify the spectral type, the redshift, and possibly
other characteristics of the galaxy. It has been proven that using spectroscopic 
information in the template-fitting procedure, by introducing a
Bayesian prior probability \citep{Benitez:1998br} or by modifying the
SED template \citep{BudaEtAl2000,Ilbert:2006dp} for example,
improves the photo-z quality. This highlights the necessity to have 
access to at least some spectroscopic data.

On the other hand, \textit{empirical methods} extract information from a 
spectroscopic training sample and are therefore generally limited to the 
spectroscopic redshift range of the sample itself. Among these, the empirical 
color-redshift relation \citep{Connolly:1995yq,Sheldon:2012} and neural networks 
\citep{Vanzella:2003ca,Collister:2003cz} are commonly used.

In this paper, we address the issue of estimating the photo-z quality 
with a survey similar to the LSST, and in particular, we introduce a
new method, the likelihood ratio statistical test, that aims 
to remove most of the galaxies with catastrophic redshift determination (hereafter 
called outliers). We utilize a galaxy photometric catalog, which is simulated for a study of 
the uncertainty that is expected from the LSST determination of the dark energy equation of state parameter 
using baryon acoustic oscillations (BAO). The related results will be presented in a 
companion paper by Abate et al.~\textit{in prep}. 

Based on a Bayesian $\chi^2$ template-fitting method, our photo-z
reconstruction algorithm gives access to the posterior probability density
functions (pdf) of the fit parameters. Using a training sample, the \textit{likelihood ratio} 
test, which is based on the characteristics of the posterior pdf and the 
colors, is calibrated and then applied to each galaxy in the photometric sample. 
The technique is tested on a spectro-photometric catalog from the T0005 data release 
of CFHTLS\footnote{Based on observations obtained with MegaPrime/MegaCam, a joint
project of CFHT and CEA/DAPNIA at the Canada-France-Hawaii Telescope (CFHT),
which is operated by the National Research Council (NRC) of Canada, the Institut
National des Science de l'Univers of the Centre National de la Recherche
Scientifique (CNRS) of France, and the University of Hawaii. This work is based
in part on data products produced at TERAPIX and the Canadian Astronomy Data
Centre as part of the Canada-France-Hawaii Telescope Legacy Survey, a
collaborative project of NRC and CNRS.} that is matched with spectroscopic catalogs from
VIMOS-VLT Deep Survey (VVDS) \citep{LeFevre:2004hx,Garilli:2008pt}, DEEP2 \citep{deep2}
and zCOSMOS \citep{Lilly:2006va}. We also outline the 
possibility to discard outlier galaxies by using photometric redshifts
estimated from both the template-fitting method and a neural network.

Finally, we illustrate the modification to the systematic and statistical uncertainties 
on the photo-z when the redshift distribution of the training sample is 
biased compared to the actual redshift distribution of the photometric catalog to 
be analyzed.

The paper is organized as follows. The LSST is presented in
Sect. \ref{Sec:LSST} and
followed by our simulation method in Sect. \ref{Sec:Method}. In the latter, the 
simulation steps employed and the physical ingredients required to produce the 
mock galaxy catalogs are described. In Sect. \ref{Sec:MockCatValidation}, the 
mock galaxy catalogs for Great Observatories Origins Deep Survey
(GOODS), CFHTLS, and LSST are presented and validated 
against data for the former two surveys.  Our template-fitting method and the likelihood ratio test are described in Sect.~\ref{Sec:PhotozTemplateFittingMethod}. The performance of our photo-z
template-fitting method is shown in Sect. \ref{Sec:photoz}. 
In Sect. \ref{Sec:NN}, the photo-z neural network technique and its 
performance in conjunction with our template-fitting method is investigated. Finally, 
we give a brief discussion of the current limitations of our
simulations in Sect. \ref{Sec:Discussion}
and conclude in Sect. \ref{Sec:Conclusion}.

Throughout the paper, we assume a flat cosmological $\Lambda CDM$ model with the
following parameter values: $\Omega_m = 0.3$, $\Omega_{\Lambda} = 0.7$, $\Omega_k
 = 0$, and $H_0 = 70 ~\unit{km/s/Mpc}$. Unless otherwise noted, all magnitudes given are in the AB magnitude system.


\section{Simulation description and verification}\label{Sec:DataSimValid}

\subsection{The Large Synoptic Survey Telescope}\label{Sec:LSST}

The LSST is a ground-based optical telescope survey designed in part to study the 
nature of dark energy. It will likely be one of the fastest and widest telescopes of the 
coming decades.  The same data sample will be used to study the four major
probes of dark energy cosmology: type 1a supernovae, weak gravitational
lensing, galaxy cluster counts, and baryon acoustic oscillations.

The LSST will be a large aperture $8.4~\unit{m}$ diameter telescope
with a 3200 Megapixel camera. It will provide unprecedented
photometric accuracy with six broadband filters
$(u,g,r,i,z,y)$. Figure~\ref{Fig:filter} shows the LSST transmission
curves, including the transmissions of the filters, the expected CCD quantum
efficiency, and the optics throughput.
The field of view 
will be $9.6~\unit{deg^2}$
and the survey should cover $30~000~\unit{deg^2}$ of sky visible from Cerro Pach\'on. 

The LSST will perform two back-to-back exposures of $2\times
15~\unit{sec}$ with a readout time of $2 \times 2 ~\unit{sec}$. The
number of visits and the $(5\sigma)$ limiting apparent magnitude in each
band for the point sources for one year and ten years of the running survey are listed in Table \ref{Tab:LimitMag}.
With such deep observations, photometric redshifts will necessarily be computed in an essentially unexplored redshift range.

The photo-z requirements, as published in the LSST Science Book 
\citep{ScienceBook}, are given in Table \ref{Tab:PZreq}.  

The final specifications of the LSST are subject to change; see \cite{Ivezic:2008fe} for
the latest numbers.

 \begin{table}[t]
 \caption{Number of visits and $5\sigma$ limiting apparent magnitudes \textbf{(point sources)}, for one year and ten
years of LSST operation \citep{Ivezic:2008fe,OpSim}.}
\begin{center}
\begin{tabular}{cccccccc} 
   \multicolumn{4}{c}{One year of observation} \\ \hline \hline
 &  $u$ & $g$ & $r$ & $i$ & $z$ & $y$ \\ \hline
 Number of visits & 5 & 8 &  18 &  18 & 16 & 16 \\
  $m_{5\sigma}$  & 24.9 & 26.2& 26.4& 25.7& 25.0& 23.7 \\ \hline
   \multicolumn{4}{c}{Ten years of observation} \\ \hline \hline
 &  $u$ & $g$ & $r$ & $i$ & $z$ & $y$ \\ \hline
 Number of visits & 56 & 80 &  184 &  184 & 160 & 160 \\
 $m_{5\sigma}$   & 26.1 & 27.4 & 27.5 & 26.8 & 26.1 & 24.9 \\ \hline
 \end{tabular}
\end{center}

\label{Tab:LimitMag}
\end{table}

 \begin{table}[t]
 \caption{LSST photo-z requirements for the high signal-to-noise
   ``gold sample" subset, which is defined as having $i<25.3$. The
   parameters are defined as follows: $\sigma_z/(1+z)$ is the
   root-mean-square scatter in photo-z;  $\eta$ is the fraction of
   3$\sigma$ outliers at all redshifts; and $e_z$ is the bias, defined as the mean of $(z_p-z)/(1+z)$ at a given $z$, where $z_p$ is the photo-z.}
\begin{center}
\begin{tabular}{|c|c|c|} 
   \multicolumn{3}{c}{}\\ \hline \hline
quantity & requirement & goal \\
\hline
$\sigma_z/(1+z)$ & $<0.05$ & $<0.02$\\
$\eta$ & $<10$\% &\\
$\mid e_z\mid$ & $<0.003$ & \\
\hline \hline
 \end{tabular}
\end{center}
\label{Tab:PZreq}
\end{table}

\begin{table}
 \caption[]{\label{nearbylistaa2}
 Number of galaxies that are both observed photometrically with the CFHTLS survey and spectroscopically.}
\begin{tabular}{lccc}
 \hline \hline
  Spectroscopic  &    CFHTLS field & No. of galaxies &   Refs. \\ 
  survey & & & \\
   \hline
VVDS Deep1 & W1 \& D1 & 2011 &  (1)\\
DEEP2  \textit{Data Release 3} & W3 \& D3 & 5483 & (2) \\
VVDS f22& W4 & 4485& (3) \\
zCOSMOS & D2 & 2289 &(4) \\
\hline
\end{tabular}\label{Tab:CFHTLS}
\tablebib{(1) \cite{LeFevre:2004hx}; (2) \citep{deep2}; (3)
\cite{Garilli:2008pt}; (4) \cite{Lilly:2006va}.}
\end{table}


\subsection{Simulation of galaxy catalogs}\label{Sec:Method}

The simulation method we employ is to draw basic galaxy attributes: we
consider redshift, luminosity, and type from observed distributions,
assign each galaxy a SED and a reddening based on those attributes,
and then calculate the observed magnitudes expected for the survey in
question.  For similar efforts, see the following work:
\cite{Dahlen:2007jd} for a SNAP\footnote{Supernova/Acceleration
Probe mission}-like mission, \cite{Jouvel:2009mh} for
JDEM\footnote{Joint Dark Energy Mission}/Euclid-like missions,
\cite{Benitez:2009} for the PAU\footnote{Physics of the
Accelerating Universe} survey.

\begin{figure}
   \centering
   \includegraphics[scale=0.45]{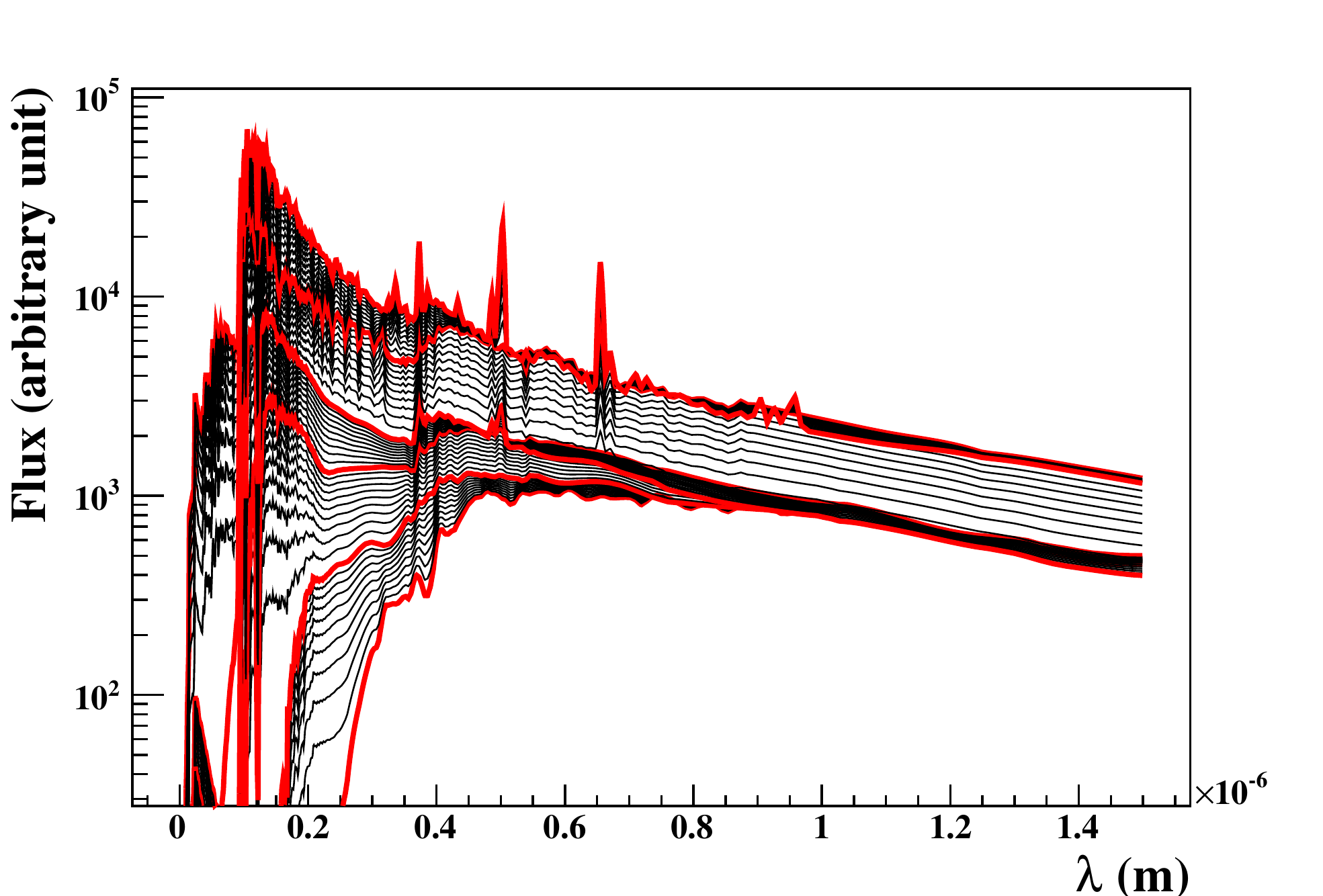} 
      \caption{ SED templates are linearly interpolated from the original six templates
      from \cite{1980ApJS...43..393C} and \cite{1996ApJ...467...38K}. The original templates are drawn in red. \label{Fig:SED}}
\end{figure}


\subsubsection{Simulating galaxy distributions}\label{Sec:GalDist}

To simulate the galaxy catalog, we first compute the total number of galaxies $N$
within our survey volume between absolute magnitudes $M_1$ and
$M_2$. Then we assign redshifts and galaxy types for each
of these $N$ galaxies.

If $\phi$ is the sum of luminosity functions over the \textit{early}, \textit{late} and \textit{starburst} galaxy types (see Sect.~\ref{Sec:LF} for more details), then the number of galaxies $N_g$ is given by
\begin{eqnarray}
N_g = \frac{c}{H_0} \int_0^6 \int_{M_1}^{M_2}  \phi(M,z)(1+z)^2d_A(z)^2
E(z)^{-1}\Omega dz dM ~,
\end{eqnarray}
where $M$ is the absolute magnitude in some band, $d_A(z)$ is the angular diameter distance, the function $E(z) = \sqrt{\Omega_m(1+z)^3+\Omega_{\Lambda}}$,
and $\Omega$ (no subscript) is the solid angle of the simulated survey.
The redshift range is chosen so as not to miss objects that may be
observable by the survey. We chose to use luminosity functions
observed from the GOODS survey in the $B$-band. The exact choice of
$M_1$ and $M_2$ is not critical, since: (i) At the bright limit, the luminosity function goes
quickly to zero; therefore the integral does not depend on $M_1$ as
long as it is less than $-24$. (ii) As long as $M_2$ is chosen to be
fainter than the maximum absolute magnitude observable by the survey,
then all galaxies that are possible to observe are included in the
integral.
We calculated this to be $M_2 = -13$.
The redshift $z_s$ of each simulated galaxy is drawn from the cumulative density function:
\begin{eqnarray}
C_z(z_s) =  \frac{\displaystyle \int_0^{z_s} \displaystyle \int_{M_1}^{M_2}
\phi(M,z') dV(z') dM
}{\displaystyle\int_0^6 \displaystyle \int_{M_1}^{M_2} \phi(M,z) dV(z) 
dM}~,
\end{eqnarray}
where $dV$ is the comoving volume element. Once the redshift of the galaxy, denoted by $z_s$, is assigned, the absolute magnitude
$M$ is drawn from the following cumulative density function
\begin{eqnarray}
C_{M}(M,z_s) = \frac{\int_{M_1}^{M} \phi(M',z_s) 
 dM'}{\int_{M_1}^{M_2} \phi(M',z_s) 
dM'} ~.
\end{eqnarray}

Finally, a broad galaxy type is assigned from the observed distribution of each type at redshift
$z_s$ and absolute magnitude $M$. This distribution is constructed from the type-dependent luminosity functions. Therefore, each galaxy is designated a broad type value of either \textit{early}, \textit{late} or \textit{starburst}.  An SED from the library is then selected for each galaxy, according to the simulation procedure described in Sect. \ref{Sec:SED}. 

\subsubsection{Simulating the photometric data}\label{Sec:AppMag}

%
The simulated apparent magnitude $m_{X,s}$$^[$\footnote{Subscript $s$ stands for the simulated value.}$^]$ in any LSST band $X$ with transmission $X(\lambda)$ for a galaxy of SED type $T_s$$^[$\footnote{Here, type $T_s$ refers to the actual SED of the galaxy and not the broad type value, e.g \textit{early}, \textit{late} or \textit{starburst}.}$^]$, redshift $z_s$, color excess $E(B-V)_s$, and absolute magnitude $M_{Y,s}$ is generated as follows: 
\begin{eqnarray}
m_{X,s} = M_{Y,s} +\mu(z_s)+K_{XY}(z_s,T_s,E(B-V)_s)~,
\end{eqnarray}
 where $\mu(z_s)$ is the distance modulus and $K_{XY}(z_s,T_s,E(B-V)_s)$ is the K-correction, defined as described in \cite{Hogg:2002yh} for spectral type $T_s$, with flux observed in observation-frame band $X$ and $M_{Y,s}$ in rest-frame band $Y$. Then, the magnitude is converted into the corresponding simulated flux $F_{X,s}$ value.
The simulated observed flux $F_{X,obs}$  is drawn from a Gaussian with
a mean $F_{X,s}$ and standard deviation $\sigma(F_{X,s})$. This is correct as long as the flux is large enough to be well distributed with a Gaussian distribution. 
The uncertainty $\sigma_X\left(m_{X,s}\right)$ on true magnitude in band $X$ is given by Eq.~\ref{Eq:SigmaMagLSST} in Sect. \ref{Sec:LSSTAppMagErrors}.
  
Note that the apparent magnitude uncertainty $\sigma_X\left(m_{X,s}\right)$ depends on the number of visits $N_{X,vis}$.  We have performed the simulation for two sets of values of $N_{X,vis}$ that correspond to one and ten years of observations with the LSST, according to the $N_{X,vis}$ given in Table \ref{Tab:LimitMag}.

Throughout the paper, the quantity $z_s$ refers to the simulated or true value of the redshift.  Here we also assume that a spectroscopic redshift obtained for one of the simulated galaxies has a value equal to $z_s$.  Therefore, the value $z_s$ can be also considered to be the galaxy's spectroscopic redshift with negligible error.


   \begin{figure}
   \includegraphics[scale=0.45]{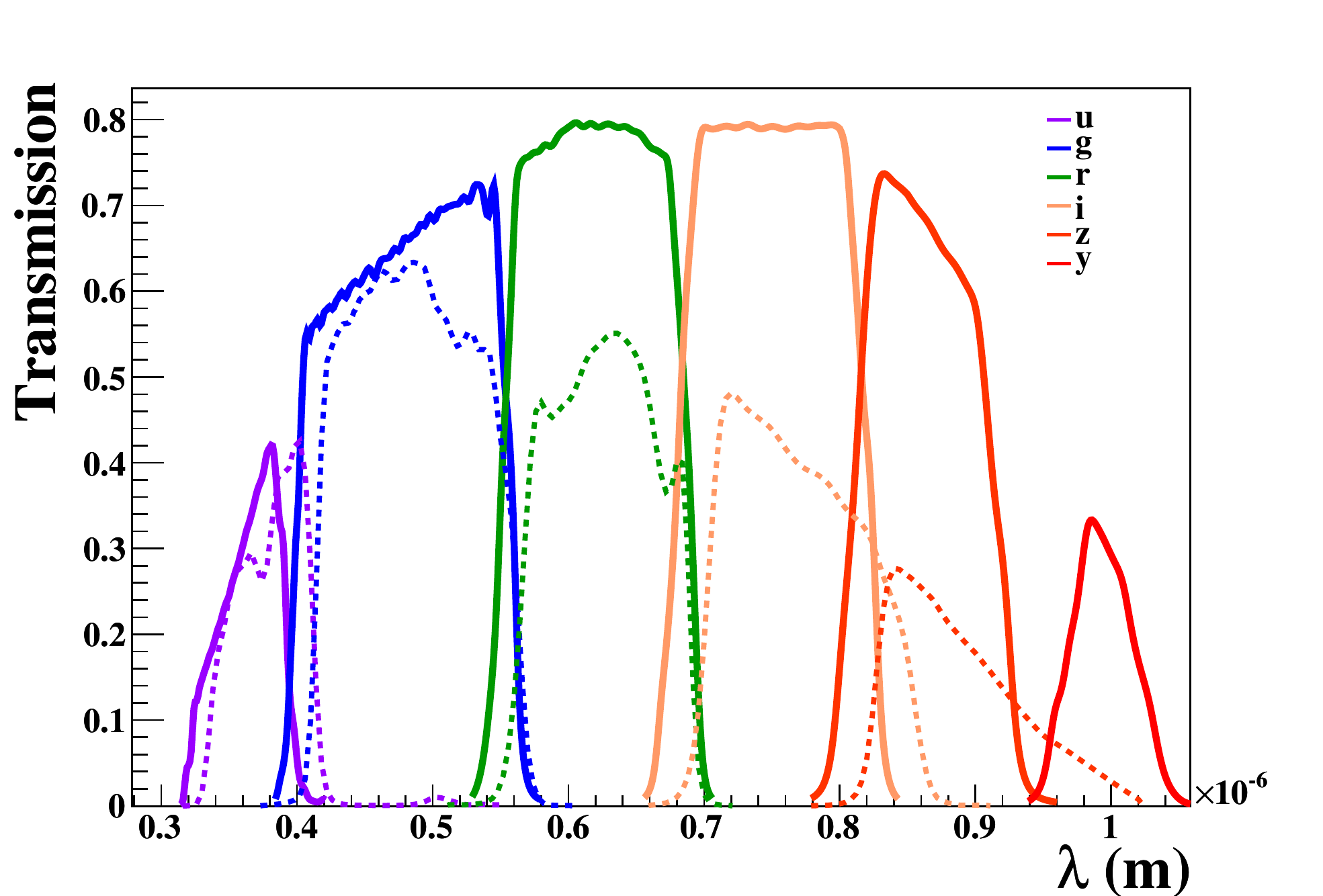}
      \caption{LSST transmission curves shown by the solid lines and CFHTLS transmissions shown by the dashed lines.
      The transmission includes the transmission of the filter
      itself, the expected CCD quantum efficiency, and the telescope optical
      throughput. \label{Fig:filter} }
   \end{figure}

\subsubsection{Luminosity function}\label{Sec:LF}

The luminosity function probabilistically describes the expected number of galaxies per unit
volume and per absolute magnitude. If the luminosity functions are redshift- and type-dependent, then they give the relative amount of galaxies for each galaxy type at a given redshift. 

We use luminosity functions measured from the GOODS survey \citep{Dahlen:2005}. The luminosity functions here are modeled by a parametric Schechter function that takes the form:
\begin{eqnarray}
\phi(M)& =&0.4\ln(10)\phi^\star y^{(\alpha+1)}\exp(-y)\\
y&=&10^{-0.4(M-M_\star)},
\end{eqnarray}
where $M$ is the absolute magnitude in the B-band of GOODS wide field
imager (WFI), and $M_\star$, $\phi^\star$, and $\alpha$ are the parameters defining the function. Their values can be obtained from \cite{Dahlen:2005}.

\subsubsection{Spectral Energy Distribution (SED) library}\label{Sec:SED}

 \begin{figure}
  \centering
  \includegraphics[scale=0.45]{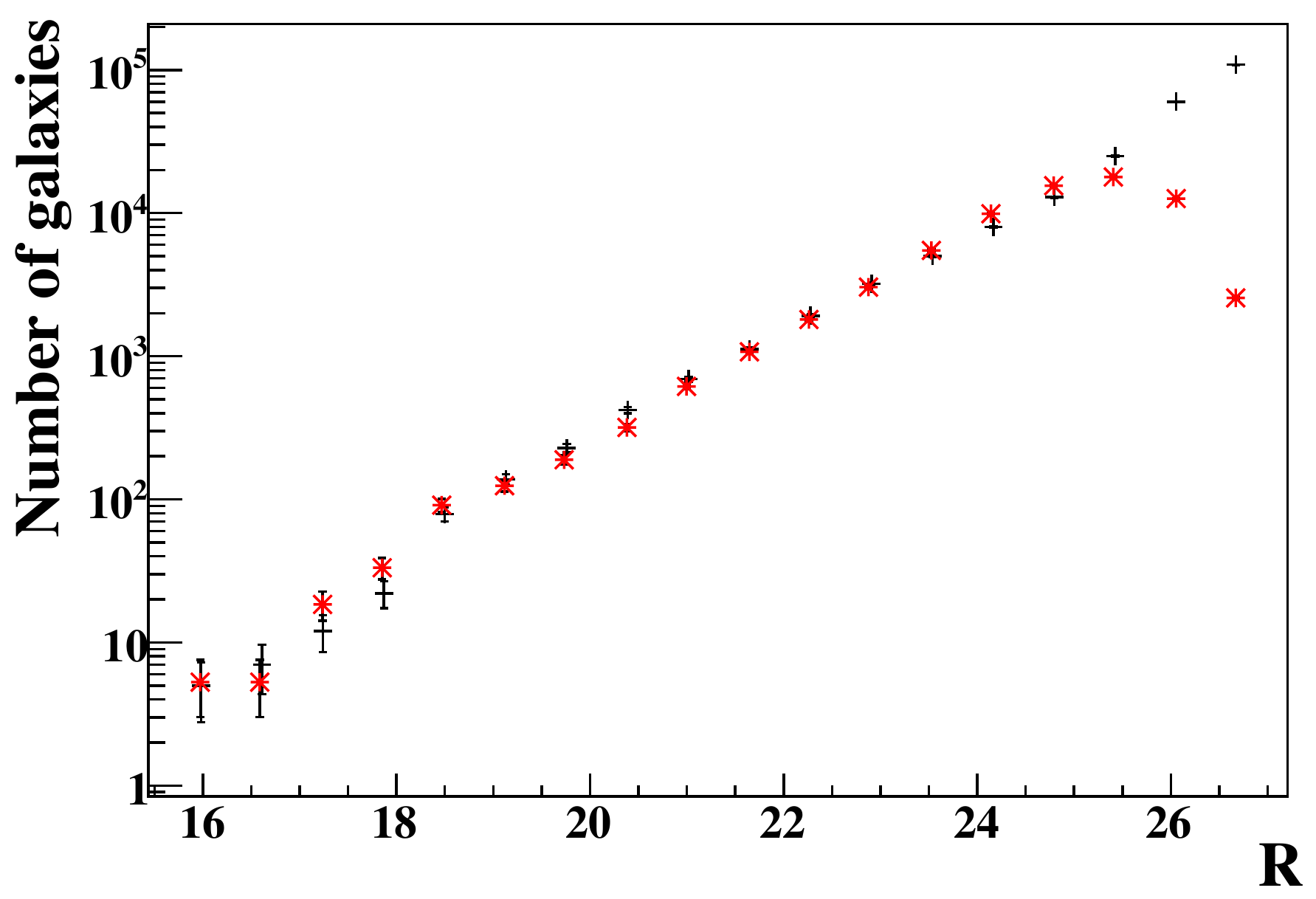}
     \caption{Histograms of the apparent magnitude in the $R$ band
     comparing the GOODS simulated data (black points with errorbars)
     to the actual GOODS data (red stars). We note that there may be a systematic shift in all data points in the $x$-axis direction up to $R\lesssim$0.05 mag due to differences between the simulation and data filter zero points.}
      \label{Fig:GOODSmag}
\end{figure}

We built a SED library composed of 51 SEDs. They were created by interpolating between six
template SEDs, as described here:
\begin{itemize}
\item the \textit{early}-type El, the \textit{late}-types Sbc, Scd and the \textit{starburst}-type Im from \cite{1980ApJS...43..393C},
\item the \textit{starburst}-types SB3 and SB2 from \cite{1996ApJ...467...38K}. 
\end{itemize}
These six original SEDs were linearly extrapolated into the UV by
using the GISSEL\footnote{Galaxy Isochrone Synthesis Spectral Evolution Library}
synthetic models from \cite{2003MNRAS.344.1000B}. The interpolated spectra of the 51 types are displayed in Fig~\ref{Fig:SED}. In the following, we denote by $T_s$ the true spectral type (SED) of the galaxy.

Each galaxy is assigned a SED from this library using a flat probability distribution based on
their broad type value, originally assigned as either \textit{early}, \textit{late} or \textit{starburst} (see Sect.~\ref{Sec:GalDist}). This way of generating the spectra may not be as optimal as
using more realistic synthetic spectra, but it has heuristic
advantages.
For example, there is an easy way to relate the galaxy type of the
luminosity function to a SED type, and additionally it is much faster,
in terms of computing time, to produce a large amount of galaxy spectra at different evolutionary stages.
We are aware,
however, that this linear interpolation may bias photometric redshifts
that are estimated using a template-fitting method, because real galaxies are probably not evenly distributed across spectrum space.  Therefore, this feature may allow the neural network method to be more effective in estimating the redshift.

\subsubsection{Attenuation by dust and intergalactic medium (IGM)}

 \begin{figure*}
 \centering
 \begin{tabular}{ccc}
  \includegraphics[scale=0.4]{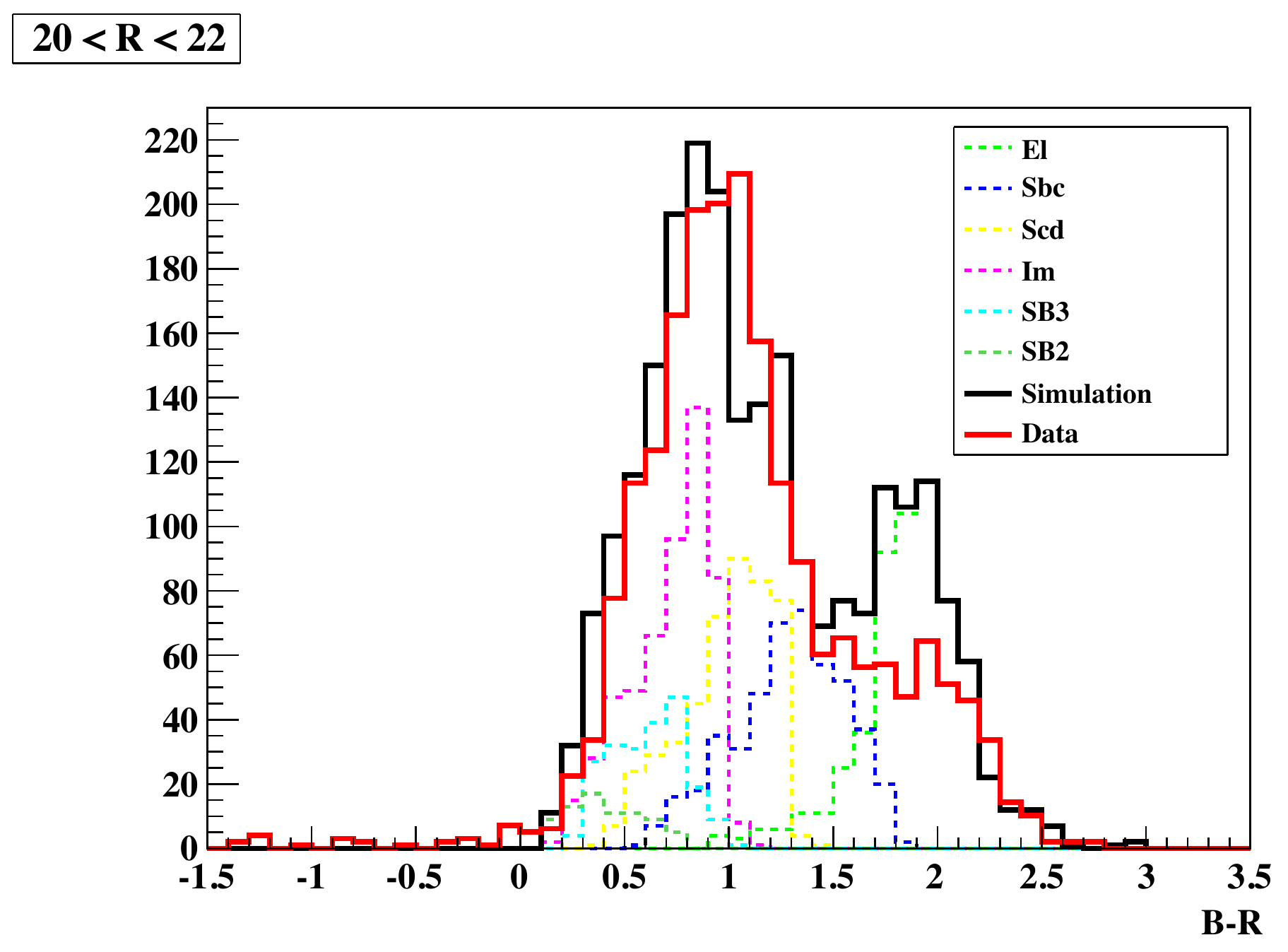} &
  \includegraphics[scale=0.4]{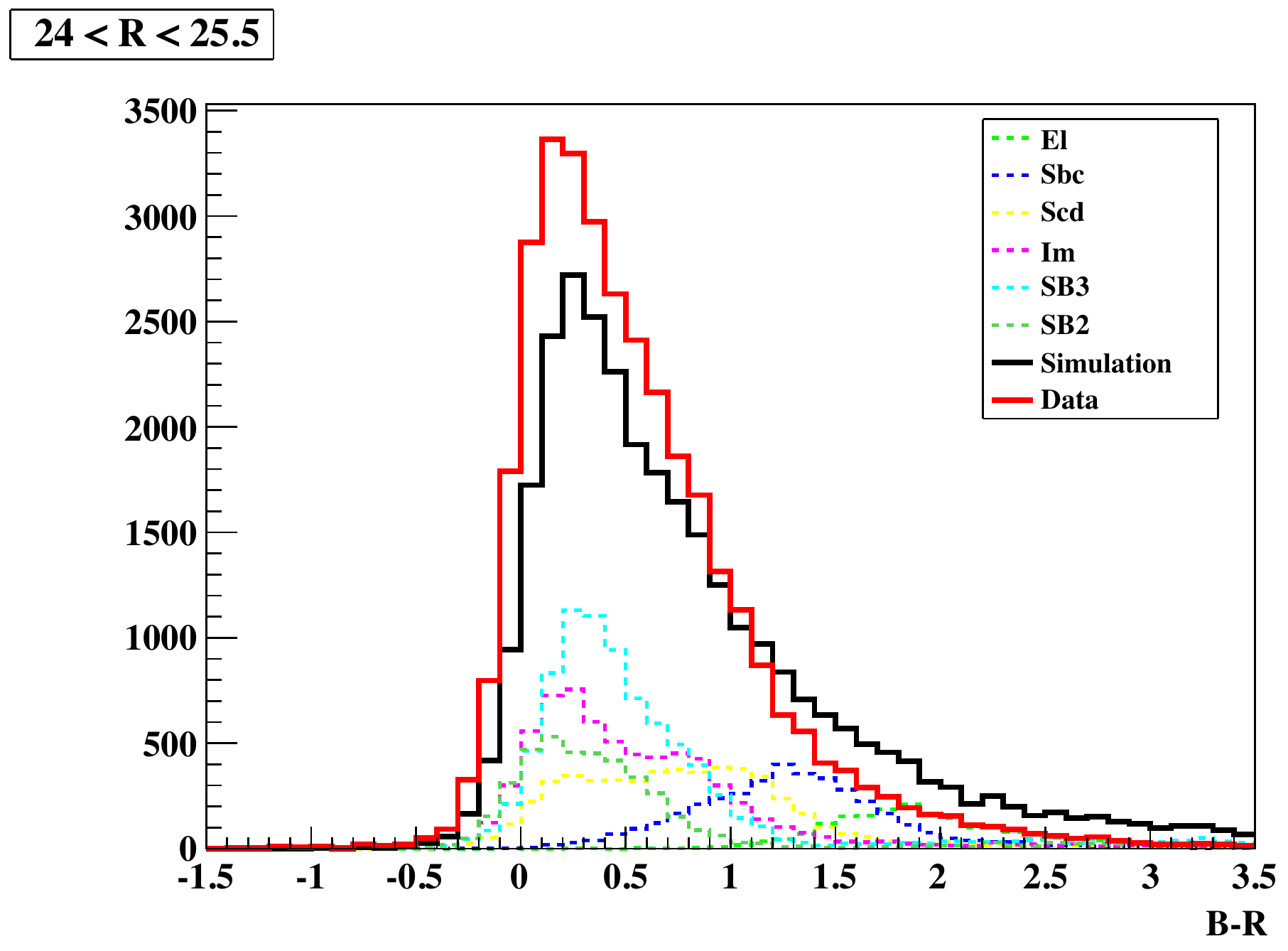}
  \end{tabular}
     \caption{Histograms of the $B-R$ term for different apparent magnitude ranges.
     Left-hand panel ($20<R<22$) corresponds to the bright galaxies and right-hand panel ($24<R<25.5$) corresponds to the very faint galaxies. The solid black lines correspond to the simulation and the solid red lines correspond to the GOODS data. The dotted colored lines correspond to the main spectral types in the simulation.}
      \label{Fig:GOODSColor}
\end{figure*}

The reddening caused by dust within the target galaxy is quantified in our simulation by the color excess term $E(B-V)$.  With this term, the Cardelli law \citep{Cardelli:1989fp} is used for the galaxies closest to the El, Sbc, and Scd spectral types, whereas the Calzetti law is used for the galaxies closest to the Im, SB3, and SB2 spectral types. 
The color excess $E(B-V)$ is drawn from a uniform distribution between 0 to 0.3 for all galaxies, except  for galaxies closest to the El type. Indeed, elliptical galaxies are composed of old stars and contain little or no dust; therefore, $E(B-V)$ is drawn only between 0 to 0.1 for these galaxies.

Another process to be considered is the absorption due to the intergalactic medium (IGM). It is caused by clumps of neutral hydrogen along the line of sight and is well-modeled by the Madau law
\citep{1995ApJ...441...18M}. As the absorption occurs at a fixed wavelength in the hydrogen reference frame, it is redshift-dependent in the observer frame. Strong features in the optical part of the SEDs are induced by the IGM at redshifts above about $z\sim2.8$ in the LSST filter set, when the Lyman-$\alpha$ forest has shifted into the LSST band passes.   Here we assume this absorption to be constant with the line of sight to the galaxy.  An investigation of the effect of the stochasticity of the IGM will be the subject of future work.

\subsubsection{Filters}

The six LSST bandpasses displayed in Fig.~\ref{Fig:filter} include the quantum efficiency of the CCD, the filter transmission, and the telescope optical throughput. The CFHTLS filter set \footnote{The CFHTLS transmissions have been downloaded from \url{http://www1.cadc-ccda.hia-iha.nrc-cnrc.gc.ca/community/CFHTLS-SG/docs/extra/filters.html.}} is also displayed in the same figure.  We expect to be able to obtain good photo-z estimates up to redshifts of about 1.2 for CFHTLS and 1.4 for LSST when the 4000 \AA\ break is redshifted out of the filter sets.  At redshifts above 2.5, the precision should improve dramatically when the Lyman break begins to redshift into the $u$-band.
 
\begin{figure}
  \centering
  \begin{tabular}{cc}
    \includegraphics[scale=0.45]{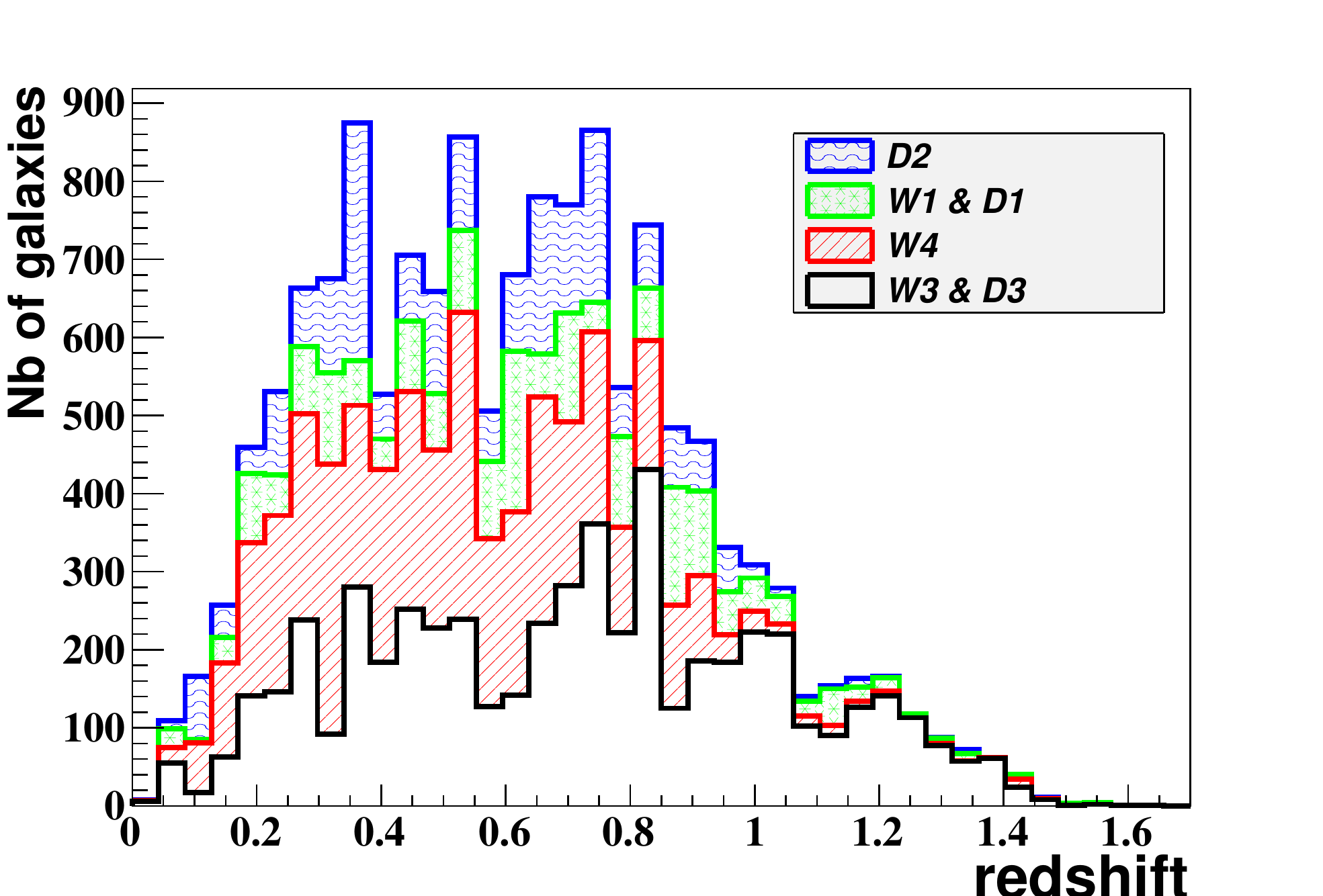} &
  \end{tabular}
  \caption{Redshift distribution of the spectroscopic sample for the different CFHTLS fields.
    The histograms are stacked. 
    \label{Fig:CFHTLSredshift}   }
\end{figure}

\subsubsection{Apparent magnitude uncertainties for the LSST}\label{Sec:LSSTAppMagErrors}

The apparent magnitude uncertainties for the LSST are computed
following the semi-analytical expression from the LSST Science Book
\citep{ScienceBook}. This expression has been evaluated from the LSST
exposure time calculator, which considers the instrumental throughput,
the atmospheric transmission, and the airmass among other physical parameters.

The total uncertainty on the apparent magnitude includes a systematic uncertainty that comes from
the calibration, such that the photometric error in band $X$ is
\begin{eqnarray}\label{Eq:SigmaMagLSST}
\sigma_X\left(m_X\right) = \sqrt{\sigma_{rand,X}^2+\sigma_{sys,X}^2}~,
\end{eqnarray}
where $\sigma_{rand,X}$ is the random error on the magnitude and $\sigma_{sys,X}$ is taken to be equal to 0.005 and is the photometric systematic
uncertainty of the LSST for a point source. We have adopted this
simple formula defined for point sources and have used it for extended
sources. A more realistic computation of this uncertainty for extended
sources will be completed in future work (see also Sect. \ref{Sec:Discussion}).

\subsubsection{Apparent magnitude uncertainties for CFHTLS and the GOODS survey}\label{Sec:SigMagCG}

An analytical expression similar to the one given by Eq.~\ref{Eq:SigmaMagLSST} does not exist
for the CFHTLS and the GOODS data. The apparent magnitude uncertainties are estimated with
algorithms and analysis techniques specific to these surveys and the
relations described in the previous Sect. does not apply. 

In the following, where simulations of photometric galaxy catalogs of both of these surveys are carried out, the simulated uncertainties on the apparent magnitudes are estimated directly from the survey data themselves. In this way, one can obtain the probability distribution of having $\sigma_X$ given $m_X$. This allows the assignment of $\sigma_X$ by randomly drawing the value, according to this probability density function, given the value of $m_X$.
\begin{figure*}
   \centering
   \includegraphics[scale=0.45]{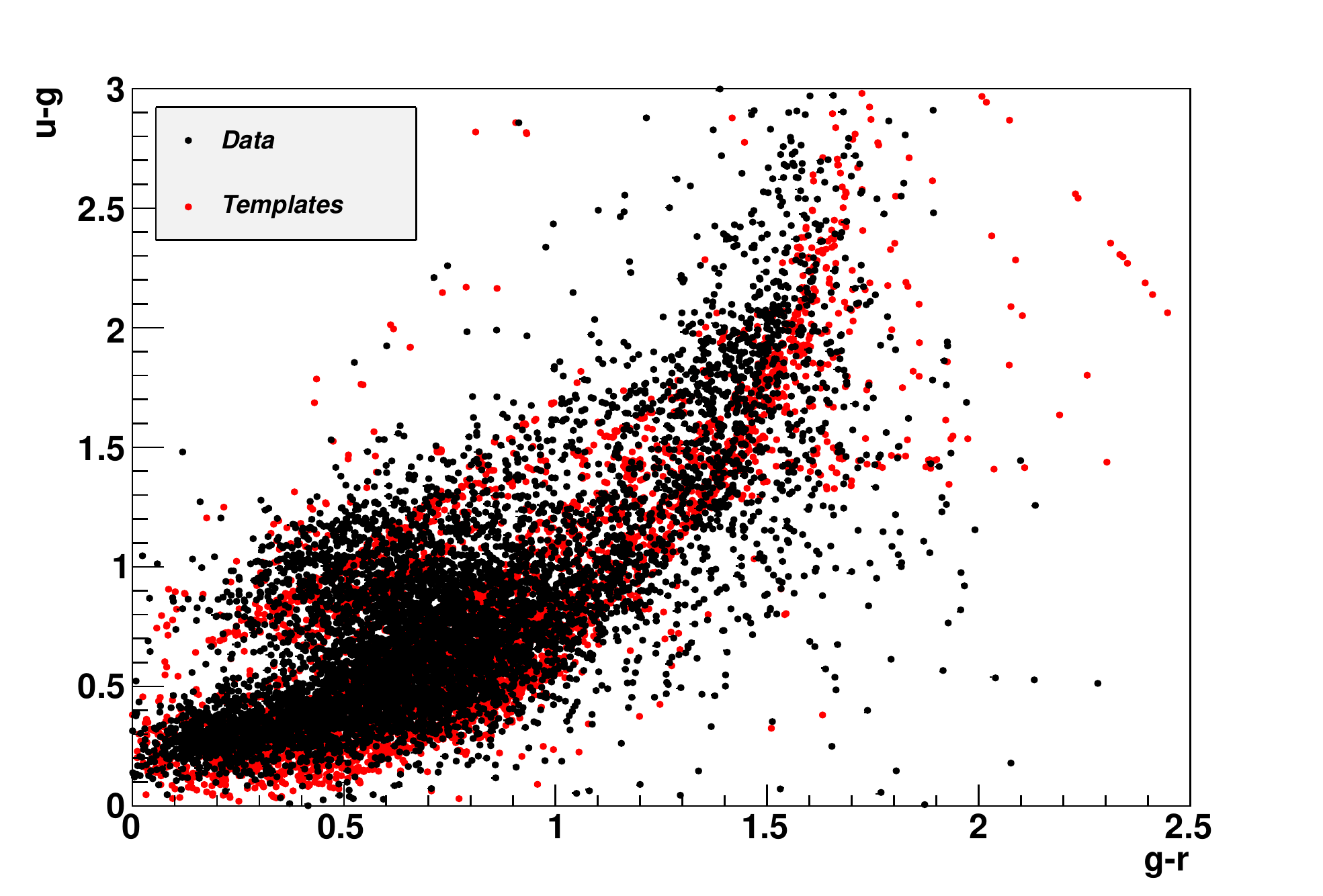}
   \includegraphics[scale=0.45]{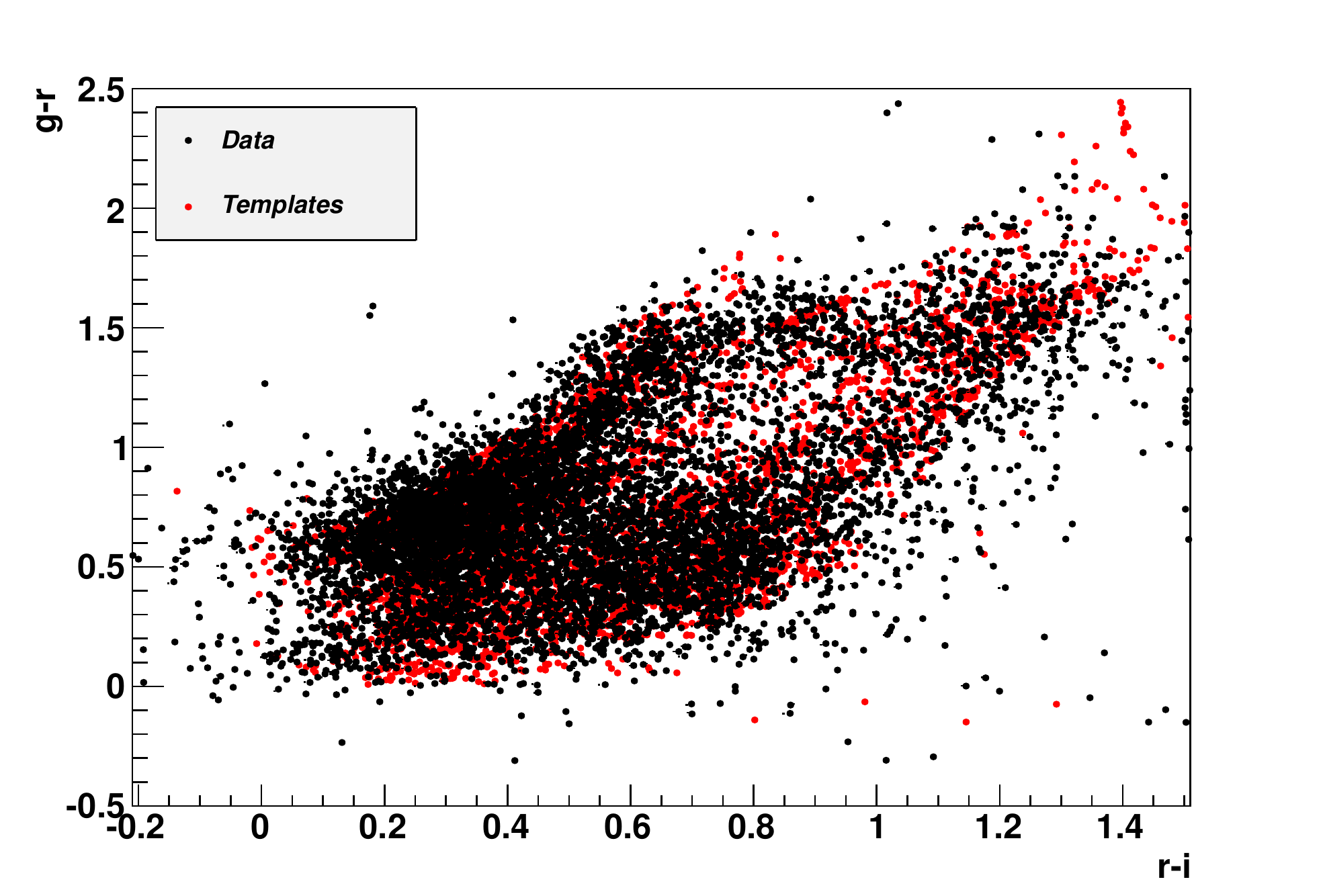}
      \caption{
      Color-color plots showing the relation between CFHTLS data (black dots) and the distribution expected from our SED template set (red dots).
      \label{Fig:CFHTLScolor} }
\end{figure*}

\begin{figure*}
   \centering
   \begin{tabular}{cc}
       \includegraphics[scale=0.45]{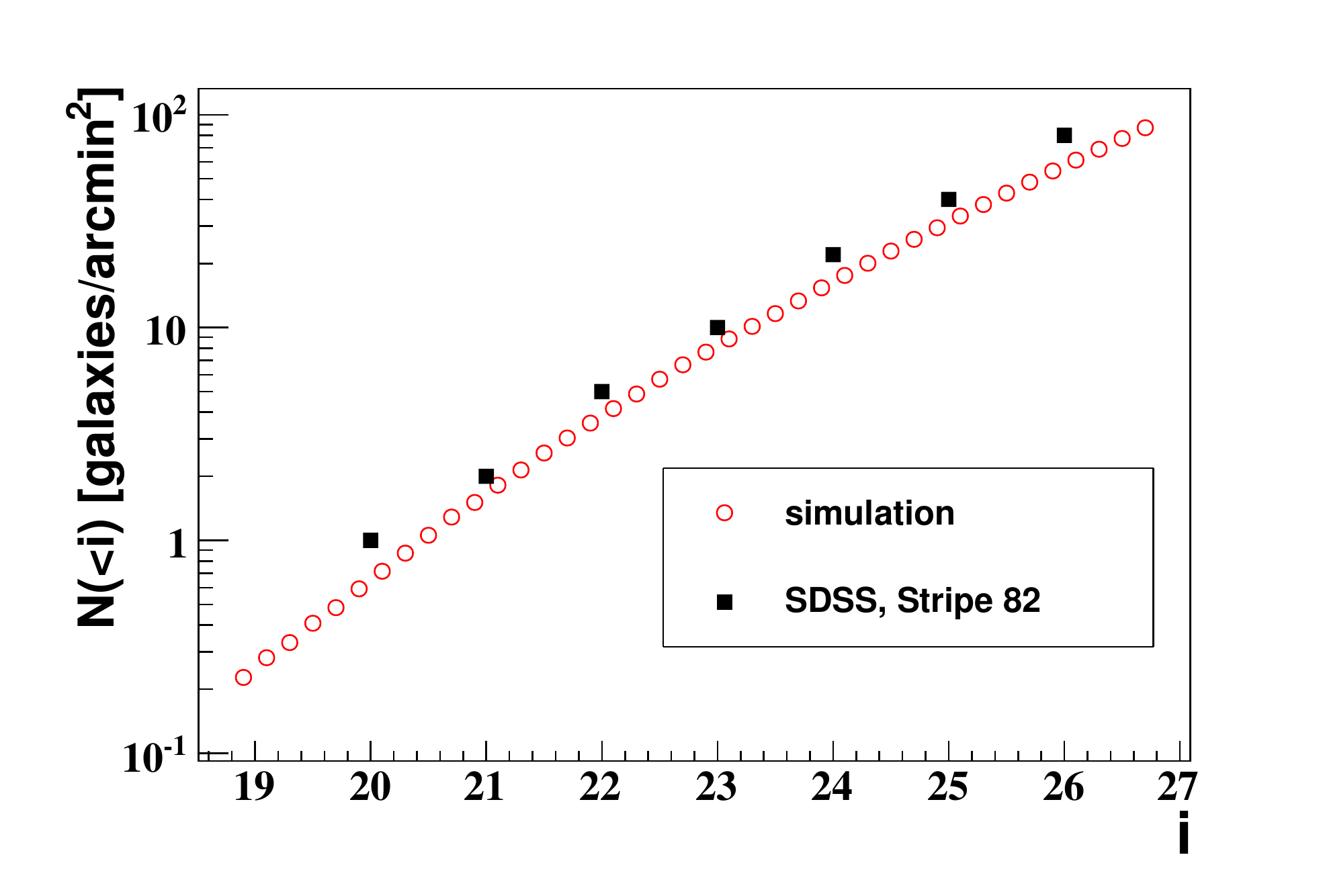} &
    \includegraphics[scale=0.45]{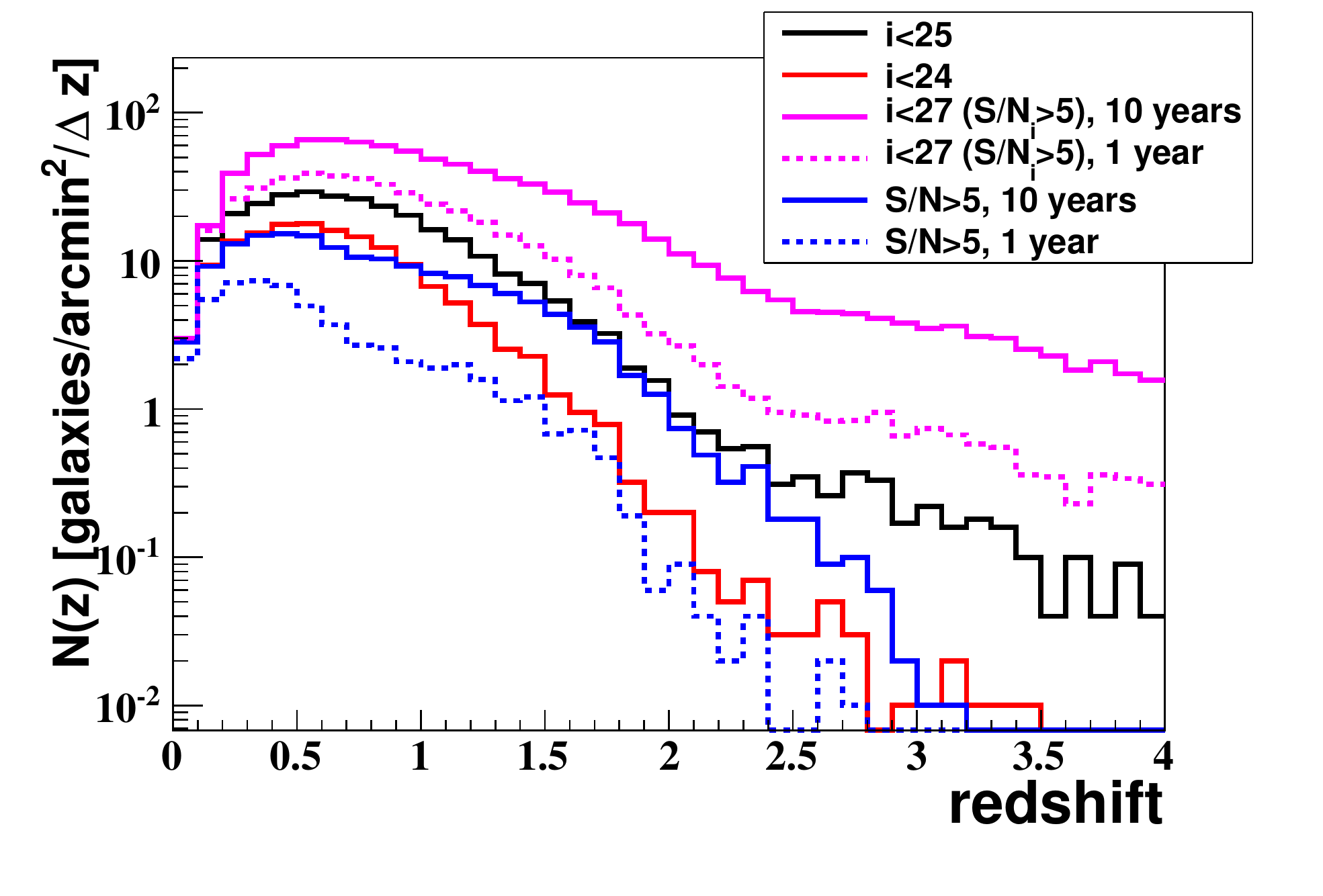}
   \end{tabular}
      \caption{\textit{Left-hand side}: Cumulative distribution of the apparent magnitude in the $i$
      band for the LSST simulation compared to the Sloan Digital Sky
      Survey (SDSS) measurement from the
      Stripe 82 region \citep[cf][]{Abazajian:2008wr}.  The statistical error bars from the data are smaller than
the dots size. \textit{Right-hand side}: Redshift
       distribution of galaxies for the LSST: $i<24$, $i<25$ with $\sigma_i<0.2$ (solid red, solid black curves respectively); $i<27$ with $S/N>5$ in at least the $i$ band for one and ten years of observations (dotted and solid magenta respectively); $S/N>5$ in all bands for one and ten years of observations (dotted and solid blue respectively).
      \label{Fig:HistoMagZspec}}
   \end{figure*}
   

\subsection{Method validation}\label{Sec:MockCatValidation}

\subsubsection{GOODS}

To validate the simulation scheme, we have performed a simulation of the GOODS WFI data\footnote{Data from the Wide Field Imager (WFI) on the 2.2-m MPG/ESO telescope at La Silla.} and compared our results  to the real photometric catalog used for the computation of the $B$-band luminosity functions reported in \cite{Dahlen:2005}. 

The simulated photometric catalog corresponds to an effective solid angle of $1100
~\unit{arcmin^2}$, which is equal to the area covered by the actual data catalog. The simulated redshift and absolute magnitude ranges are respectively $[0,6]$ and $[-24,-13]$. The apparent magnitudes are computed for the WFI $B$-band and $R$-band. The apparent magnitude uncertainty is now given by the distribution of $\sigma_X$ given $m_X$ computed from the real data (see \ref{Sec:SigMagCG}). If $m_{X,s}$ is the simulated apparent magnitude in any $X$-band, the uncertainty is randomly drawn from the distribution $\mbox{Prob}(\sigma_X|m_{X,s})$ found from the data. The observed apparent magnitude and its uncertainty are then simulated as detailed in Sect. \ref{Sec:AppMag}.

Figure~\ref{Fig:GOODSmag} shows the very good agreement of the galaxy
number counts in the $R$-band between the simulation and the real
data, except for the very faint galaxies at $R \ge 25$, where the
selection effect in the real data becomes important. The agreement is
also very good for the $B$-band (not shown here). As displayed in
Fig.~\ref{Fig:GOODSColor}, the distributions of colors from our GOODS
simulation (black lines) reasonably agree with the ones from the real
data (red lines) for bright and faint galaxies.  At bright magnitudes,
the fitted luminosity function seems to predict a larger fraction of
elliptical galaxies than the data. This feature probably comes from
the SED templates and their linear interpolation. We have chosen to
interpolate linearly between the SED templates with an equal number of
steps between each template. With real galaxies, it could easily be
that this is not the case; perhaps, for example, the distribution of
the galaxy's SEDs is not uniform between the El and the Sbc
galaxy type. Instead it could be more probable for an intermediate SED
to be more
similar to the Sbc. This reasoning could explain our excess. Since El
galaxies exist only in significant numbers at low redshift and
photo-z's are well estimated at low redshift, this excess should not
have any impact on our conclusions.  In any case, the overall shape of
the distributions indicates that our simulated photometric catalog represents reality. This is expected because the luminosity functions were computed from the real data sample used for the comparison to the simulation.
 
\subsubsection{CFHTLS}\label{Sec:CFHTLS}

The different photo-z reconstruction methods were tested on real data, namely on galaxies observed both photometrically (from CFHTLS T0005) and spectroscopically (from either VVDS, DEEP2 or zCOSMOS surveys). We have followed the procedure described in detail by \cite{Coupon:2009}.

The CFHTLS T0005 public release contains a photometric catalog of objects observed in five bands
$(u,g,r,i,z)$ from the D1, D2, D3, W1, W3, and W4 fields. Among these, some galaxies are also spectroscopically observed by the VVDS, DEEP2 Redshift Survey, and zCOSMOS surveys. The numbers of galaxies in the CFHTLS fields that were matched to spectroscopic observations are listed in Table \ref{Tab:CFHTLS}. To perform the matching, the smallest angle between each
galaxy in the CFHTLS catalog and a galaxy from the spectroscopic catalogs was computed. Only galaxies for which the angle is less than $0.7~\unit{arcsec}$ (the order of the PSF) are grouped into the spectro-photometric catalog. Since we also required the galaxies to be detected in all CFHTLS bands, the spectro-photometric sample is not as large as direct matching between the catalogs would produce.

The spectroscopic redshift distribution of the spectro-photometric catalog is shown in Fig.~\ref{Fig:CFHTLSredshift}.
A simulation of the CFHTLS data was also performed. This was to enable us to evaluate whether the statistical test described in Sect. \ref{Sec:LR} could be calibrated with a simulation and then applied to real data, for which the spectroscopic redshifts of galaxies are not known. This procedure would be useful if no spectroscopic sample is available to calibrate the prior probabilities or the likelihood ratio statistical test presented in Sect.~\ref{Sec:LR}. The same argument stands for the neural network analysis described in Sect. \ref{Sec:NN}, which could also benefit from a simulated training sample.

Because the selection function of the spectro-photometric catalog is not only based on
the detection threshold, the redshift and color distributions of the simulation and real data cannot
be rigorously compared. The photometric selection criteria is not homogeneous over this data sample due to the differing selection of VVDS, DEEP2, and zCOSMOS, and, therefore, does not match the simulation. However, we can qualitatively compare the color distribution domains of galaxies from the CFHTLS catalog with the ones obtained from the simulations.


Figure~\ref{Fig:CFHTLScolor} shows two color-color distributions,
$(u-g)$ vs $(g-r)$ and $(g-r)$ vs $(r-i)$, for the CFHTLS data, on
which we have superimposed (red) the distribution obtained from
calculating the \textit{expected} magnitudes, given the fitted values
of photo-z, galaxy template, and reddening, which are obtained by
running the photo-z algorithm on the data (as described in
Sect. \ref{Sec:PhotozTemplateFittingMethod}).  It can be seen that
there is a satisfactory agreement between the domains covered by the
data and our simulations, indicating that our templates represent the
galaxies well. We have also checked the compatibility of color
variations with the redshift for data and simulations, especially for early and late type galaxies.


\setlength{\tabcolsep}{0.1cm}
\renewcommand{\arraystretch}{1.4} 
 \begin{table}[t]
 \caption{Values of the fit parameters of Eqs.~\ref{Eq:PTI} and ~\ref{Eq:PzTI} . }
\begin{center}
\begin{tabular}{cccc} 
Spectral family: & $Early$ & $Late$ & $Starburst$  \\\hline \hline
\multicolumn{4}{c}{LSST} \\ \hline 
$\alpha$ &  $2.97\pm 0.01$ & $1.88\pm 0.00$ &$1.32\pm0.00$  \\
$z_0$ & $0.00\pm 0.00$ & $0.12\pm 0.02$ & $0.00\pm
 0.00$ \\ 
$k_m$ & $0.13\pm 0.00$ & $0.08 \pm 0.00$ &  $0.08 \pm 0.00$ \\
$f_t$ & 0.22 & 0.41 & \ldots\\
$k_t$  &$0.27\pm 0.00$& $0.10\pm 0.00$& \ldots \\ \hline  
   \multicolumn{4}{c}{CFHTLS} \\ \hline
$\alpha$ & $3.51 \pm 0.02$ & $2.77\pm0.01$ & $1.89 \pm 0.00$ \\
$z_0$ & $ 0.00\pm0.00$ & $0.18\pm 0.02$ & $0.02\pm0.01$ \\
$k_m$ & $0.13\pm 0.00$ & $0.02\pm0.01$ & $0.08\pm0.00$ \\
$p_m$ &0 & $0.18\pm0.02$& $0.02 \pm 0.01$ \\
$\beta$ &\ldots &$1.71\pm 0.08$ &  $3.12 \pm 0.23$ \\
$f_t$ & 0.23 & 0.43 & \ldots \\
$k_t$ & $0.27\pm 0.00$ & $0.11\pm 0.00$ & \ldots \\
 \end{tabular}
\end{center}
\label{Tab:FitParamPZT}
\end{table}
\renewcommand{\arraystretch}{1.1} 

\begin{figure*}
 \includegraphics[scale=0.85]{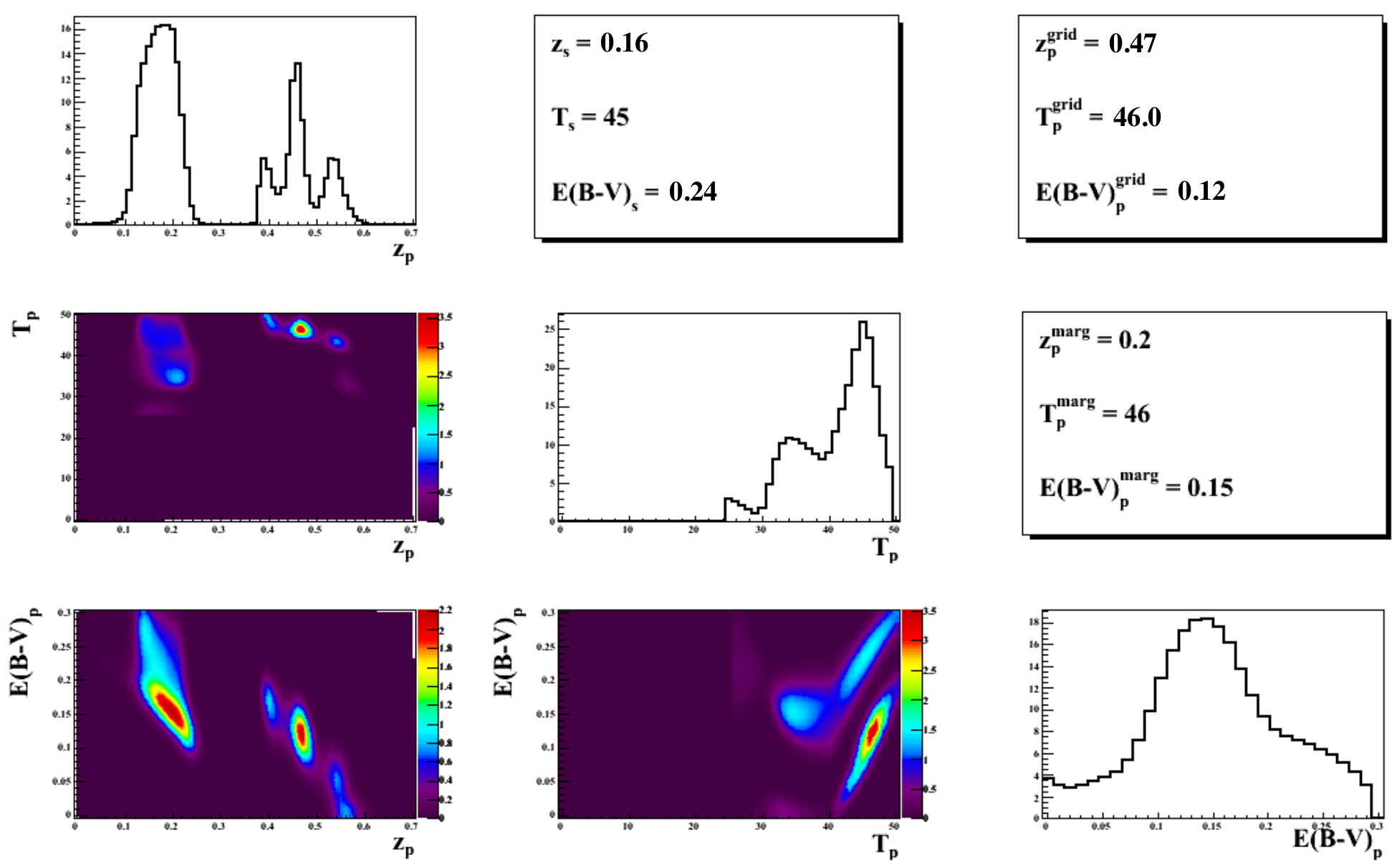}
  \caption{ Example of photometric computation for a simulated galaxy
  observed with the LSST in six bands at $5\sigma$ for ten years of observation. The
  2D distributions correspond to the posterior probability density functions
  marginalized over the remaining parameter, and the 1D distributions correspond
  to the posterior probability density functions marginalized over the two
  remaining parameters. The top middle box corresponds to the value of the input
  parameters. On the top right-hand panel, the index $grid$ denotes the
  parameters that maximize the 3D posterior probability density function on the grid, and on the
  middle right-hand panel, the index $marg$ denotes the parameters that maximize
  the 1D posterior probability density functions. The size of
  the grid cells has been reduced, and the $z_p$ axis has been
  shortened compared to the size of the grid that is usually used to compute the likelihood function.
  \label{Fig:Landscape}}
\end{figure*}

\subsubsection{LSST forecasts}

The simulated LSST sample considered here is the same as the one used
in the companion paper Abate et al. \textit{in prep.}. It is generated
with a solid angle of $7850~\unit{deg^2}$ and a redshift range of
$[0.1,6]$. This upper redshift limit is large enough to include all
observable galaxies. The total number of galaxies is $\sim
8\times10^9$. For the purpose of this paper, namely the reconstruction
of photometric redshifts, such a large number of galaxies is not
necessary. Therefore, the following analysis is done with a smaller
subsample of galaxies, as seen Sect.~\ref{Sec:LSSTSim}.

The expected cumulative number counts per unit of solid angle and per $i$ band apparent magnitude
for the LSST is shown on the left-hand side of
Fig.~\ref{Fig:HistoMagZspec}, and is compared  to the SDSS
measurements made on the Stripe 82 region
\citep[cf][]{Abazajian:2008wr}.  The LSST galaxy count is generally
below that of SDSS with a more pronounced effect at high and low end
magnitudes ($m\sim26$ and $m\sim20$). The discrepancy
(Fig.~\ref{Fig:HistoMagZspec}, left-hand side) might be due to a
systematic zero point magnitude error in the simulation due to an imperfect filter model, excessive absorption (extinction) in the simulation, or uncorrected differences between the selection of SDSS galaxies and GOODS galaxies.

\begin{figure*}
   \centering
   \begin{tabular}{cc}
	\includegraphics[width=14cm,height=8cm]{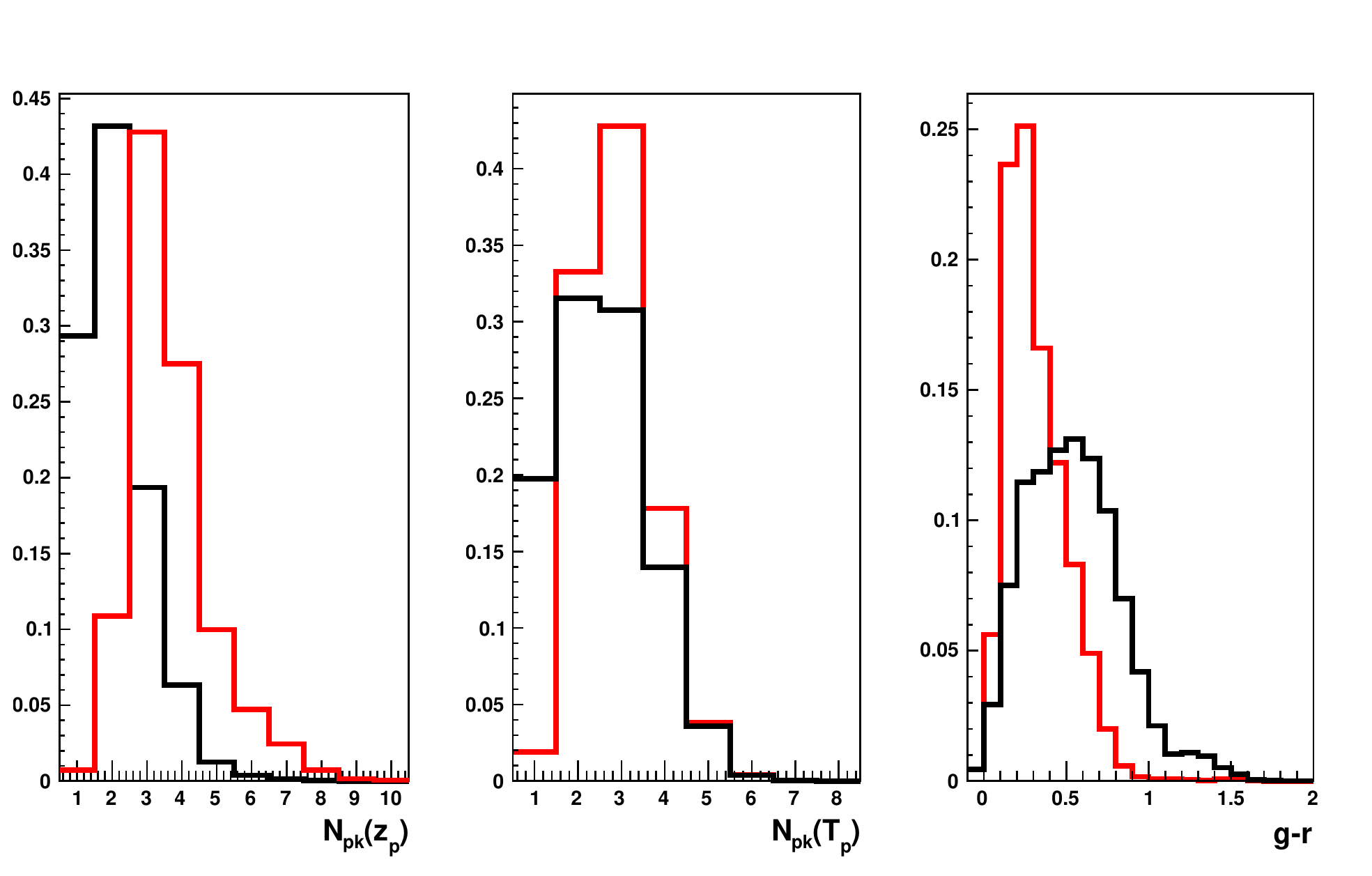}	
	\end{tabular}
      \caption{Probability density function of the reduced variables $N_p(z_p)$, $N_p(T_p)$, and
      $g-r$. The black lines correspond to $P(\mu_i|G)$ and the red lines to $P(\mu_i|O)$.}
         \label{Fig:LRvar}
\end{figure*}

The expected number of galaxies per unit of solid angle and per redshift is shown on the right-hand side of
Fig.~\ref{Fig:HistoMagZspec} for different cuts: $i$ magnitude cuts of
$i<24$, $i<25$, and $i<27$ with $\sigma_X<0.2$ for all $X$ bands for
both one (dotted lines) and ten years (solid lines) of
observations. The LSST gold sample is defined as all galaxies with
$i<25.3$, so that it will contain around 4 billion galaxies that have
up to a redshift of 3 and is expected to produce high quality photometric redshifts.

The number of galaxies with S/N$>5$ in all six bands at ten years is fairly small because these constraints are strong for the low acceptance $u$ and $y$ bands.  Low S/N in the $u$-band is expected from both its shallower depth and dropout galaxies at higher redshift, where hydrogen absorption removes all flux blueward of the Lyman break, and non-detections above $z>3$ are expected. The low S/N in the $y$-band is expected just from its shallower depth.


\section{Enhanced template-fitting method}\label{Sec:PhotozTemplateFittingMethod}

\subsection{Maximum of the posterior probability density function}

In this section, our template-fitting method for estimating photometric redshifts is presented. The algorithm follows the approach
developed by \cite{Ilbert:2006dp}, \cite{Bolzonella:2000js}, or \cite{Benitez:1998br}. 

Basically, the template-fitting method consists in finding the photo-z $z_p$, the SED template $T_p$, the color excess $E(B-V)_p$\footnote{Subscript $p$ refers to the best-fit parameters.}, and the SED normalization $N$ that give the fluxes in each band that best fit to the observed values. Following \cite{Benitez:1998br}, the normalization parameter is marginalized over, so that the parameters of interest are given by the minimum of 
a $\chi^2$ statistic, whose expression is given later in this section.

A Bayesian prior probability can be used to improve the \photoz reconstruction. It is defined as  the probability of
having a galaxy of redshift $z$ and type $T$, given its apparent magnitude. It was introduced by
\cite{Benitez:1998br}.  Bayes' theorem indicates that this probability can be expressed as
the product of the probability of having a galaxy of type $T$ given the apparent magnitude $i$,
$P(T|i)$ times the probability of having a galaxy of redshift $z$, given the type and the apparent
magnitude, $P(z|T,i)$. In other words,
\begin{eqnarray}
P(z,T|i) = P(z|T,i)\times P(T|i)~.
\end{eqnarray}
The two terms are well described
by the functions,
\begin{eqnarray}\label{Eq:PTI}
 P(T|i) = f_t e^{-k_t(i-20)}~,
\end{eqnarray}
and 
\begin{eqnarray}\label{Eq:PzTI}
P(z|T,i) & \propto& z^{\alpha} \exp\left[  -\left(\frac{z}{z_m
}\right)^\alpha \right]~,\nonumber \\
z_m &=& z_0+k_m(i-16) + p_m(i-16)^{\beta}~.
\end{eqnarray}
Here, $T$ represents the spectral family (broad type) instead of the spectral type (the exact SED).
That is, galaxies with a spectral type that is lower than 5 belong to
the \textit{early}-type, those with spectral type between $6$ and $25$
belong to the \textit{late}-type and the rest belong to the
\textit{starburst}-type. This parametrization follows a general model
for galaxy number counts with redshift and is improved to account for
higher redshifts in CFHTLS by the addition of the $p_m$ and $\beta$
parameters, as compared to Eqs. 22, 23 and 24 in \cite{Benitez:1998br}. 
The parameters of $P(T|i)$ and $P(z|T,i)$ are found from fitting
Eq.~\ref{Eq:PTI} and Eq.~\ref{Eq:PzTI} to the simulated magnitude-redshift
distributions for both the LSST and CFHTLS surveys. The value of the parameters in
Eq.~\ref{Eq:PTI} and Eq.~\ref{Eq:PzTI} are given by Table \ref{Tab:FitParamPZT}.   The fitted $p_m$ parameter for the LSST is compatible with zero. While it is meaningful for CFHTLS (see Table \ref{Tab:FitParamPZT}), it is therefore set to zero when representing the prior probability for the LSST simulations.
There is no value for the parameters $f_t$ and $k_t$ for Starburst galaxies, whose probability distribution is set by the condition that the sum of probabilities over all galaxy types must be equal to 1.

 When prior probabilities are taken into account, the $\chi^2$ is extended and is defined as
\begin{eqnarray}\label{Eq:chi2CFHTLS}
\chi^2(z, T, E(B-V)) &=&
	\displaystyle\sum_{X}^{N_{band}}  \left(\frac{ F_{X,obs} -
	    A F_{X,pr}(z,T,\ebv)}{\sigma\left(F_{X,obs}\right)} \right)^2 \\
	&& -	2\ln\left( B \right) 2\ln\left(P(T|i)\times P(z|T,i)\right)~,\nonumber
\end{eqnarray}
where $F_{X,obs}$ is the observed flux in the $X$-band;
$F_{X,pr}(z,T,E(B-V))$ is the expected flux; $N_{band}$ is equal to $5$ for
the CFHTLS and is equal to $6$ for the LSST; and $\sigma\left( F_{X,obs}\right)$ is the
observed flux uncertainty. The terms $A$ and $B$ come from the analytical
marginalization over the normalization of the SEDs; they are defined as follows:
\begin{eqnarray}
A = \sum_{X = 1}^{N_{band}}
\frac{F_{X,obs}F_{X,pr}}{\sigma^2(F_{X,obs})} /  \sum_{X =
1}^{N_{band}}
\frac{F_{X,pr}F_{X,pr}}{\sigma^2(F_{X,obs})}\nonumber \\
B = \sum_{X = 1}^{N_{band}}
\frac{F_{X,pr}F_{X,pr}}{\sigma^2(F_{X,obs})}.
\end{eqnarray}

In the following, the 3D posterior pdf is defined as $\mathcal{L}~=~\exp\left[ -\chi^2/2
\right]$. It is computed for each galaxy on a 3D grid of $100\times25\times5$ nodes
in the  ($z$, $T$, $\ebv$) parameter space. The values of the parameters $z,T,\ebv$ lie in the
intervals: $[0,4.5]$, $[0,50]$, and $[0,0.3]$ respectively.  Since we are controlling the domain of possible parameter values to match the ranges used to make the simulation,
we are probably reducing the number of possible degeneracies in $z,T,\ebv$ space.
The SED library used for the photo-z reconstruction is the
CWW+Kinney library, as described in Sect. \ref{Sec:SED}.
However, the templates have been optimized following the technique developed by
\cite{Ilbert:2006dp} when considering
the CFHTLS spectro-photometric data; therefore, they naturally match the data better.

\begin{figure}
  \centering
  \includegraphics[width=9cm]{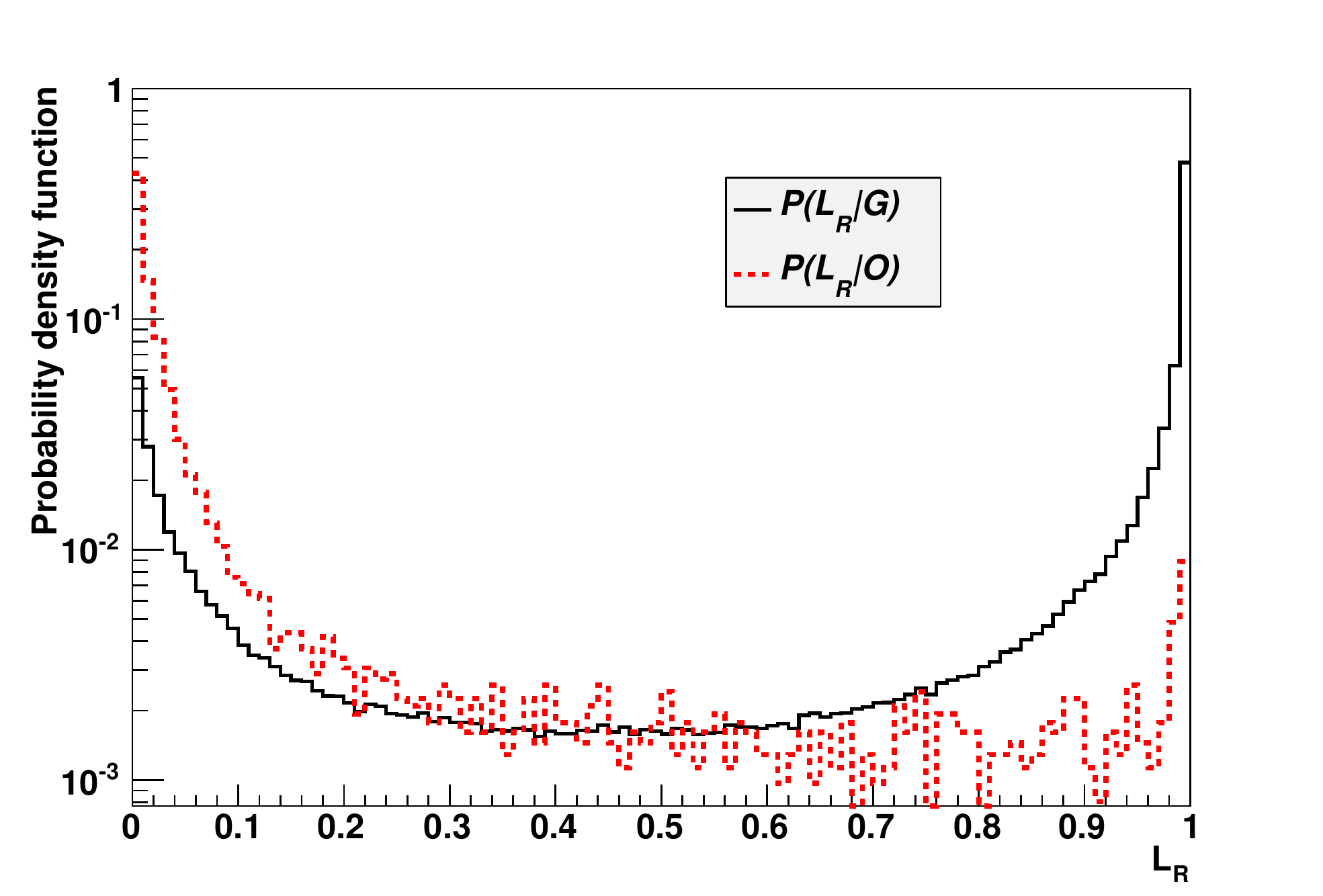}
  \caption{Likelihood ratio distribution from the LSST simulation training sample. The probability density
    $P(L_R|G)$ in solid black and $P(L_R|O)$ in dashed red.}
  \label{Fig:LR}
\end{figure}

The probability distribution is a function of three parameters. To
derive the information on just one or two of the parameters, we
integrate the distribution over all values of the unwanted
parameter(s) in a process called marginalization.  The marginalized 2D
probability density functions of the parameters $(z,T)$, $(z,\ebv)$,
$(T,\ebv)$ and the marginalized 1D \pdf of each parameter are computed
in this manner. Figure~\ref{Fig:Landscape} shows an example of these
\pdfs for a galaxy with a true redshift $z_s = 0.16$, true type $T_s =
45$ (starburst), and true excess color $\ebv_s = 0.24$. In many cases,
the 3D posterior pdf is highly multimodal: therefore minimizing the $\chi^2$ with traditional algorithms, such as \textsc{Minuit}, often misses the global minimum. A scan of the parameter space is better suited to this application. Even a Markov Chain Monte Carlo (MCMC) method, which is usually more efficient than a simple scan, is not well suited to a multimodal 3D posterior pdf. Moreover, the production of the chains and their analysis in a 3D parameter space is slower than a scan. This example, where $(z_p -  z_s)/(1+z_s) = 0.27$, corresponds to a catastrophic reconstruction. In this case, the parameters $\left(z_{pk}^{grid},T_{pk}^{grid},\ebv_{pk}^{grid}\right)$ that maximize the 3D posterior pdf grid do not coincide with the ones that maximize the individual posterior probability  density functions, namely the parameters $\left(z_{pk}^{marg},T_{pk}^{marg},\ebv_{pk}^{marg}\right)$ . 

\begin{figure}
  \centering
  \includegraphics[width=8cm]{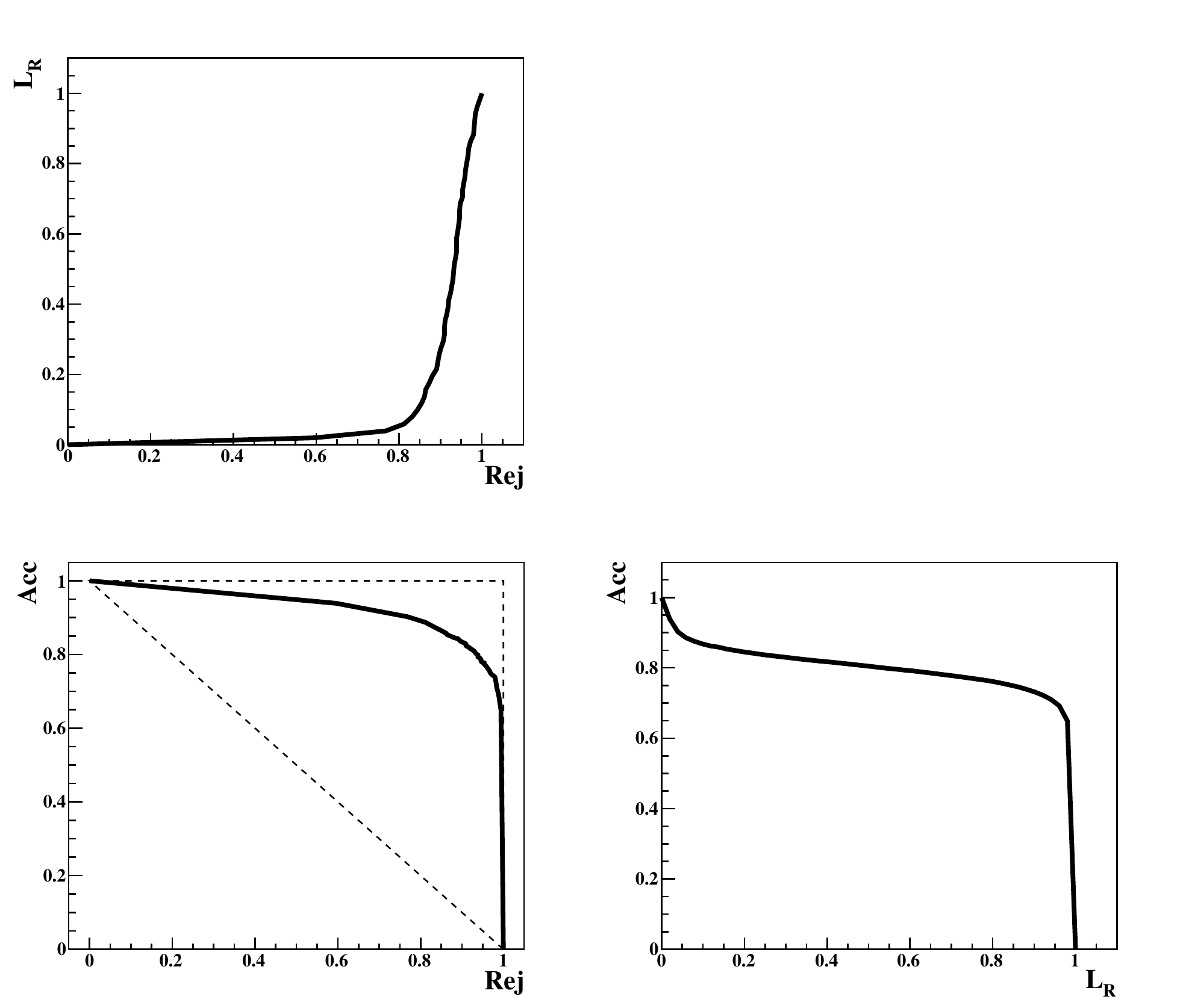}
  \caption{From the LSST simulation training sample. Top panel: Evolution of the rejection
    \textit{vs.} $L_R$. Bottom left-hand panel: Evolution of the acceptance \textit{vs.} the
    rejection.  Bottom right-hand panel: Evolution of the acceptance
    \textit{vs.} $L_R$.}
  \label{Fig:AccRej}
\end{figure}

\subsection{Statistical test and rejection of outliers}\label{Sec:LR}

In this section, we outline a statistical test that aims at rejecting some of
the outlier galaxies, where $|z_p-z_s|/(1+z_s)>0.15$. It is based on the
characteristics of the 1D posterior probability density functions $P(z_{pk})$, $P(T_{pk})$, and
$P(E(B-V)_{pk})$. The test is calibrated with a training sample, for which the true
redshift is known (or the spectroscopic redshift in the case of real data). 
We use a subsample of our data (simulated or CFHTLS) for training. See Sect. \ref{Sec:cfhtresults} and \ref{Sec:LSSTSim} for full details.
In the following, the LSST simulation for ten years of observation is considered
to illustrate the method.

\subsubsection{The probability density function characteristics}

The variables considered to establish the statistical test are the following:

\begin{itemize}
	\item The number of peaks in the marginalized 1D posterior probability density functions
	denoted by $N_{pk}(\theta)$, where $\theta$ is either $z$, $T$, or $\ebv$;
	\item When $N_{pk}>1$, the logarithm of the ratio between the height of the secondary peak over the primary peak in the 1D posterior probability
	density functions, denoted by $R_L(\theta)$;
	\item When $N_{pk}>1$, the ratio of the probability associated with the secondary
	peak over the probability associated with the primary peak in the 1D posterior probability
	density functions, denoted by $R_{pk}(\theta)$. The probability is defined as the integral between two
	minima on either side;
	\item The absolute difference between the value of $z_{pk}$ and $z_{pk}^{marg}$,
	as denoted by $D_{pk} = |z_{pk} - z_{pk}^{marg}|$, where $z_p^{marg}$ is the redshift
	that maximizes the posterior probability density function $P(z)$;
	\item The maximum value of $\log(\mathcal{L})$;
	\item The colors, $\vec{C} = (u-g,g-r,r-i,i-z)$, in the case
          of CFHTLS, with an extra $z-y$ term in the case of the LSST.
\end{itemize}

We denote the galaxies that are considered as outliers by $O$ and the
galaxies for which the redshift is well reconstructed by $G$ in the following way:
\begin{itemize}
\item O: $|z_p-z_s|/(1+z_s)>0.15$ 
\item G: $|z_p-z_s|/(1+z_s)<0.15$.
\end{itemize}
The set of variables defined in the list above are denoted by the vector $\vec{\mu}$. From a given training sample, we compute the distributions $P(\mu_i|O)$ and $P(\mu_i|G)$. For convenience, we adopt reduced variables that are renormalized to lie between 0 and 1. Distributions of some of the reduced variables are plotted in Fig.~\ref{Fig:LRvar}. It is clear that the distributions $P(\mu_i|G)$ and $P(\mu_i|O)$ are different. The probability that an outlier galaxy $O$ presents more than three peaks in its posterior probability density function $P(z_p)$ is larger than for a well reconstructed galaxy $G$. A combination of these different pieces of information leads to an efficient test to distinguish between good and catastrophic reconstructions.


\subsubsection{Likelihood ratio definition}
To combine the information contained in the densities $P(\mu_i|G)$
and $P(\mu_i|O)$, we define the likelihood ratio variable $L_R$:
\begin{eqnarray}
L_R(\vec{\mu}) = \frac{P(\vec{\mu}|G)}{P(\vec{\mu}|G) + P(\vec{\mu}|O)}~,
\end{eqnarray}
where 
\begin{eqnarray}
P(\vec{\mu}|G) = \prod_{i =1}^{N_{\mu}} P(\mu_i|G)~,\\
P(\vec{\mu}|O) = \prod_{i=1}^{N_{\mu}} P(\mu_i|O)~,
\end{eqnarray}
where $N_{\mu}$ is the number of components of $\vec{\mu}$. Here, the variables
$\mu_i$ are assumed to be independent. The correlation matrix of the $\mu_i$'s indeed shows
a low correlation between the parameters. We approximate the two probabilities, $P(G|\vec{\mu})$ and $P(O|\vec{\mu})$, as the product of $P(\vec{\mu}|G)$ and $P(\vec{\mu}|O)$, neglecting the correlations, as our aim is just to define a variable for discriminating the two possibilities.
The probability density functions, $P(L_R|G)$ and $P(L_R|O)$, are computed from the training sample and are displayed in Fig.~\ref{Fig:LR}.  The results shown here are from the LSST simulation for ten years of observations with $m_X<m_{5,X}$. As expected, the two distributions are very different.

\begin{figure}
  \centering
  \includegraphics[width=9cm]{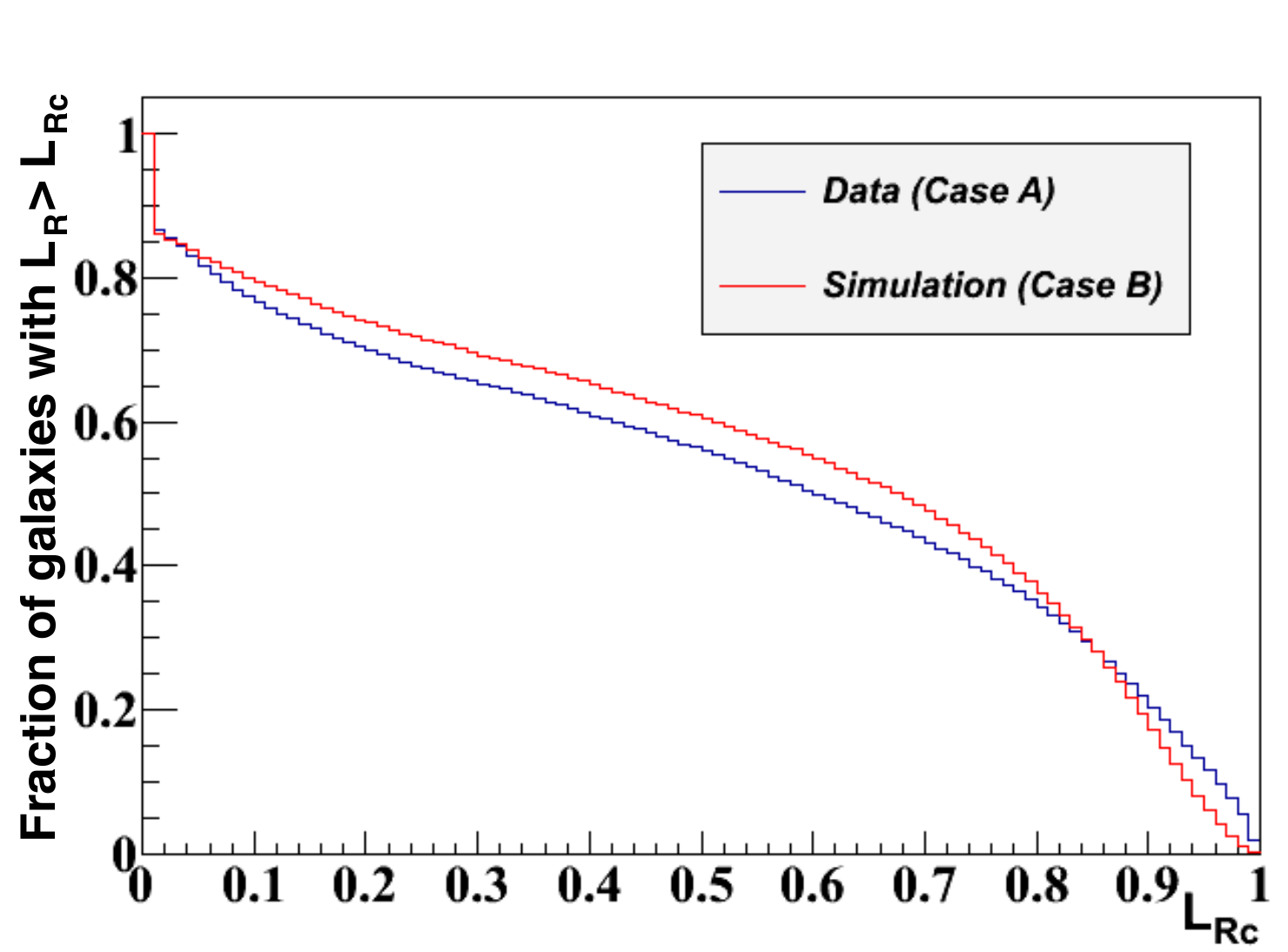}
  \caption{Histogram of the number of galaxies from the CFHTLS sample with
    $L_R\geq L_{R,c}$ as a function of $L_{R,c}$. The black curve has been
    obtained with densities $P(\vec{\mu}|G)$ and $P(\vec{\mu}|O)$,
    as computed from the CFHTLS data themselves, whereas the red curve relies on
    densities obtained from the CFHTLS simulation.}
  \label{Fig:NGalLR}
\end{figure}

\begin{figure*}
  \centering
  \includegraphics[width=8.9cm]{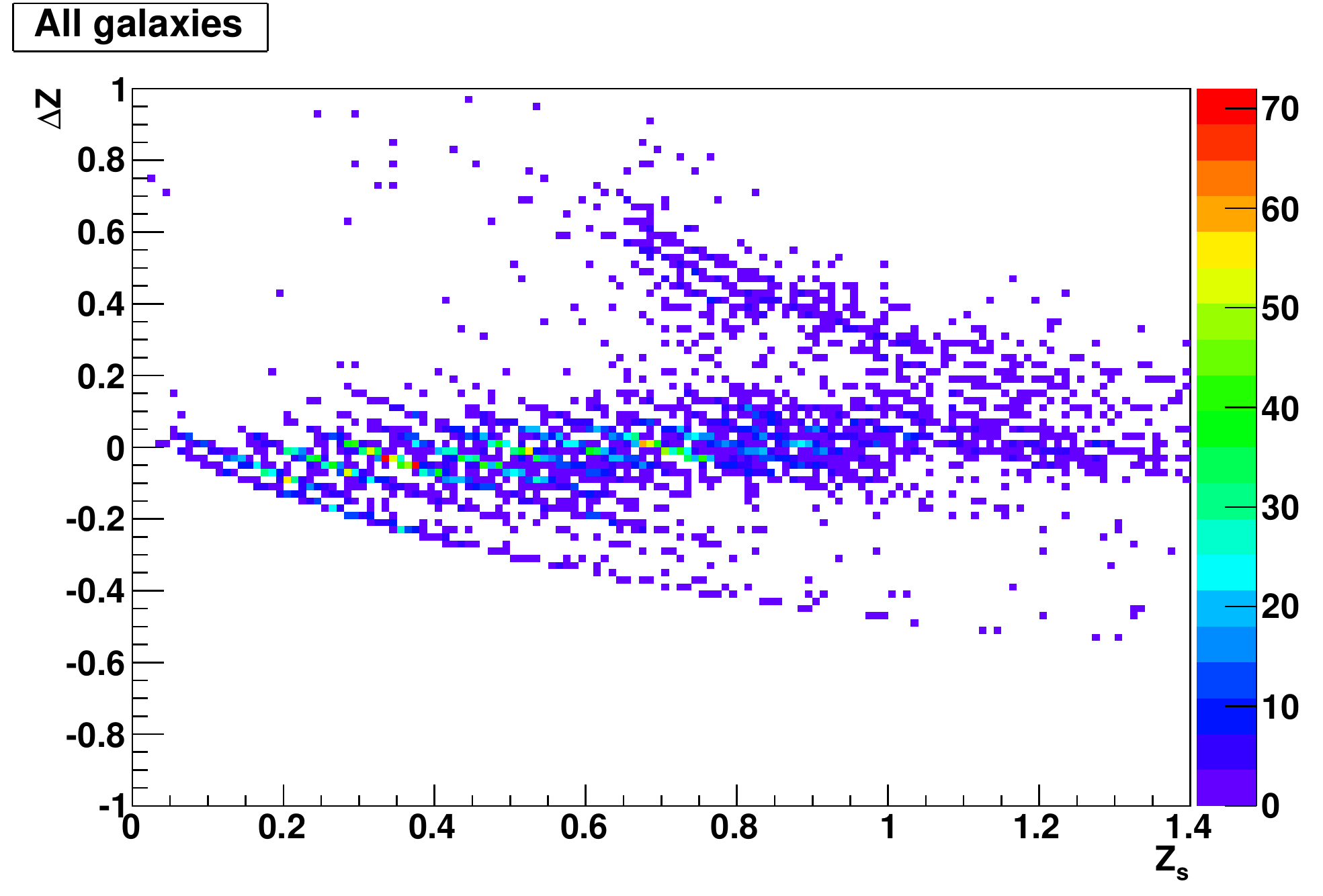}
  \includegraphics[width=8.9cm]{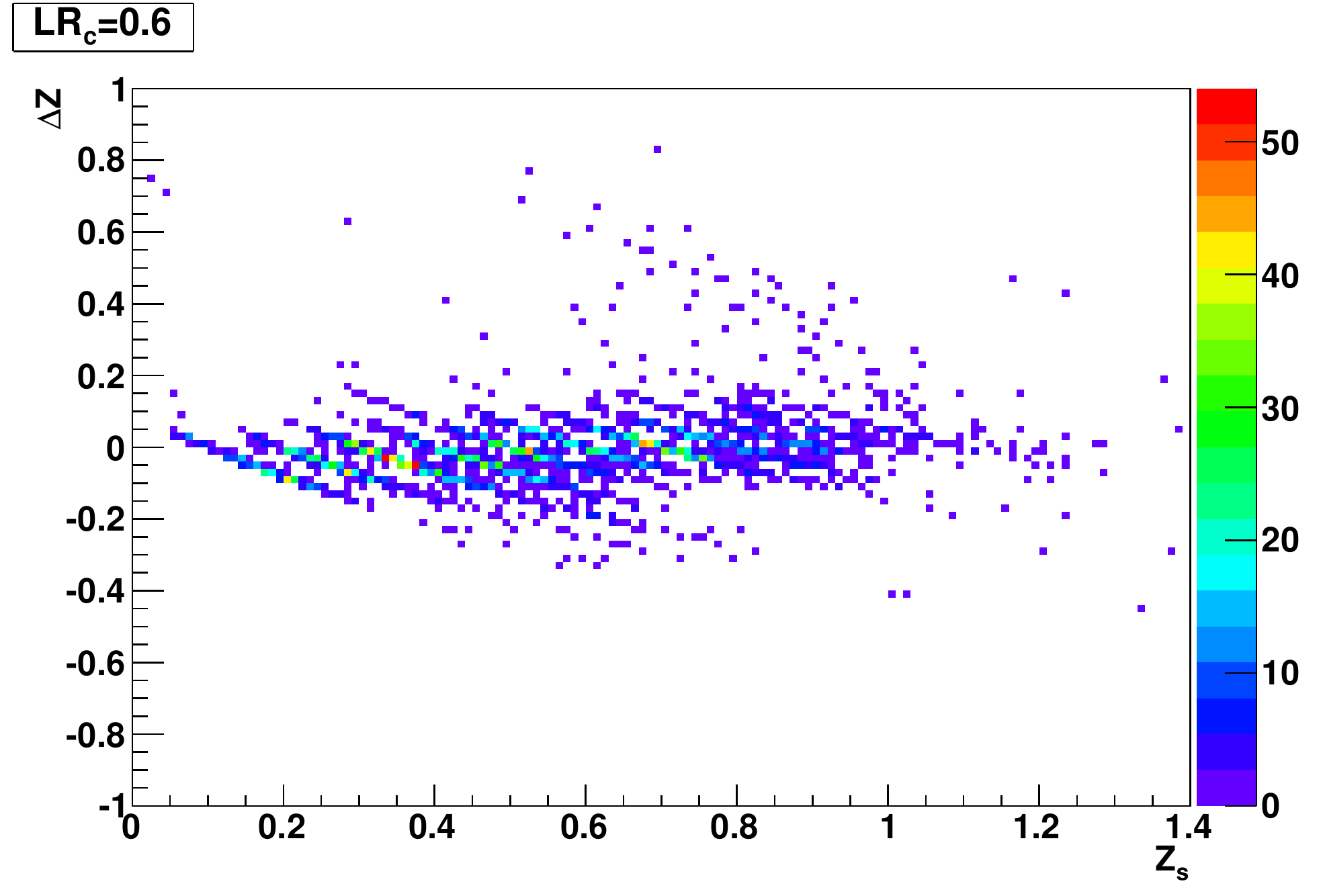}
  \includegraphics[width=8.9cm]{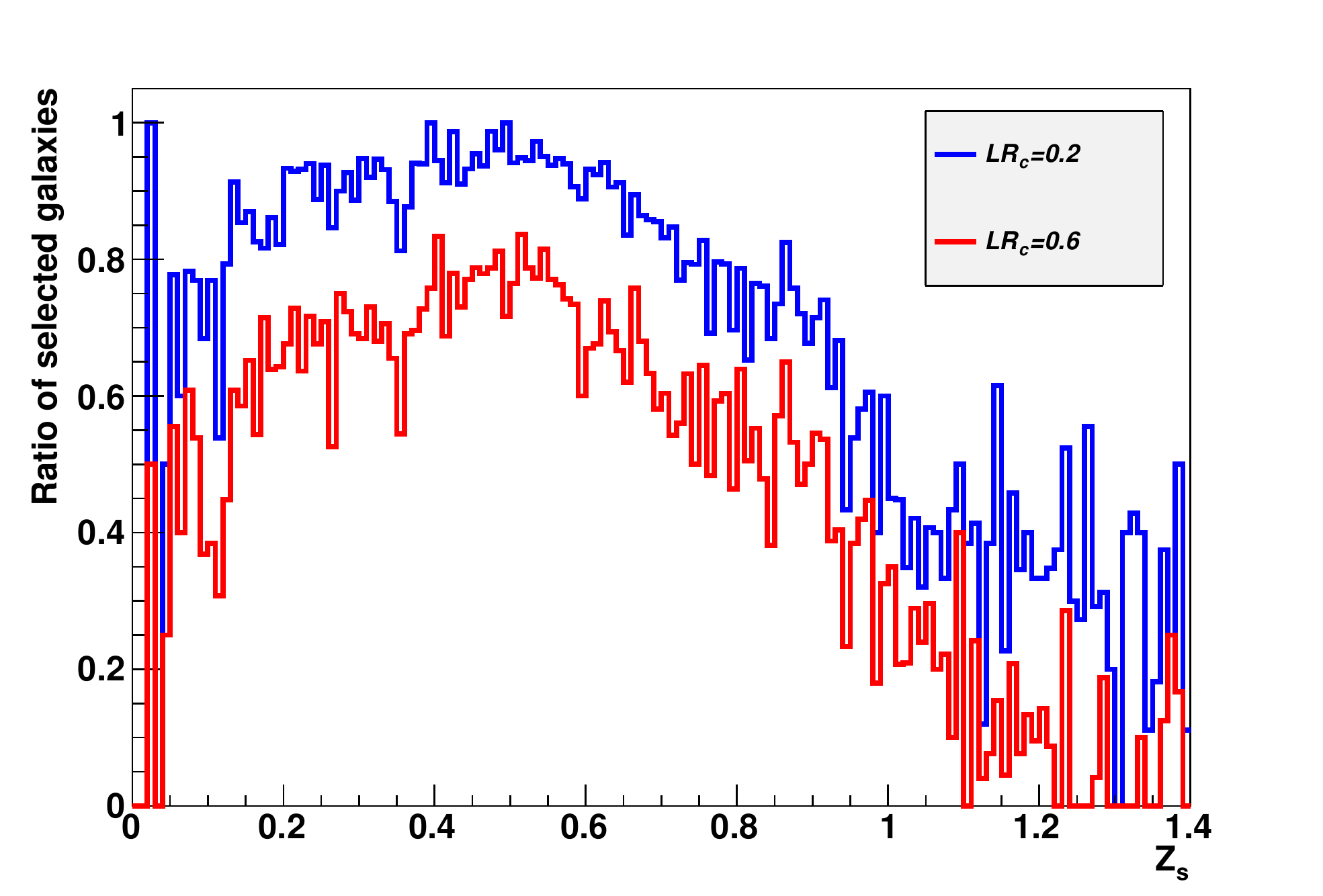}
  \includegraphics[width=8.9cm]{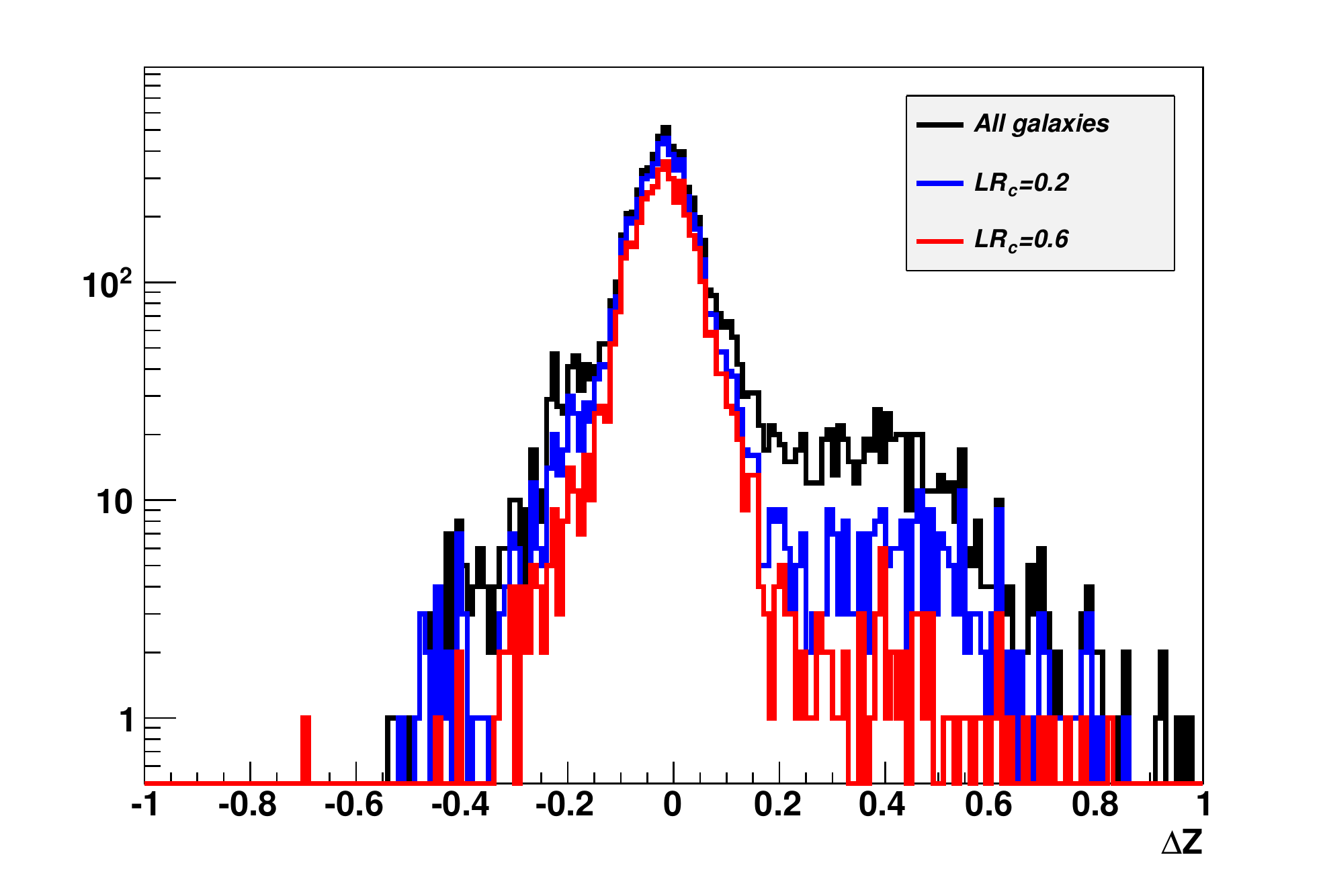}
  \caption{CFHTLS spectro-photometric data: \textit{Top}:  $\Delta z =
    z_s - z_p$ as a function of $z_s$ distribution for all galaxies (left) and for galaxies satisfying $L_R> L_{R,c} = 0.6$. \textit{Bottom left}: Fraction of galaxies satisfying $L_R >  L_{R,c} = 0.2$ (blue), $0.6$ (red). \textit{Bottom right}: $\Delta z$ distribution for all galaxies and for galaxies satisfying the various $L_R$ cuts.
  \label{Fig:DeltazParameterLR_CFHTLS} }
\end{figure*} 

\begin{figure*}
  \centering
  \includegraphics[width=8.9cm]{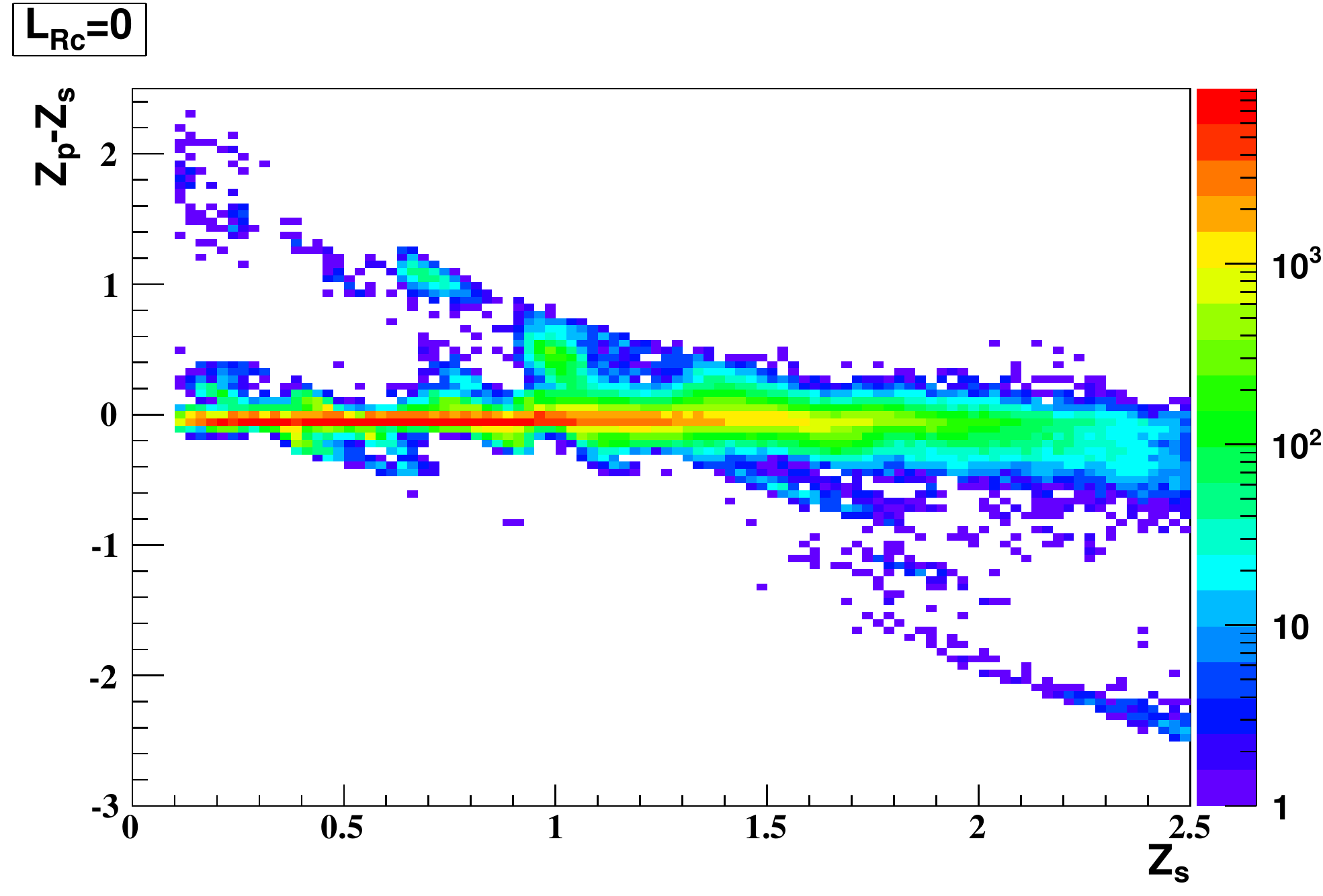}
  \includegraphics[width=8.9cm]{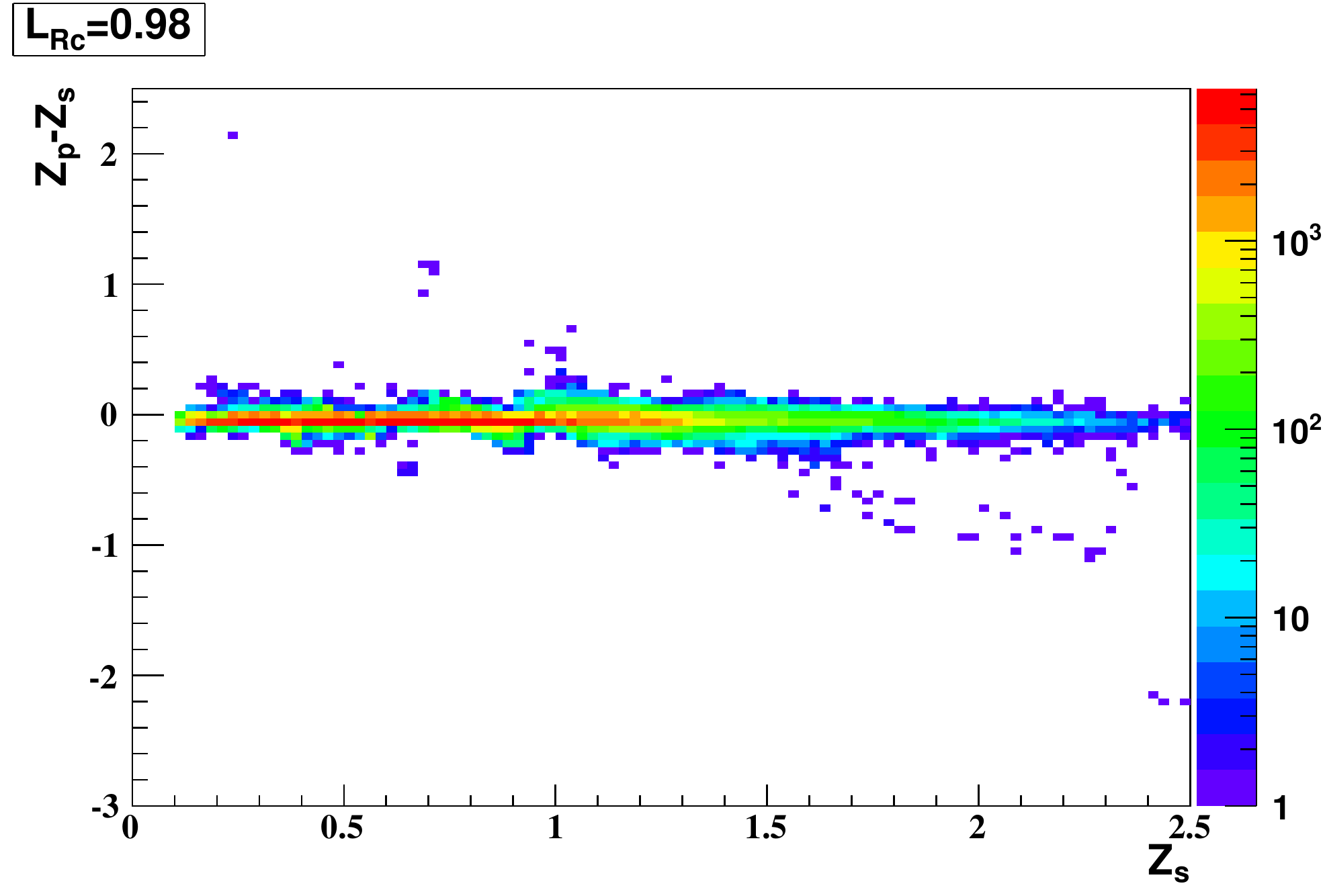}
  \caption{Distribution of $z_p - z_s$ versus $z_s$ for a simulated
    LSST catalog for all galaxies (left) and for galaxies with a likelihood ratio $L_R$ greater than 0.98 (right).}
  \label{Fig:DeltazLR}
\end{figure*}

The quality of a discrimination test, such as $L_R>L_{R,c}$, can be quantified by the acceptance $Acc$ and rejection $Rej$ rates:
\begin{eqnarray}
Acc(L_R) = \int_{L_R}^{1} P(L_R'|G)dL_R'~,\\
Rej(L_R) = \int^{L_R}_{0} P(L_R'|O)dL_R'~.
\end{eqnarray}
The evolution of $Acc$ and $Rej$ as functions of $L_R$ and $Acc$ as a function
of $Rej$ is displayed in Fig.~\ref{Fig:AccRej}. The larger the difference between the curve $Acc$
\textit{vs.} $Rej$ and the curve $Acc = 1-Rej$, the higher the rejection power.  Figure~\ref{Fig:AccRej} shows that the method should work because the solid line lies far from the diagonal dotted line in the bottom left panel. A high value of $L_R$ is necessary to discard outliers; however, it should be chosen so a minimum of well-reconstructed galaxies are removed. 
The plots in Figs. \ref{Fig:DeltazParameterLR_CFHTLS} and \ref{Fig:DeltazLR}, discussed below, show that there is a significant improvement when a cut on $L_R$ is applied.

\section{Photo-z performance with template-fitting} \label{Sec:photoz}

\begin{figure*}
  \centering
  \includegraphics[width=9cm]{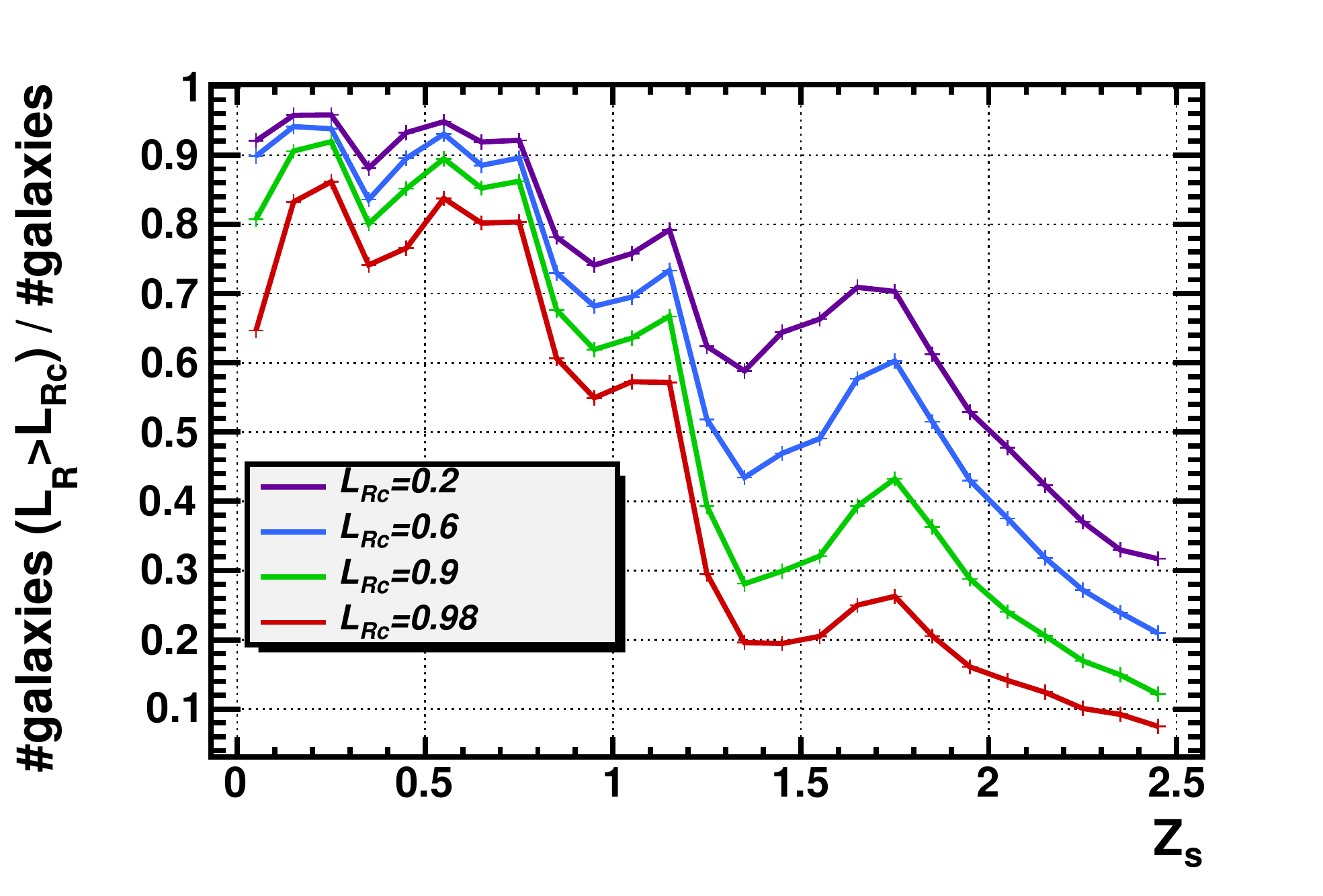}
  \includegraphics[width=9cm]{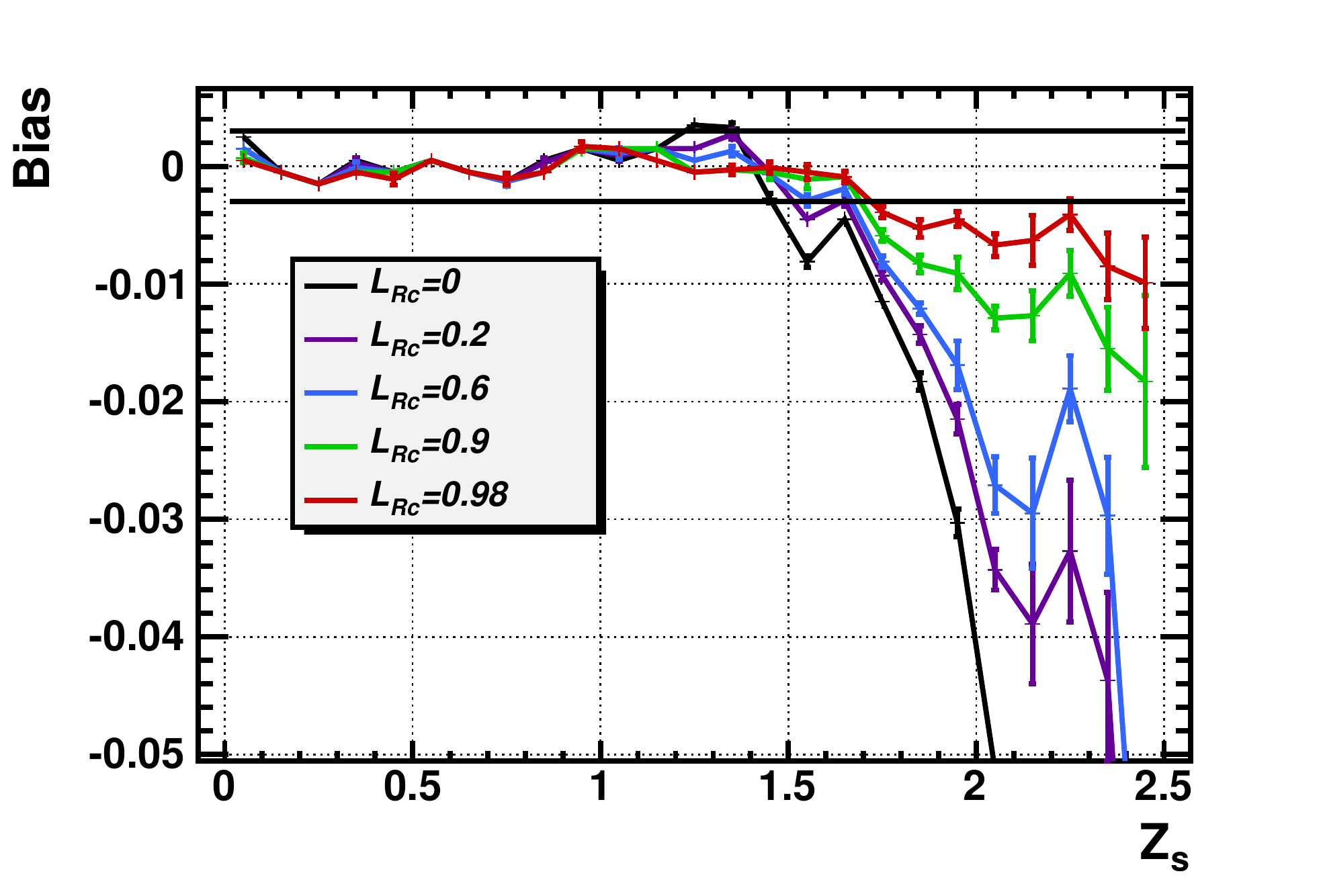}
  \includegraphics[width=9cm]{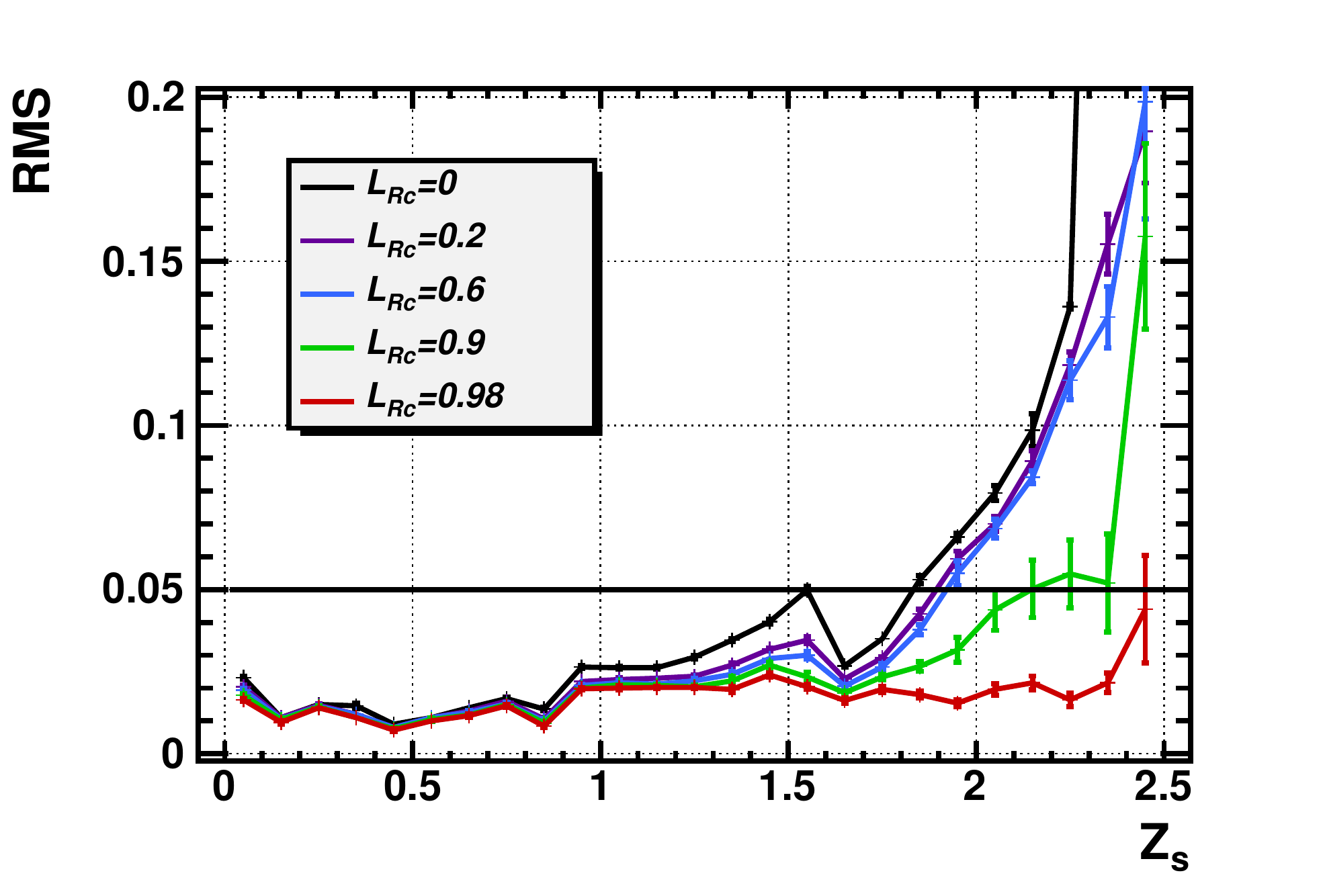}
  \includegraphics[width=9cm]{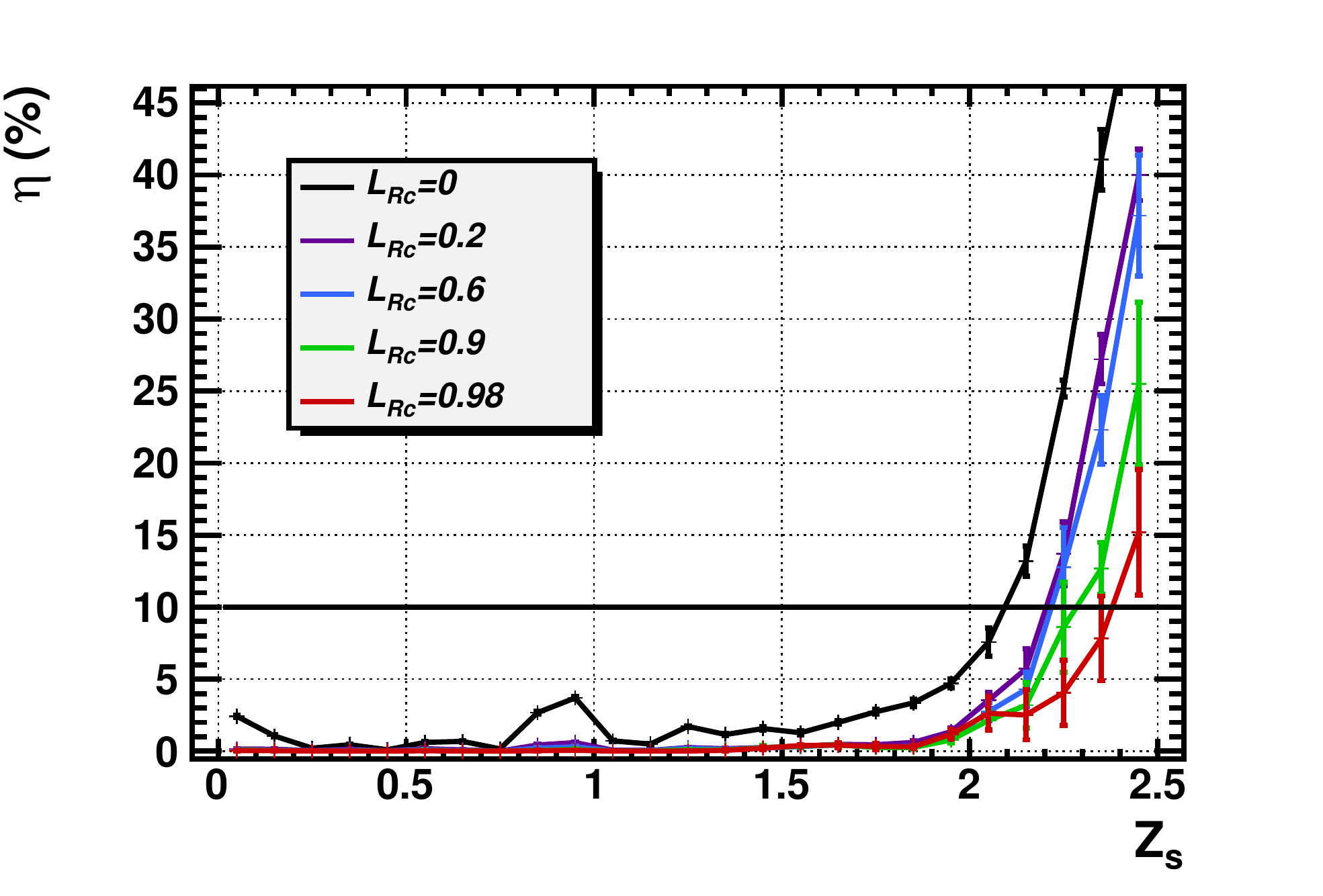}
  \caption{LSST $L_R$ selection.  \textit{Top}: Evolution of the fraction of detected galaxies satisfying $L_R>L_{R,c}$ and the bias with $z_s$.  
    \textit{Bottom}: Evolution of the rms and $\eta$ as a function of $z_s$. The thick black lines represent the LSST requirements given in Table \ref{Tab:PZreq}. Ten years of observations with the LSST is assumed. The values of $L_{R,c}$ are reported in the legend.}
  \label{Fig:DeltazParameterLR}
\end{figure*}

In the following sections, the ability of the statistical test, which
is based on the $L_R$ variable, to construct a \textit{robust sample}
of galaxies with well-reconstructed redshifts is investigated in more
detail for both the CFHTLS spectro-photometric data and the LSST
simulation. The efficiency of the photo-z reconstruction is quantified
by studying the distribution of $(z_p - z_s)/(1+z_s)$ through the following:
\begin{itemize}
\item \textbf{bias}: median that splits the sorted distribution in two equal samples.
\item \textbf{rms}: the interquartile range (IQR)\footnote{The interquartile range is the interval spanning the second and third quartiles.}. If the distribution is Gaussian, it is approximately equal to $1.35 \sigma$, where $\sigma$ is the standard deviation.
\item {\boldmath$\eta$}: the percentage of outliers for which $|z_p - z_s|/(1+z_s)>0.15$. 
\end{itemize}
Table \ref{Tab:PZreq} gives the LSST requirements for these values. Note that we use a different definition for the rms than the standard to match the definition stated for the LSST photo-z quality requirements.

\subsection{Results for CFHTLS}\label{Sec:cfhtresults}
In this section, the reconstruction of the photometric redshifts and the consequences
of the selection on $L_R$ for the spectro-photometric data of CFHTLS are presented to
validate the method. Two cases are considered:
\begin{itemize}
\item \textbf{Case A:} The distributions  $P(\vec{\mu}|G)$ and
  $P(\vec{\mu}|O)$ are computed from the data themselves. 
\item \textbf{Case B:} The distributions  $P(\vec{\mu}|G)$ and
  $P(\vec{\mu}|O)$ are computed from a simulation of the CFHTLS data, as explained in Sect.
  \ref{Sec:CFHTLS}.
\end{itemize}
In both cases, the photo-z is computed from the real CFHTLS data.  Figure~\ref{Fig:NGalLR} shows that the $L_R$ cut has a similar behavior if the distributions are determined from the simulation or from the data.  

Figure \ref{Fig:DeltazParameterLR_CFHTLS} shows the performance of the
template fitting photo-z reconstruction applied to the CHTLS data
sample and the efficiency of the likelihood ratio $L_R$ cut. The
results shown here have been obtained using the distributions for the
likelihood ratio as computed from the data (case A); the results are
similar if the distributions are computed from the simulations (case
B). The 2D distributions of $\Delta z$ as a function of spectroscopic redshift $z_s$ represented on the top part of the figure show that our simulation process using galaxy templates, K-corrections,
reddening and the filter passbands do represent correctly the data, yielding reasonable photo-z reconstruction with no significant bias. One can also see a significant fraction of outliers, specially for galaxies with redshifts $z_s >~ 0.7$. The likelihood ratio cut, $L_R >0.6$ here, removes most of the CHTLS outliers, as can be seen on Fig. \ref{Fig:DeltazParameterLR_CFHTLS}, top-right,
although with the cost of a low selection efficiency for high redshift galaxies ($z >~ 0.9$). The fraction of galaxies retained by the $L_R$ cut is shown on the lower left part of the Fig. \ref{Fig:DeltazParameterLR_CFHTLS}, while the $\Delta z$ distribution before and after $L_R$ cut is shown on lower-right part. Using the likelihood ratio criterion enhances significantly the photo-z performance since the RMS decreases from 0.16 to 0.09 and the outlier fraction from 12\% to 2.8\% between the full galaxy sample and the sub-sample of galaxies with $L_R>0.6$. However, the overall photo-z reconstruction performance and the $L_R$ cut efficiency is significantly below that on the LSST simulated data (see following subsection).

In Fig.~\ref{Fig:CFHTLSredshift} we see that the spectroscopic sample redshift coverage barely extends beyond $z>1.4$.  This means that when the $L_R$ selection was calibrated on the CFHTLS data, the sample was missing outlier galaxies with high redshifts that could be estimated erroneously to be at low redshifts, e.g. those subject to the degeneracy causing the Lyman break to be confused for the 4000 \AA\ break.  The LSST simulated data contains these degeneracies, however it would be good to test the $L_R$ selection method on real data out to higher redshifts in the future.


\subsection{Results for LSST}\label{Sec:LSSTSim}

We use a total of 50 million galaxies in our simulated catalog.
This catalog is divided into 5 different sets.
Each set is separated into a test sample (2 million galaxies) and an analysis sample
(8 million galaxies). 
In each set, the statistical test is performed on "observed" galaxies within the test sample, then
the densities $P(\vec{\mu}|G)$ and $P(\vec{\mu}|O)$ are used to compute the value of $L_R$
for "observed" galaxies in the analysis sample.
Performing the reconstruction on the 5 independent sets give us a measure of the fluctuation
from a set to another and thus an estimate of the error on our reconstruction parameters.
We performed the same analysis with 10 sets of twice less galaxies and measured the fluctuations
to be very similar.

\subsubsection{Observation in six bands}\label{Sec:LSSTSim6b}

To test the method with best photometric quality we require each galaxy to be "observed" in each band with good precision $m_X<m_{5,X}$.
This requirement leaves us with about 125~000 galaxies in the test sample and 500~000 in the analysis sample.

Fig. \ref{Fig:DeltazLR} shows the 2D distributions from the LSST simulation of $z_p - z_s$ as
a function of $z_s$ for all the galaxies in the sample compared to the same distribution after
performing an $L_R$ selection. It is clear that selecting on the likelihood ratio enhances the photometric
redshift purity of the sample. 
 
Fig.~\ref{Fig:DeltazParameterLR} (the same as Fig.~\ref{Fig:DeltazParameterLR_CFHTLS} for CFHTLS) shows the evolution with $z_s$ of the number of galaxies retained in the LSST sample, for each of the parameters listed above (bias, rms, $\eta$). This indicates the quality of the photo-z, for different values of $L_{R,c}$. 

The LSST specifications on the bias and rms (see Table \ref{Tab:PZreq})
 are fulfilled up to $z_s = 1.5$ with only a low value of $L_{R,c}>0.6$.
For redshifts greater than $1.5$, a higher value of $L_{R,c}$ is required
to reach the expected accuracy. There are two main reasons for this. Firstly, only a small percentage of the galaxies with $z_s>1.5$ are used to calibrate the densities $P(\vec{\mu}|G)$ and $P(\vec{\mu}|O)$, therefore the
high-redshift galaxies do not have much weight in the calibration test. Secondly, the ratio between the height of the distribution $P(L_R|G)$ at $L_R = 0$ and at $L_R = 1$ tends to increase with the
redshift, meaning that the purity of the test is degraded.  Finally, only a very low value of $L_{R,c}>0.2$ is needed to enable $\eta$ to meet LSST specifications for $z_s>2.2$.

\begin{figure*}
  \centering
  \includegraphics[width=9cm]{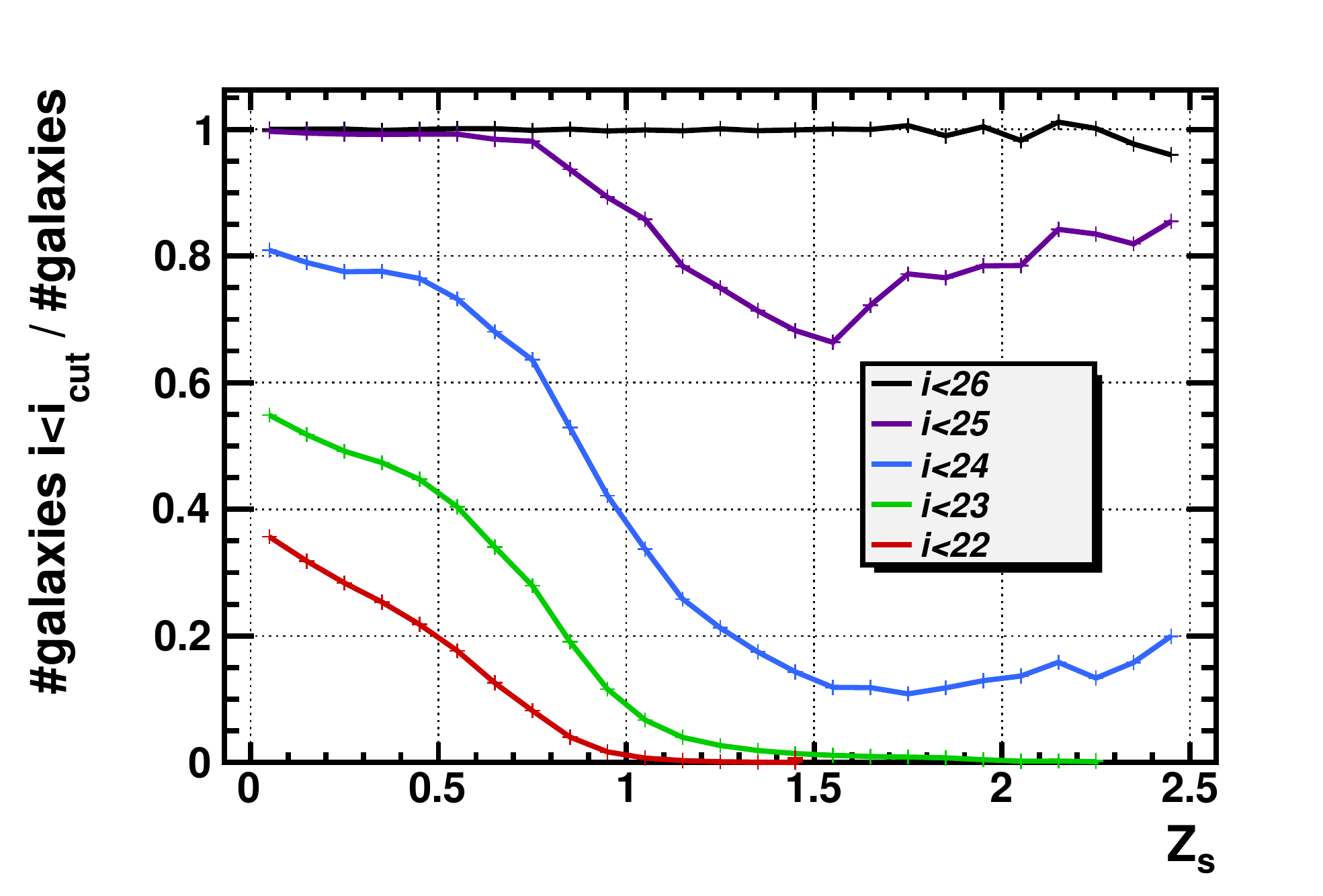}
  \includegraphics[width=9cm]{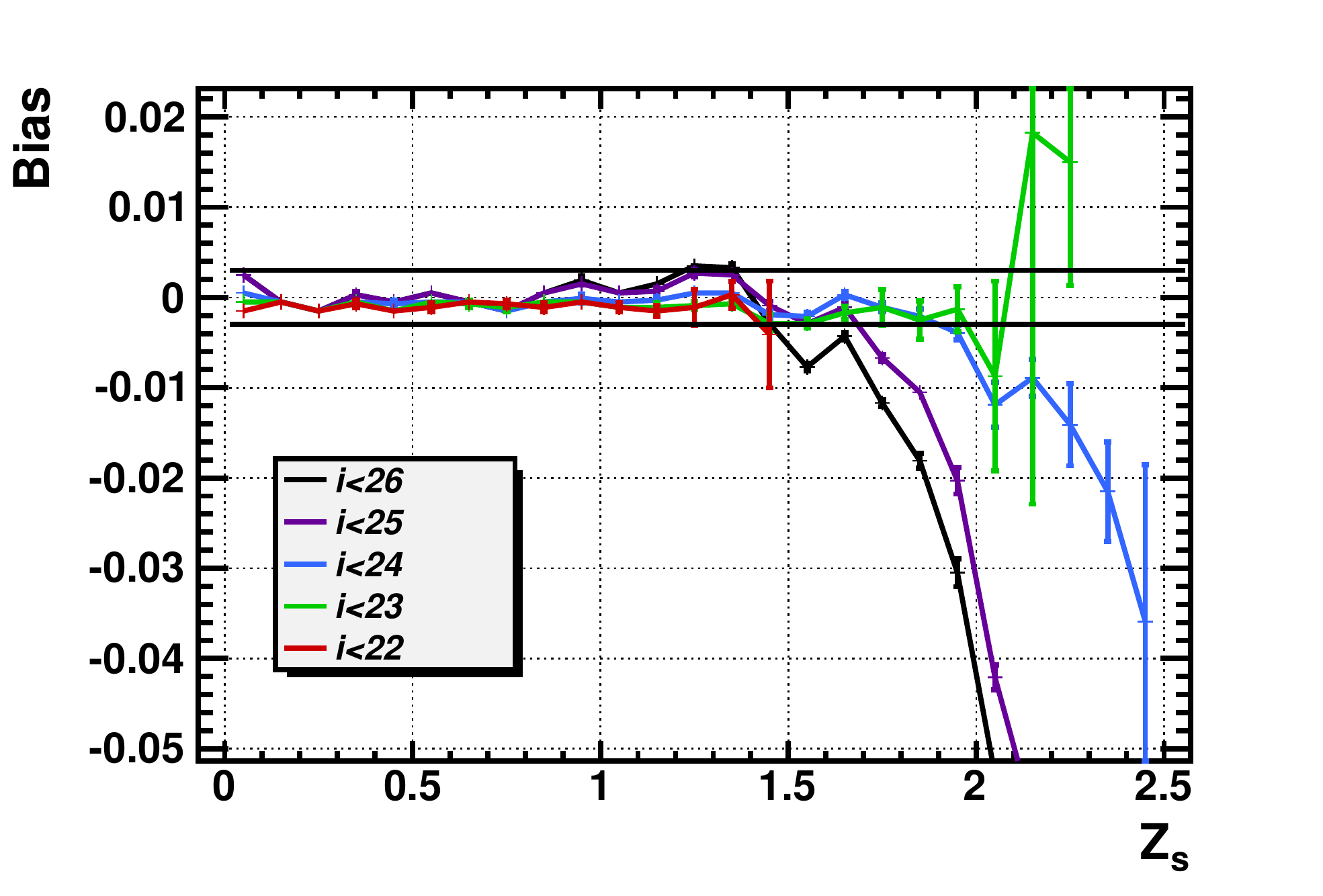}
  \includegraphics[width=9cm]{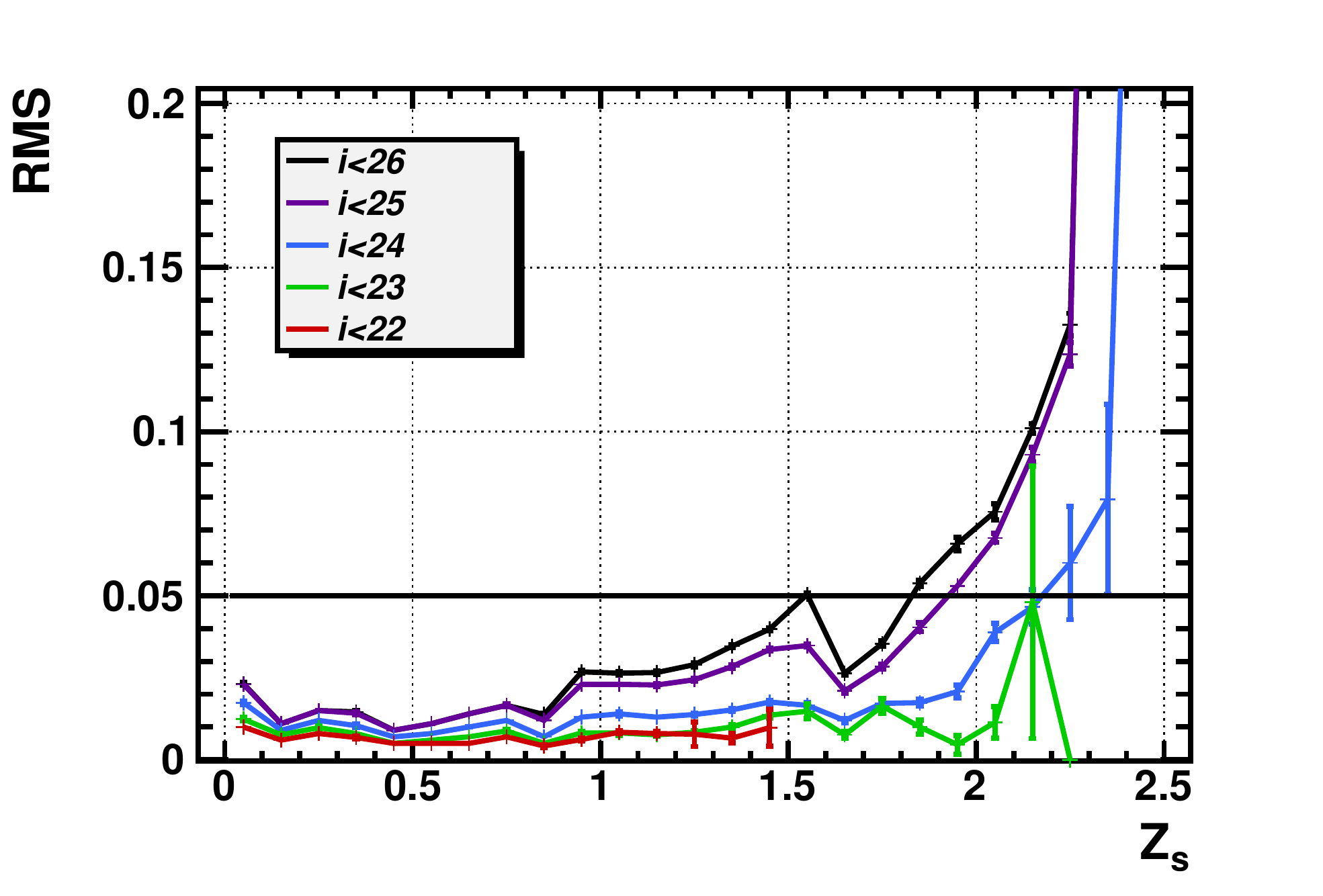}
  \includegraphics[width=9cm]{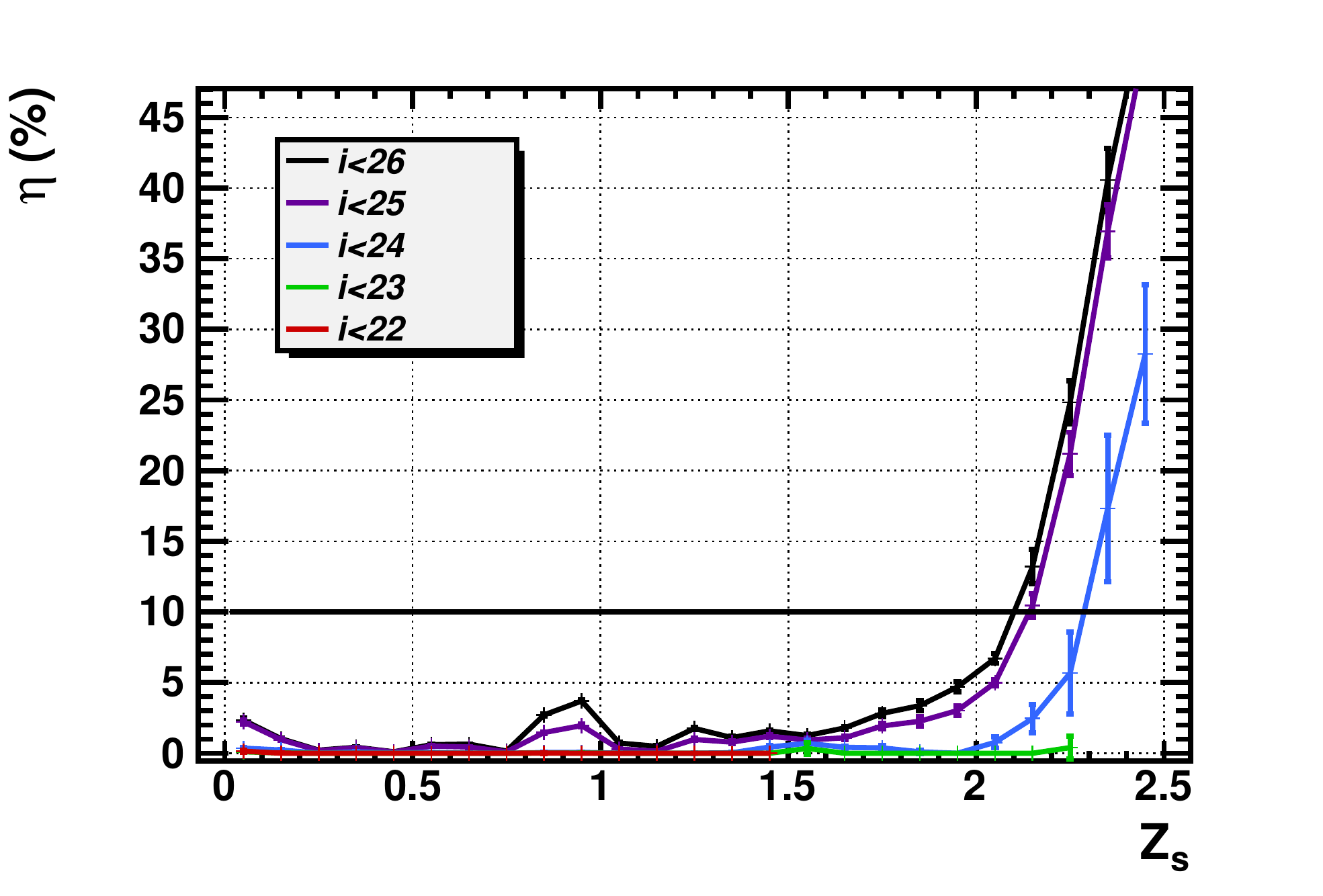}
  \caption{LSST $i$ magnitude selection. \textit{Top}: Evolution of the fraction of detected galaxies with $i<i_{cut}$ and the bias with $z_s$.  
    \textit{Bottom}: Evolution of the rms and $\eta$ as a function of $z_s$. The thick black lines
    represent the LSST requirements given in Table \ref{Tab:PZreq}. Ten years 
    of observations with the LSST is assumed. The values of the $i$-magnitude cuts are reported in the legend. }
  \label{Fig:DeltazParameterMag}
\end{figure*}

The effect of a selection on the likelihood ratio $L_R$ can be compared to the effect of a selection on the apparent
magnitude in the $i-$band, as shown in Fig.~\ref{Fig:DeltazParameterMag}. 
The increase in the $i$-magnitude selection efficiency at large z shown in Fig.~\ref{Fig:DeltazParameterMag} (top) is due to the value of $i_{cut}$ approaching the detection threshold.  Performing a selection on a quantity other than magnitude, such as the likelihood ratio, ensures that ``well measured" but faint galaxies are still included in the sample.

\begin{figure*}
   \centering
	\includegraphics[width=9cm]{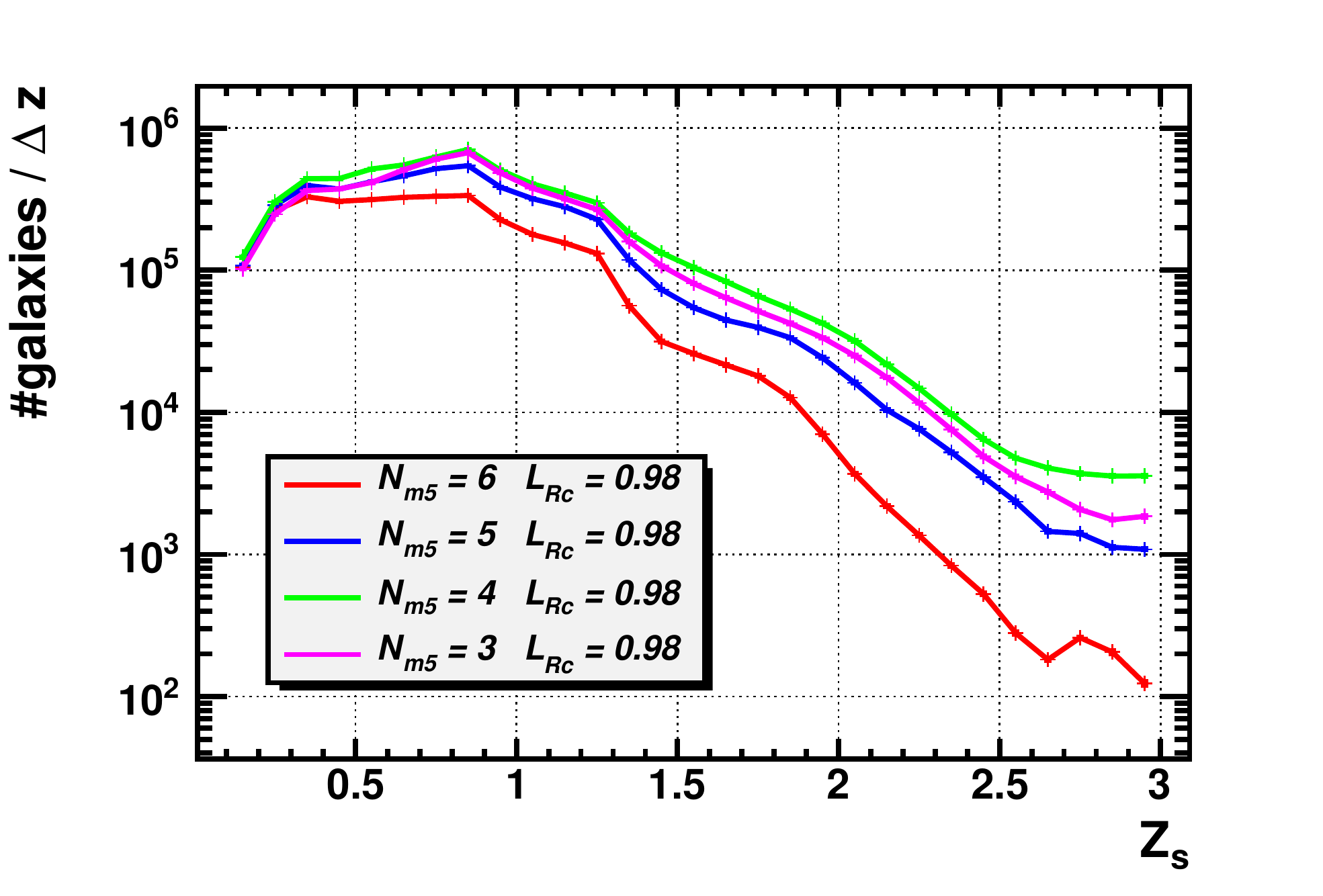}
	\includegraphics[width=9cm]{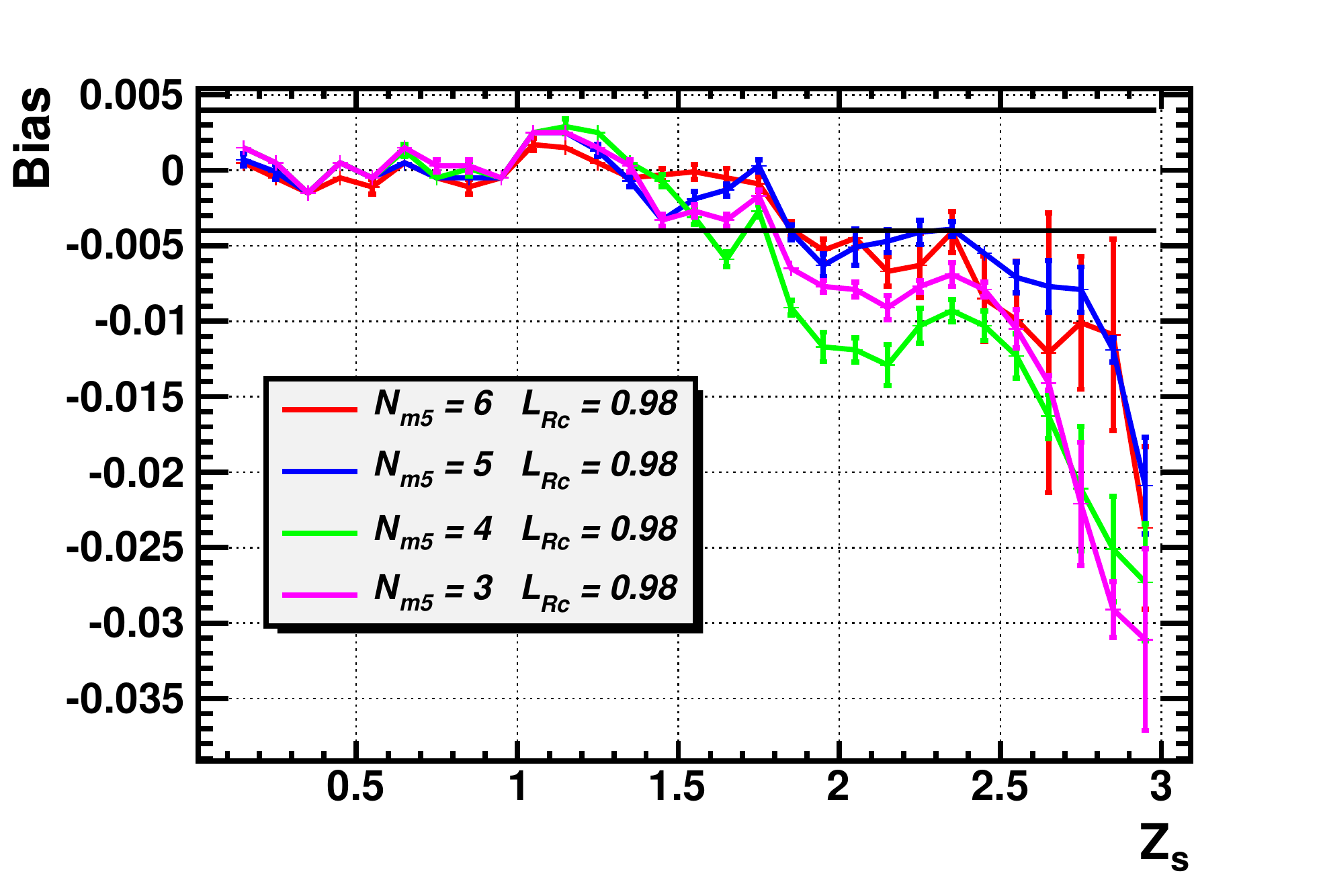}
	\includegraphics[width=9cm]{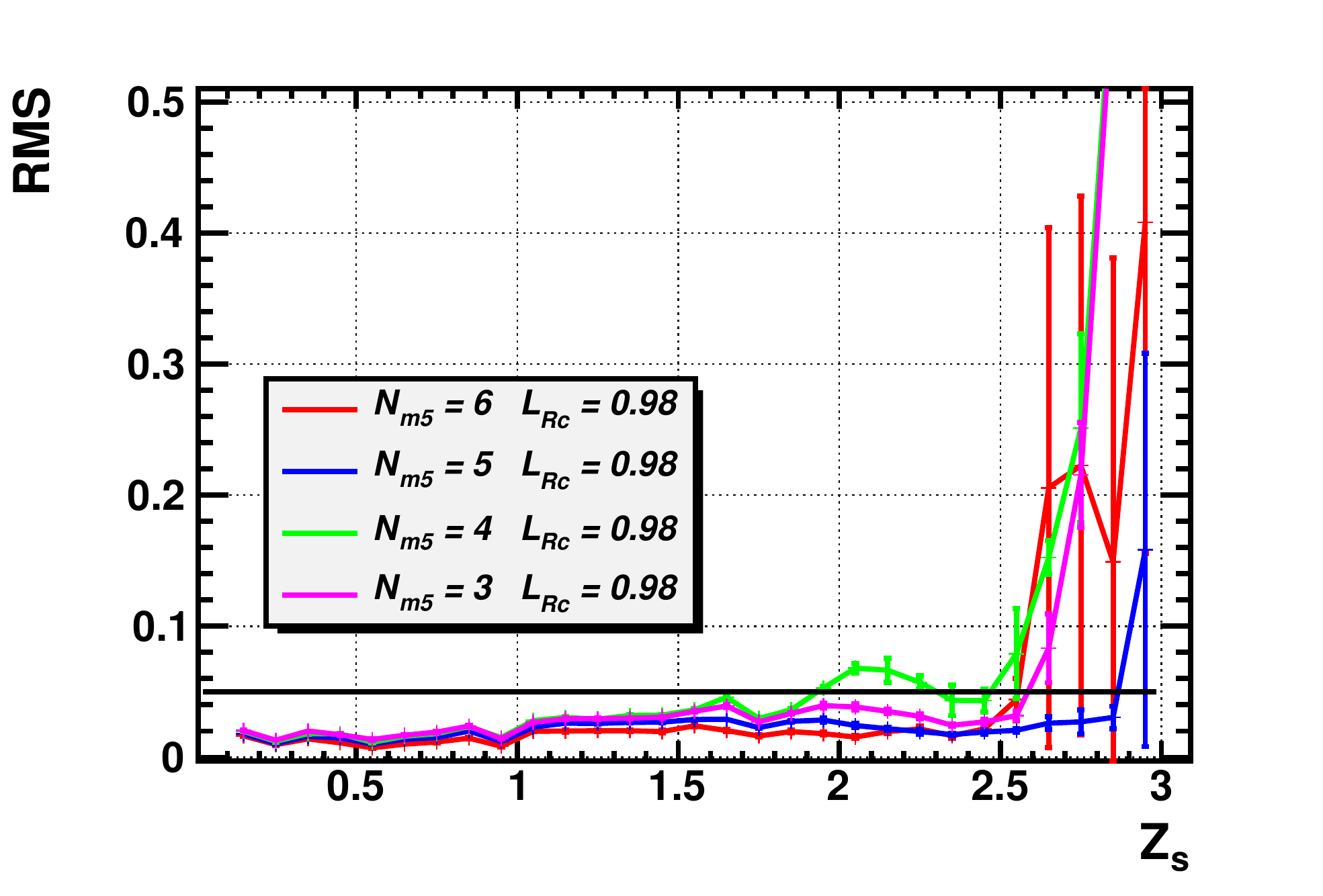}
	\includegraphics[width=9cm]{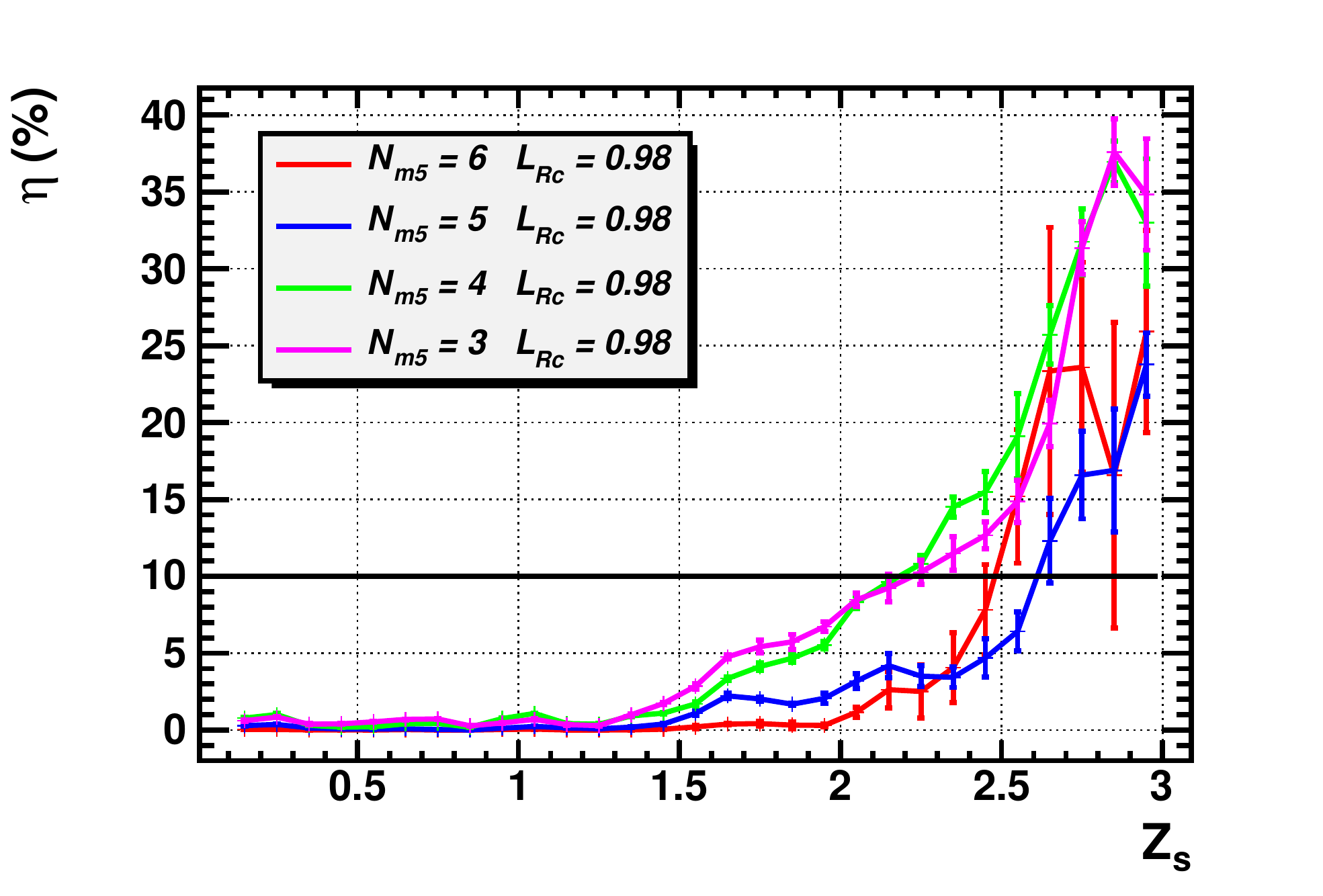}
      \caption{LSST reconstruction for different $N_{m5}$ requirement.
\textit{Top}: Evolution of the fraction of detected galaxies that satisfy the cuts (in the inset) and the bias with $z_s$.  
\textit{Bottom}: Evolution of the rms and $\eta$ as a function of $z_s$.
The thick black lines represent the LSST requirements given in Table \ref{Tab:PZreq}.
Ten years of observations with the LSST is assumed.
The values of $N_{m5}$ and $LR_{c}$ are reported in the legend. }
\label{Fig:PZNbands}
\end{figure*}

Of course requiring observation in six bands will exclude the $u$-band
drop-out galaxies at high redshift, so the photo-z performance at these redshifts will be greatly affected by this requirement.  The next section investigates the photo-z performance when observations are required in less than six bands.

\subsubsection{Observation in five bands (and less)}\label{Sec:LSSTSim5b}
 
The previous subsection demonstrated our results for good photometric data with $m_X<m_{5,X}$ in all six bands of the LSST. To detect more galaxies and extend our reconstruction at higher redshift,
we release the constraint on the number $N_{m5}$ of ``well observed" bands having $m_X<m_{5,X}$.
Both test and analysis sample are made of galaxies with $m_X<m_{5,X}$ in at least $N_{m5}$ bands.
We decreased $N_{m5}$ from $6$ to $5$, $4$ and $3$ and performed similar analysis to what was presented in the previous subsection.
The comparison of the results indicates that the selection with $N_{m5}=5$ gives the best results.
As we can see in Fig.~\ref{Fig:PZNbands}, lowering $N_{m5}$ from $6$ to $5$ greatly increases
the number of galaxies we keep in our sample without significantly degrading the reconstruction performance.
The gain in the number of galaxies is presented in Table \ref{Tab:Ngal} and increases with redshift as expected.

 \begin{table}[t]
 \caption{LSST number of galaxies, comparison between the selection on $L_{R}$ and the selection on $i$ band magnitude. The numbers in parenthesis indicate the fraction of galaxies retained after the $L_R$ cut. Observations were required in at least 5 bands.}
\begin{center}
\begin{tabular}{|c|c|c|c|c|} 
 \hline
& All redshifts & $0 \leq z<1$ & $1 \leq z<2$  & $2 \leq z<3$\\ \hline
$L_{Rc}=0$ & 895346 & 537026 & 326661 & 29623\\
$L_{Rc}=0.98$ & 468027 (0.52) & 342807 (0.64) & 120021 (0.37) & 4876 (0.16)\\
$i < 24$ & 314461 & 263207 & 50084 & 1171 \\
\hline
 \end{tabular}
\end{center}
\label{Tab:NgalLRi}
\end{table}

For galaxies ``observed" in only 5 bands, the band $X$ which has $m_X>m_{5,X}$ or is
not observed at all (noise level) is the $u$ band in 95\% of the cases.

When requiring less than 5 bands, the results are worse or similar: To reconstruct
decent photometric redshifts in this case we need to apply such a large value of $L_R$ for the selection that we reject nearly all the galaxies gained from the weaker requirement and even discard well-measured galaxies.

\begin{figure*}
  \centering
  \includegraphics[width=9cm]{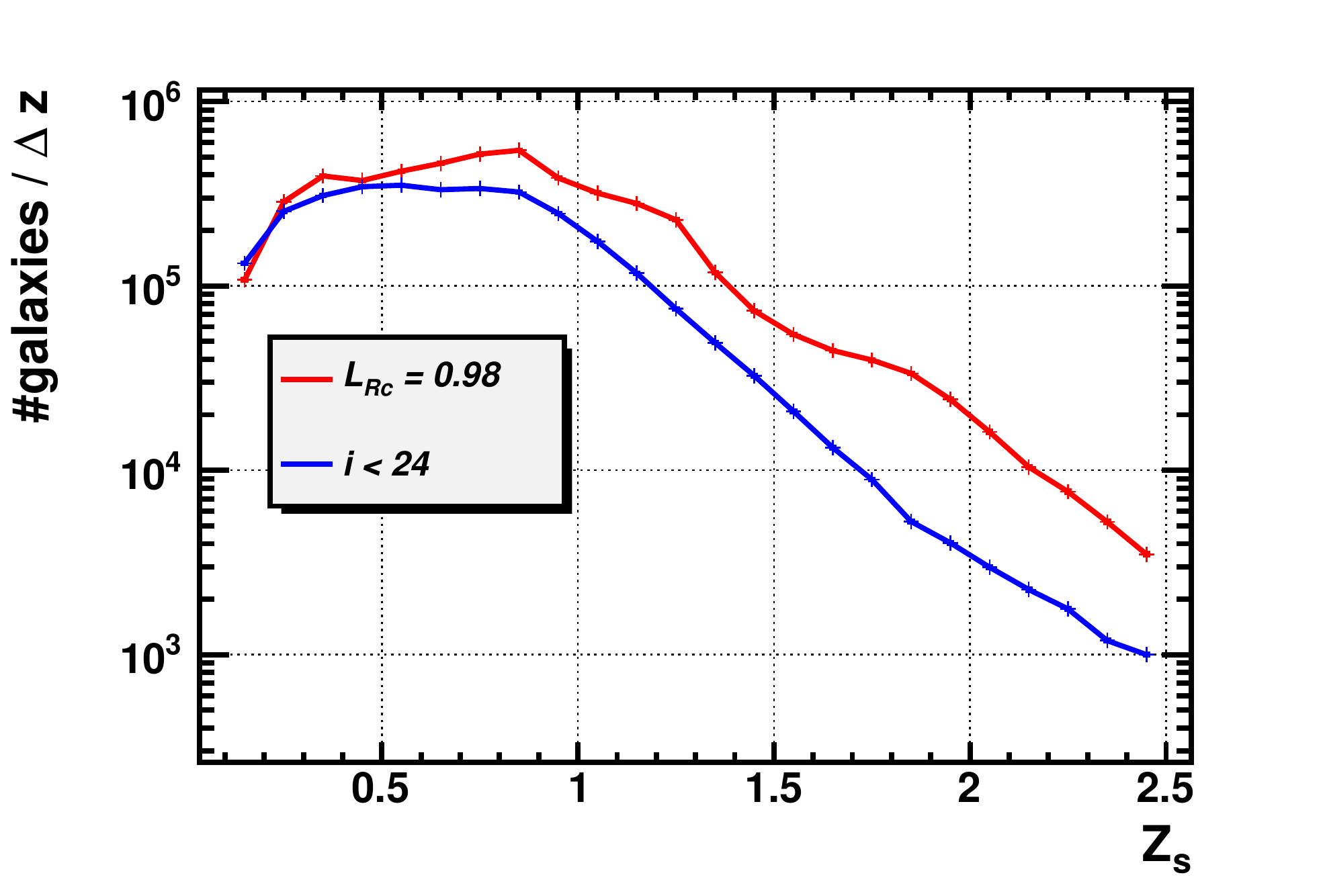}
  \includegraphics[width=9cm]{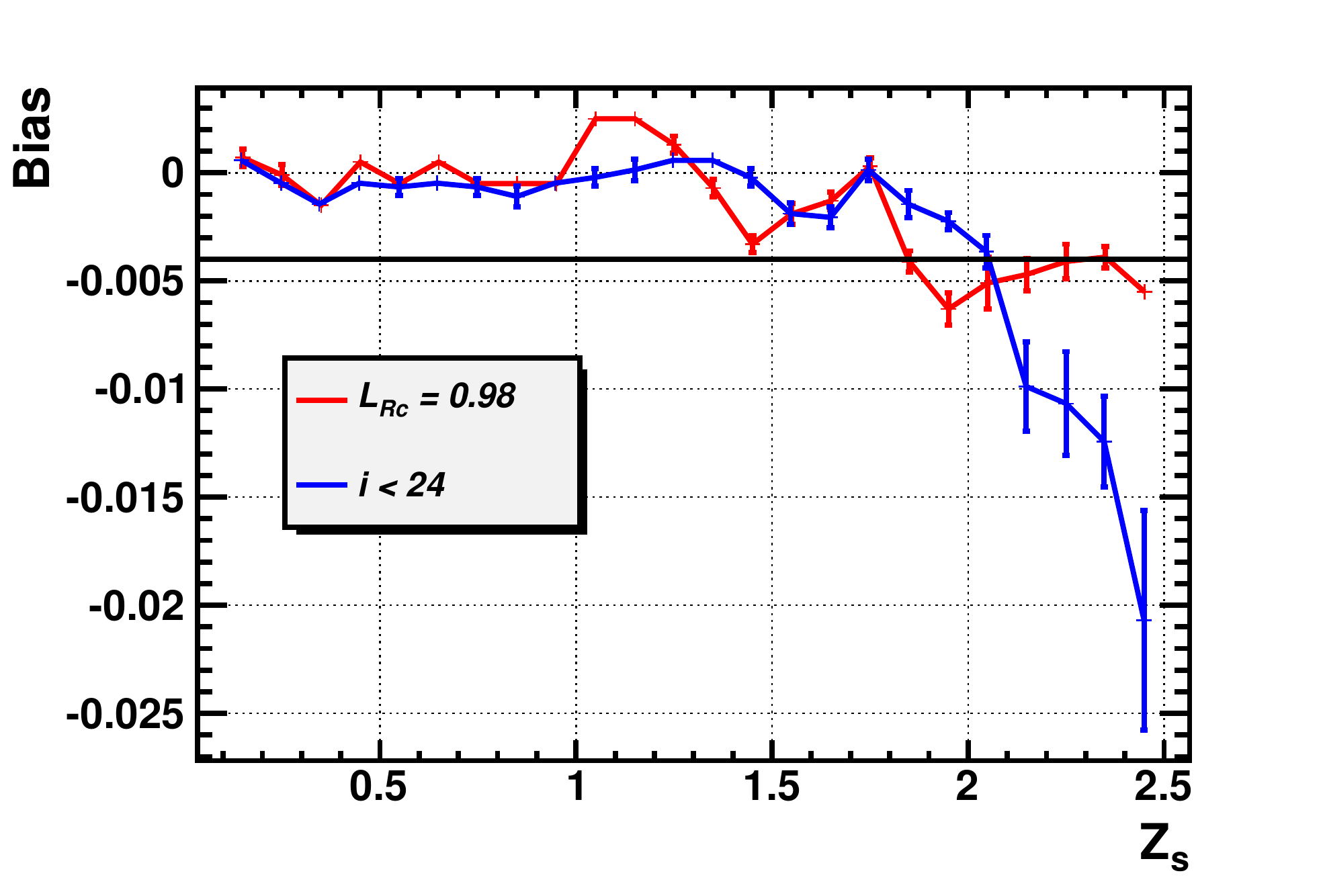}
  \includegraphics[width=9cm]{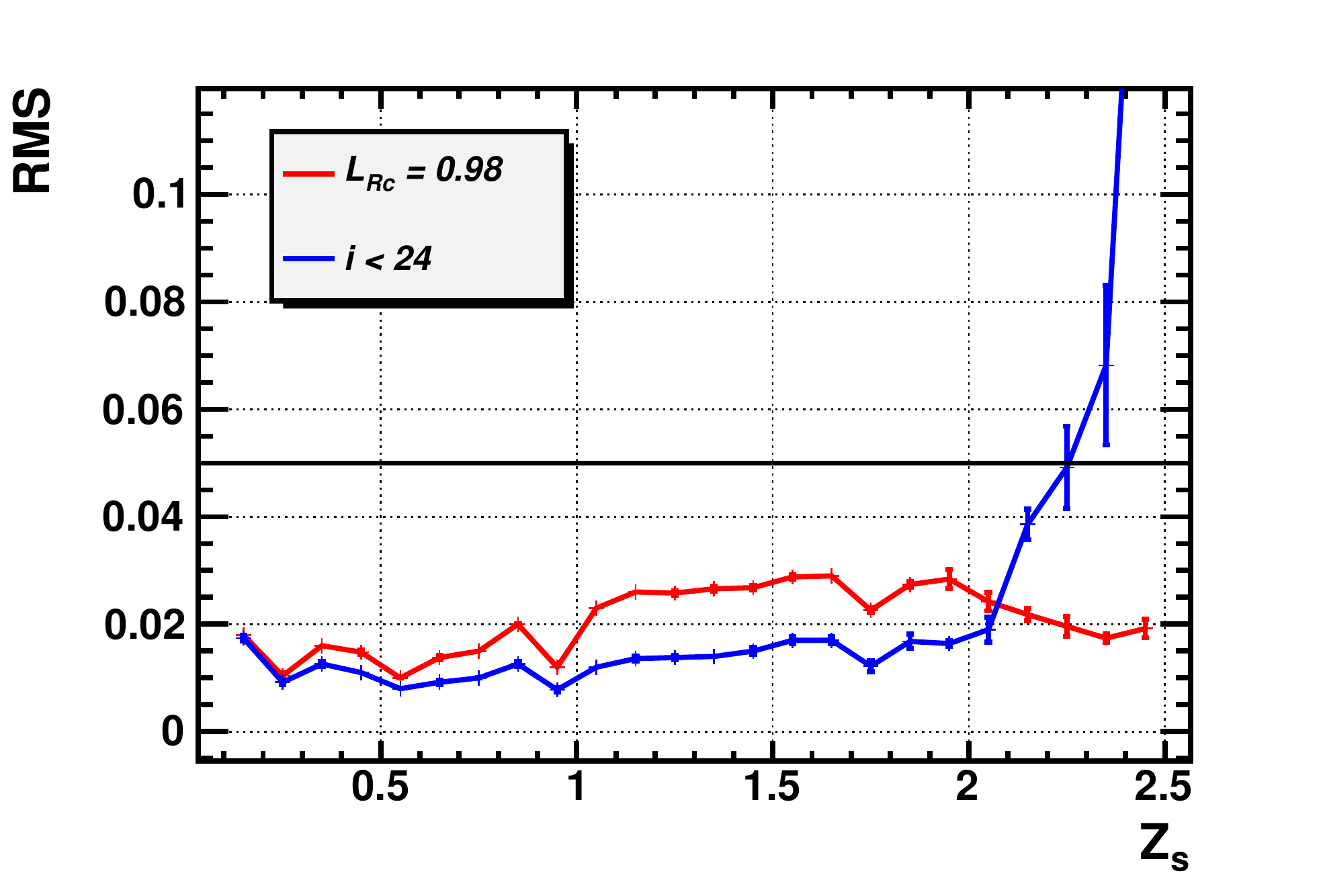}
  \includegraphics[width=9cm]{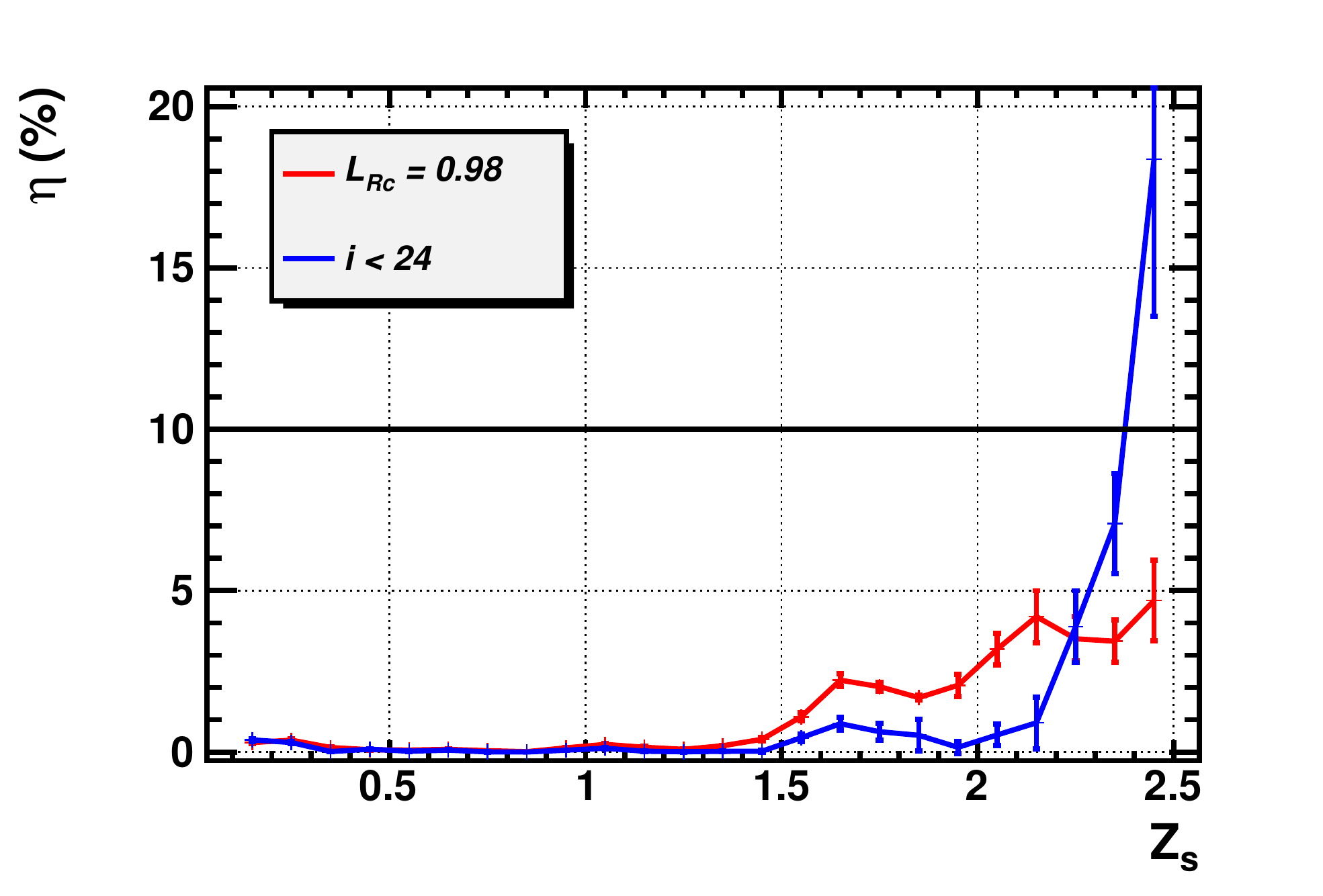}
  \caption{LSST comparison between $L_{Rc}=0.98$ and $i<24$ cuts.
\textit{Top}: Evolution of the fraction of galaxies and the bias with $z_s$.  
\textit{Bottom}: Evolution of the rms and $\eta$ as a function of $z_s$.
The thick black lines represent the LSST requirements given in Table \ref{Tab:PZreq}.
Ten years of observations with the LSST is assumed.}
\label{Fig:magLR5bands}
\end{figure*}

Figure \ref{Fig:magLR5bands} shows the comparison of the number of galaxies and photo-z performance (rms, bias, $\eta$) for a $L_R$-selected sample ($L_R>0.98$) and a magnitude-selected sample ($i<24$). While both samples satisfy the LSST science requirement given in Table \ref{Tab:PZreq}, the $L_R$ selection is more efficient, since it retains a significantly larger number of galaxies, specifically for $z>1$ (see Table \ref{Tab:NgalLRi}).  We do not present a comparison with a sample selected by a magnitude cut of $i<25.3$, since it would not satisfy the LSST photo-z requirements, according to our simulation and photo-z reconstruction (as can be seen in Fig.~\ref{Fig:DeltazParameterMag}).

Our choice of the value $L_{R,c}=0.98$, is a preliminary compromise between the quality of the redshift reconstruction and the number of measured galaxies. Increasing this threshold would lead to a smaller sample of galaxies with improved photometric performances. The final tuning will be driven by physics, depending on the impact of the cut on cosmological parameter determination and needs a more detailed and dedicated study.
%

 \begin{table}[t]
 \caption{LSST number of galaxies ($L_{R,c}=0.98$). The numbers in parenthesis indicate the fraction of galaxies retained after the $L_R$ cut.}
\begin{center}
\begin{tabular}{|c|c|c|c|c|} 
 \hline
& All redshifts & $0 \leq z<1$ & $1 \leq z<2$ & $2 \leq z<3$\\ \hline
$N_{m5}=6$ & 313473 (0.64) & 249080 (0.76) & 63451 (0.40) & 922 (0.13)\\
$N_{m5}=5$ & 468027 (0.52) & 342807 (0.64) & 120021 (0.37) & 4876 (0.16)\\
$N_{m5}=4$ & 605134 (0.42) & 421619 (0.54) & 171790 (0.31) & 10416 (0.09)\\
$N_{m5}=3$ & 536664 (0.30) & 377108 (0.42) & 150260 (0.22) & 7872 (0.04)\\
\hline
 \end{tabular}
\end{center}
\label{Tab:Ngal}
\end{table}


\section{Photo-z performance with neural network}\label{Sec:NN}

It has been shown using the public code ANNz by
\cite{Collister:2003cz} that the photometric redshifts can be
correctly estimated via a neural network. This technique, along with other empirical methods, requires a spectroscopic sample for which the apparent magnitudes and the spectroscopic redshifts are known. 

The toolkit for multivariate analysis \citep[TMVA]{tmva} provides a
ROOT-integrated environment for the processing, parallel evaluation,
and application of multivariate classification and multivariate
regression techniques. All techniques in TMVA belong to the family of
Òsupervised learningÓ algorithms. They make use of training events,
for which the desired output is known, to determine the mapping
function that either describes a decision boundary or an approximation
of the underlying functional behavior defining the target value. The
mapping function can contain various degrees of approximations and may
be a single global function, or a set of local models.  Among
artificial neural networks, many other algorithms, such as boosted
decision trees or support vector machines, are available. An advantage
of TMVA is that different algorithms can be tested at the same time in a very user-friendly way.

\begin{figure*}
   \centering
	\includegraphics[width=6cm]{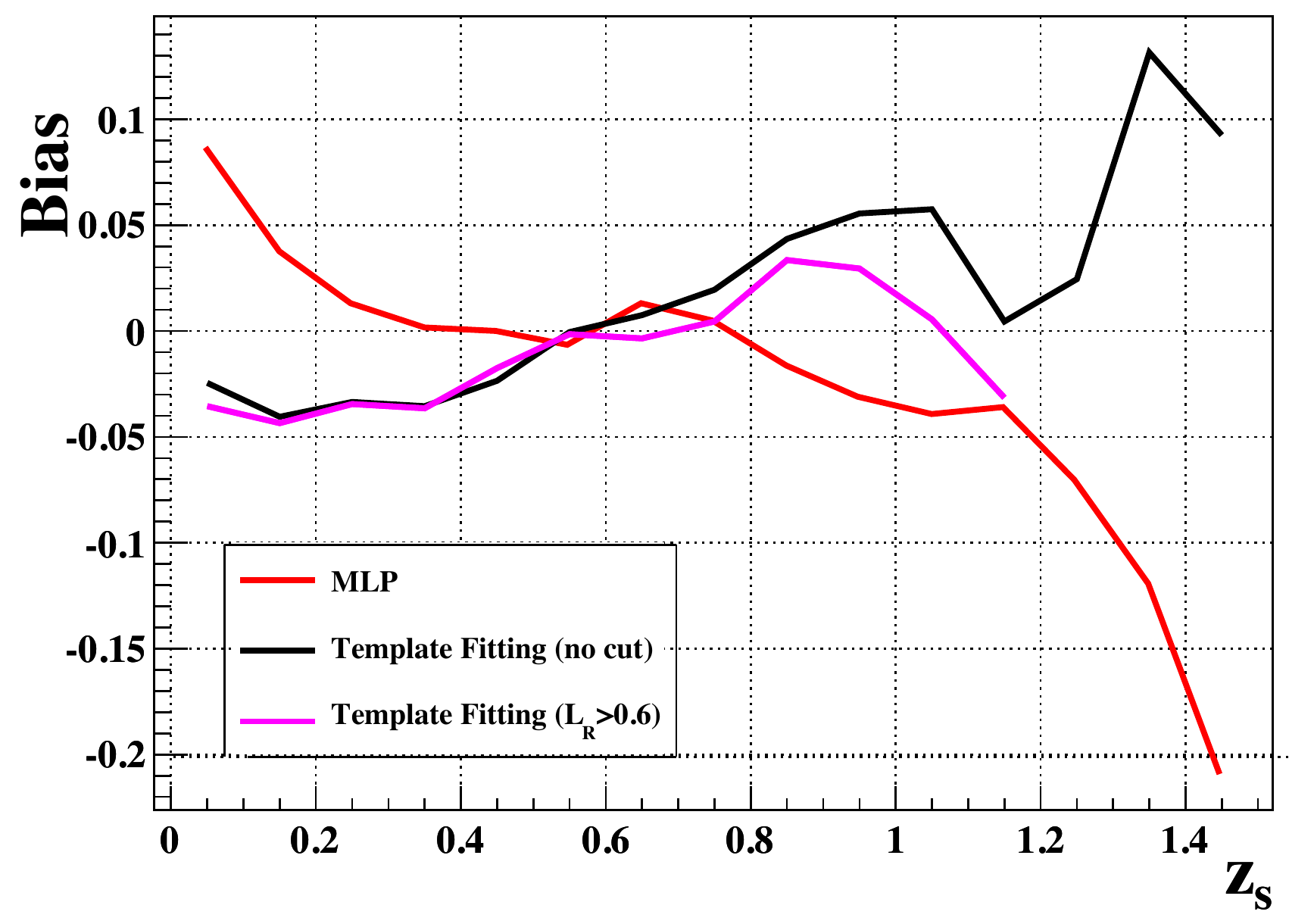}
	\includegraphics[width=6cm]{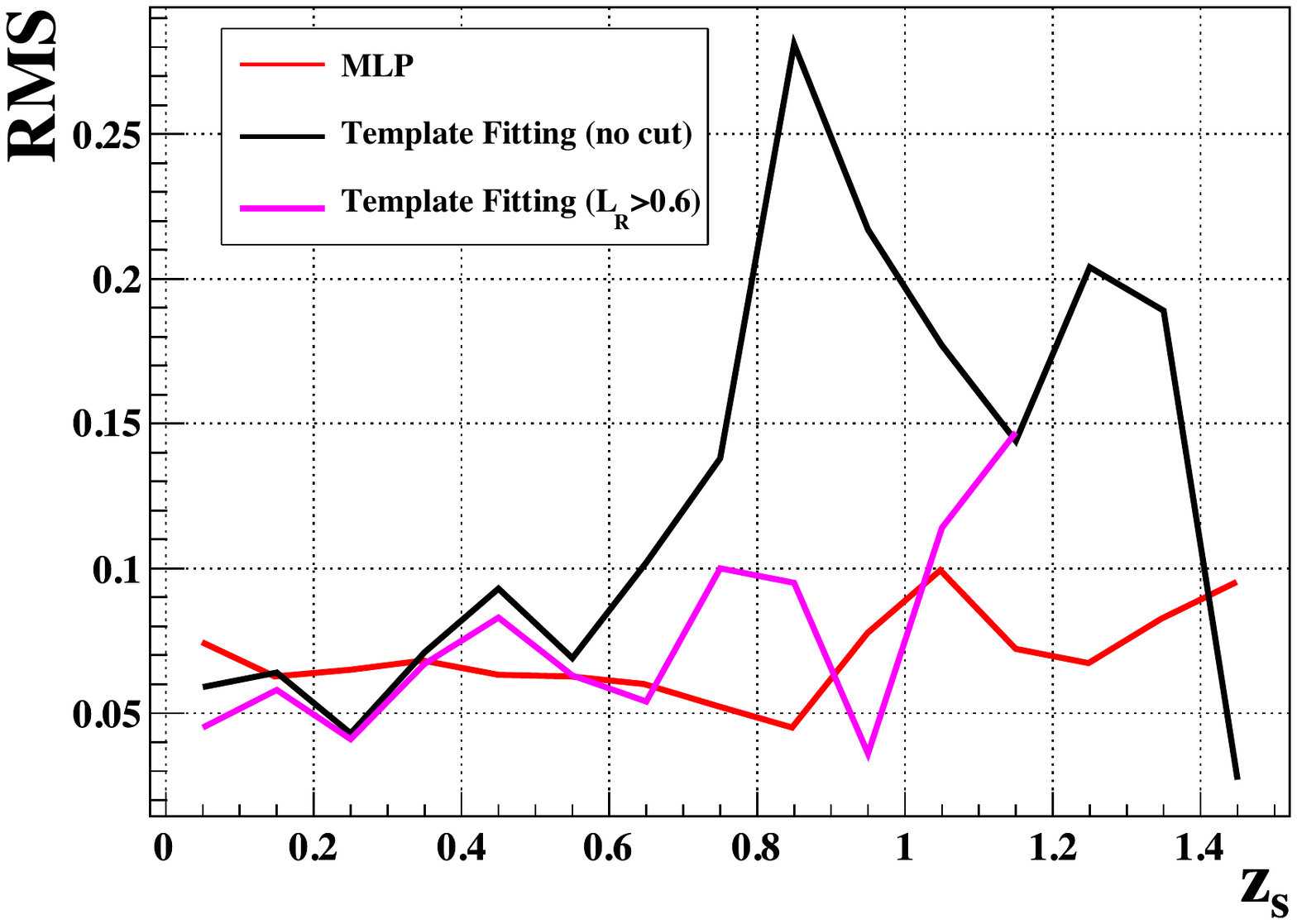}
	\includegraphics[width=6cm]{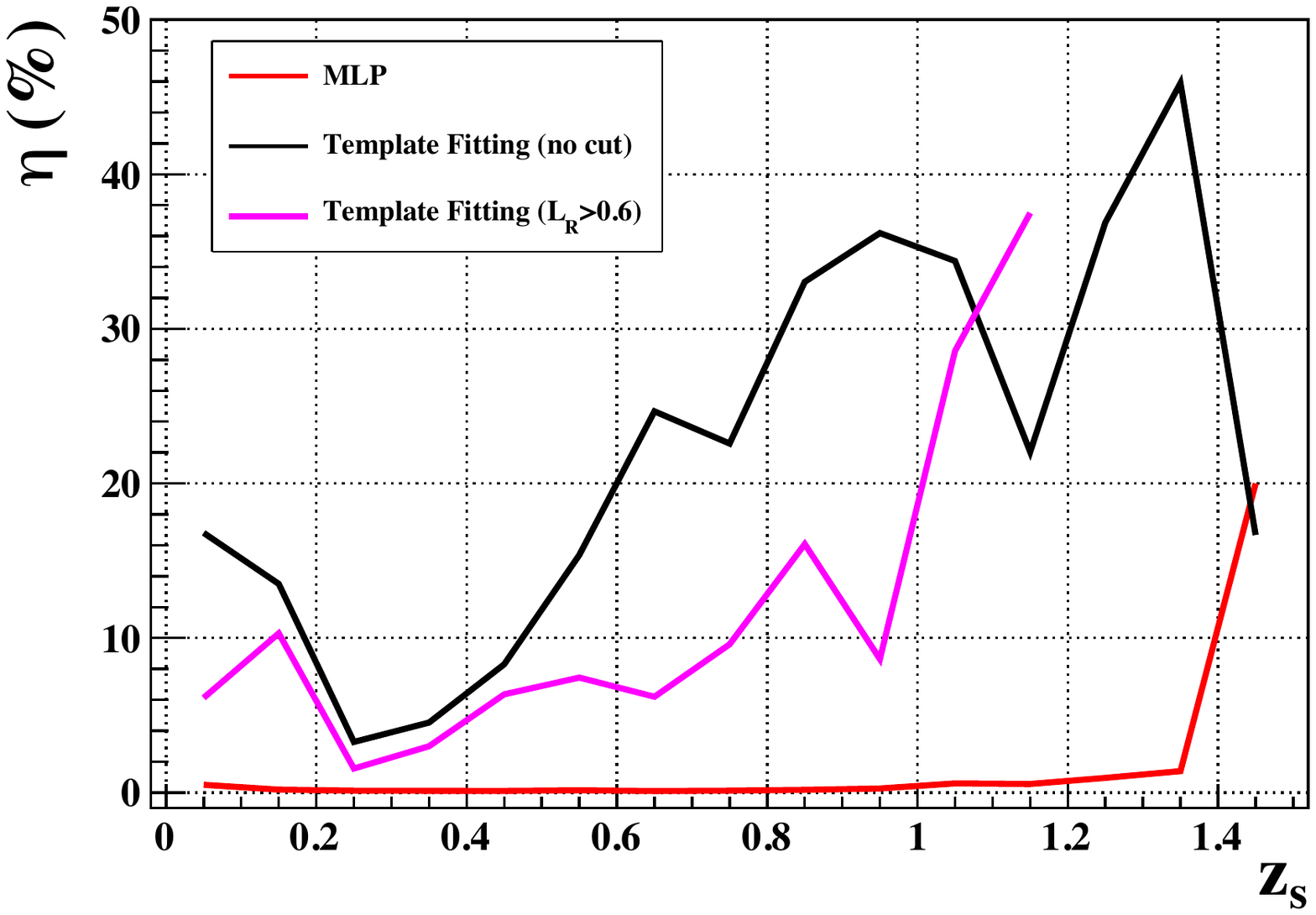}
      \caption{Comparison between the template-fitting method and the neural
      network for CFHTLS data.  The bias, the rms of the distribution of
      $\Delta z/(1+z_s)$, and the parameter $\eta$ are displayed as functions of the
      true redshift for the CFHTLS data. Data points are reported only if the number of
      galaxies in the sample is greater than ten.}
         \label{Fig:CompareNNChi2}
\end{figure*}

\subsection{Method}

The MultiLayer Perceptron (MLP) neural network principle is simple. It builds up
a linear function that maps the observables to the target variables,
which are the redshifts in our case. 
The coefficients of the function, namely the weights, are such that
they minimize the error function which is the sum over all galaxies in
the sample of the difference between the output of the network and the
true value of the target. Two samples of galaxies, the training and the test sample, are necessary. The latter is used to test the convergence of the network and to evaluate its performance. It usually prevents overtraining of the network, which may arise when the network learns the particular feature of the training sample.

A neural network is built with layers and nodes. There are at least two layers, one for the input observables $\vec{x}$ and one for the target $z_{MLP}$. Each node of a layer is related to all nodes from the previous layer with a weight $w$, which is the coefficient associated with the activation function $A$ of each connection. The value $y_j^{i+1}$ of the node $k$ from the layer $i+1$ is related to the values of all $y_{j}^{i}$ from the layer $i$:
\begin{eqnarray*}
y^{i+1}_k &=& A\left(\sum_{j = 1}^n w_{jk}^{i}y^{i}_{j} \right)~,
\end{eqnarray*}
where $n$ is the number of neurons in the $i$ layer.
For the purpose of this paper, the activation function $A$ is a
sigmoid. 
As an example, we examine the case
where there is only one intermediate layer. Then the photometric
redshift of the $g^{\textit{th}}$ galaxy is estimated as follows:
\begin{eqnarray*}
z_{MLP,g} = \sum_{j = 1}^{n}  A\left( \sum_{i = 1}^{n_{band}} w^1_{ij}x_{g,i}
\right) \times w_{j1}^2~. 
\end{eqnarray*}
The error function is simply defined by
\begin{eqnarray*}
E(\vec{w}) &=& \frac{1}{2}\sum_{g = 1}^{n_{train}} E_g(\vec{x}_g,\vec{w})
\end{eqnarray*}
with
\begin{eqnarray*}
E_g(\vec{x}_g, \vec{w})  &=& z_{MLP,g}(\vec{x}_g,\vec{w}) - z_{s,g}~,
\end{eqnarray*}
where $n_{train}$ is the number of galaxies in the training sample and $g$ denotes the
$g^{\textit{th}}$ galaxy.  At the first iteration, the weights have random values. The gradient
descent method, which consists of modifying the weight value according to the derivative of $E$ with
respect to the weight, is used to minimize $E$. For example, we have after one iteration:
\begin{eqnarray*}
w^{2}_{j1} &\rightarrow& w^{2}_{j1} + \Delta w^{2}_{j1}~,\\
\Delta w^{2}_{j1} &=& -\alpha \sum_{g=1}^{n_{train}}\frac{\partial E_g}{\partial
w^{2}_{j1}}~.\\
\end{eqnarray*}
The parameter $\alpha$ is the learning rate and has to be determined for each specific case. It must not
be too large; otherwise, the steps are so large that the minimum of $E$ is never reached. It must not
be too small either; otherwise, too many iterations are required. The testing sample is used as a
convergence and performance test. Indeed, the errors decrease with the number of iterations in the
training sample but reach a constant value on the testing sample. Weights are finally kept
when the errors on the testing sample reach a constant value.

For this non-exhaustive study on CFHTLS data, we have chosen the
observables $\vec{x} = (\vec{m},\vec{\sigma(m)})$
and two layers of ten nodes each. The training sample contained 8000 randomly picked galaxies and the testing sample contained the remaining 6268 galaxies of the CFHTLS spectro-photometric catalog.  In Figure~\ref{Fig:CompareNNChi2}, the bias, rms, and the
outlier rate $\eta$ are compared for the template-fitting method and for the neural
network. It is clear that the outlier rate is much smaller at all
redshifts when the photo-z is estimated from the neural network. The
dispersion of the photometric redshifts is also smaller for the neural network when
compared to the template-fitting method. These characteristics are expected because
the training sample is very similar to the test sample, whereas the template-fitting method
uses only a small amount of prior information. Moreover, the apparent
magnitudes in the template-fitting method are fitted with a model with
SED templates, whereas no theoretical model has to be assumed to run
the neural network. However, these attributes may be reversed if the
two samples are different, as illustrated in
Sect. \ref{Sec:LSSTNN}. The overestimation of photo-z at low redshifts
and underestimation at high redshifts, shown by the downward slope in
the bias, can be attributed to \textit{attenuation bias}.  This is the
effect of the measurement errors in the observed fluxes, resulting in
the measured slope of the linear regression to be underestimated on
average; see \cite{Freeman:2009} for a full discussion of this
bias. We note that the photo-z bias obtained from the template-fitting
method has an opposite sign and is of the same amplitude to that
obtained from the neural network method.  Since we have reason to
expect that the neural network has a downward slope in the bias, this indicates that the two estimators can be used complementarily. This is investigated in the next subsection.

\subsection{Results for CFHTLS}

Here, a possible combination of photo-z estimators (from
the template-fitting method on the one hand and from the neural network on the other hand)
is outlined. Even if the fraction of outlier galaxies is smaller with
a neural network method than for the template-fitting method, using
only the neural network to estimate the photo-z appears not to be
sufficient to reach the stringent photo-z requirements for the LSST,
especially when the spectroscopic sample is limited. Neural networks
seem to produce a photo-z reconstruction that is slightly biased at both ends of the range (see Fig.~\ref{Fig:CompareNNChi2}). This is due to the galaxy under-sampling at low and very high redshifts in the training sample. When spectroscopic redshifts are available, it is therefore worth combining both estimators.

\begin{figure}
  \centering
  \includegraphics[width=9cm]{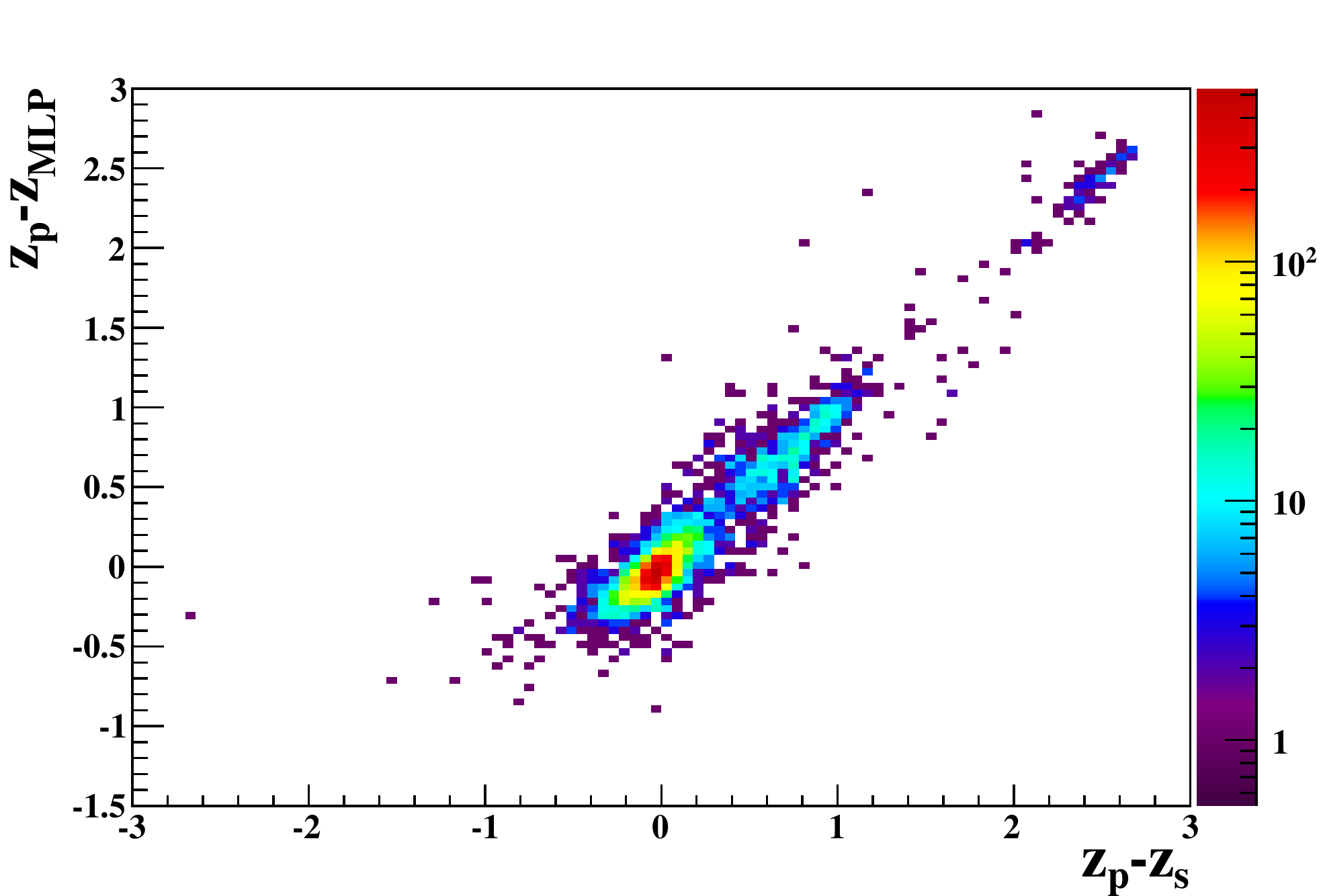}
  \caption{
    2D histogram of $z_{p} - z_{MLP}$ \textit{vs.} $z_p - z_s$ for
    the test sample of the CFHTLS data.}
  \label{Fig:CorrelationMLPTemplateFitting}
\end{figure}

With the CFHTLS data, Fig.~\ref{Fig:CorrelationMLPTemplateFitting}
shows that there is a correlation between $z_p- z_{MLP}$
and $z_p-z_s$, where $z_p$ is the photo-z that is estimated with the
template-fitting method.
This correlation could be used to remove some of
the outlier galaxies for which the difference between $z_p$ and $z_s$ is large, for example, by
removing galaxies with around $|z_p-z_{MLP}| \ge 0.3$. This
correlation appears because the neural network is well trained, and therefore
the photo-z is well estimated and $z_{MLP}$ becomes a good proxy for $z_s$.

\begin{figure}
  \centering
  \includegraphics[width=9cm]{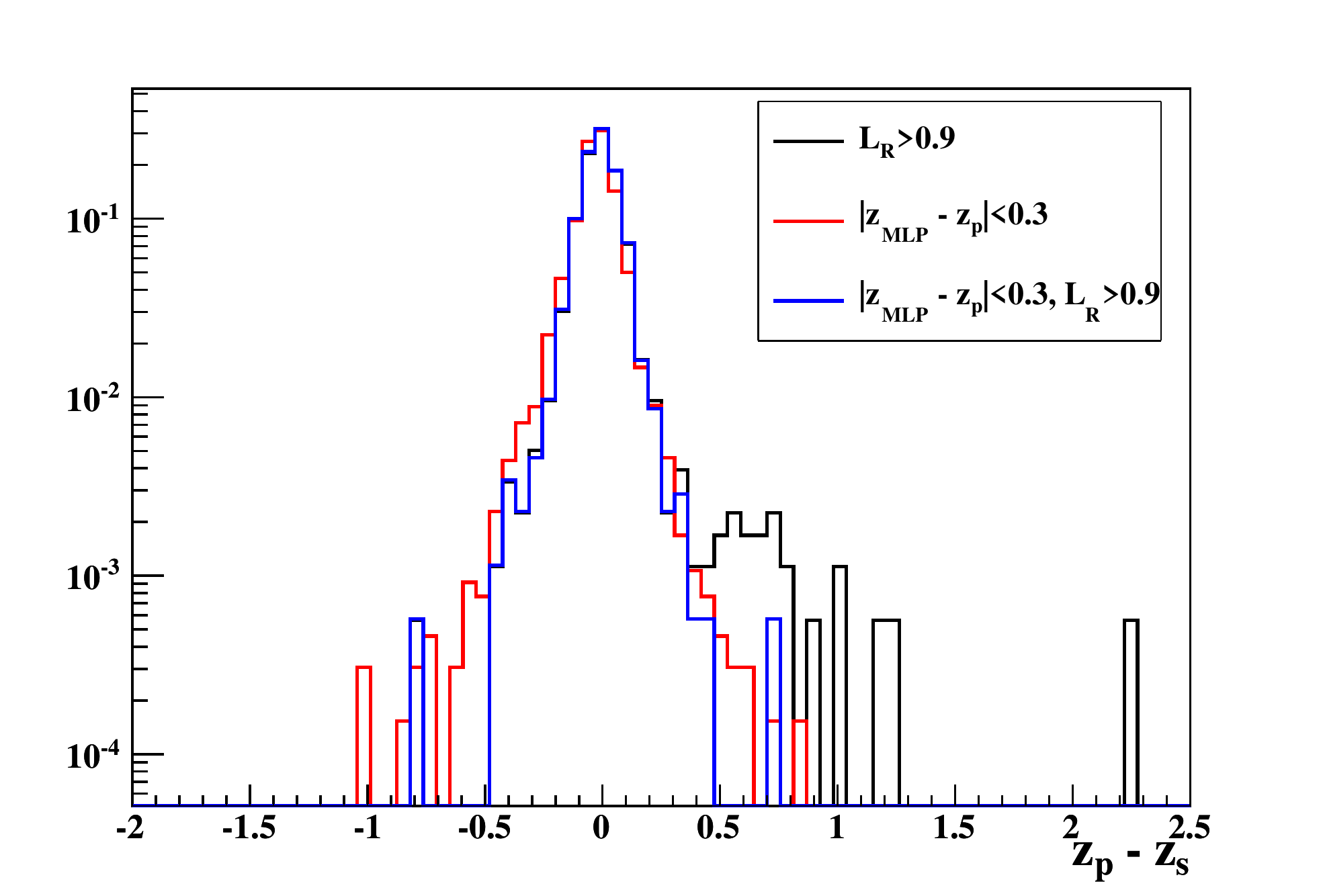}
  \caption{
    Normalized histograms of $z_p - z_s$ with $L_R>0.9$ (black curve), $|z_{MLP} - z_{p}|<0.3$
    (red curve) and both cuts (blue curve) for the CFHTLS data.}
  \label{Fig:CombineCutLRDeltaMLP}
\end{figure}

One can see an example of the impact of using both estimators $z_p$
and $z_{MLP}$ in Fig.~\ref{Fig:CombineCutLRDeltaMLP}. The distribution
of $z_p - z_s$ is plotted for three cases: $L_R>0.9$ only,
$|z_p-z_{MLP}|<0.3$ only and both cuts. There are fewer outlier
galaxies from the first to the third case.
By selecting with both variables, $|z_p-z_{MLP}|$ and $L_R$, we
improve the photo-z estimation when compared to a selection based only
on $L_R$, or (to a lesser extent) only on $|z_p-z_{MLP}|$.
This shows that neural
networks have the capability to tag galaxies with an outlier
template-fitted photo-z if the training sample is representative of
the photometric catalog.  However, this is difficult to achieve in
practice because the training sample is biased in favor of bright, low
redshift galaxies, which are most often the ones selected for spectroscopic observations.

\subsection{Results for the LSST}\label{Sec:LSSTNN}

For the LSST simulation, the network was composed of 2 layers of 12 nodes each; the training sample was composed of 10~000 galaxies and the testing sample of 20~000 galaxies.  We found that increasing the size of the training sample above 10~000 showed no improvement in the precision of the training. We attribute this to the regularity of the simulation: the galaxies were drawn from a finite number of template SEDs. As soon as the sample represents all the galaxy types in the simulation, adding more galaxies does not help in populating the parameter space any longer.

\begin{figure}
   \centering
   	\includegraphics[width=9cm]{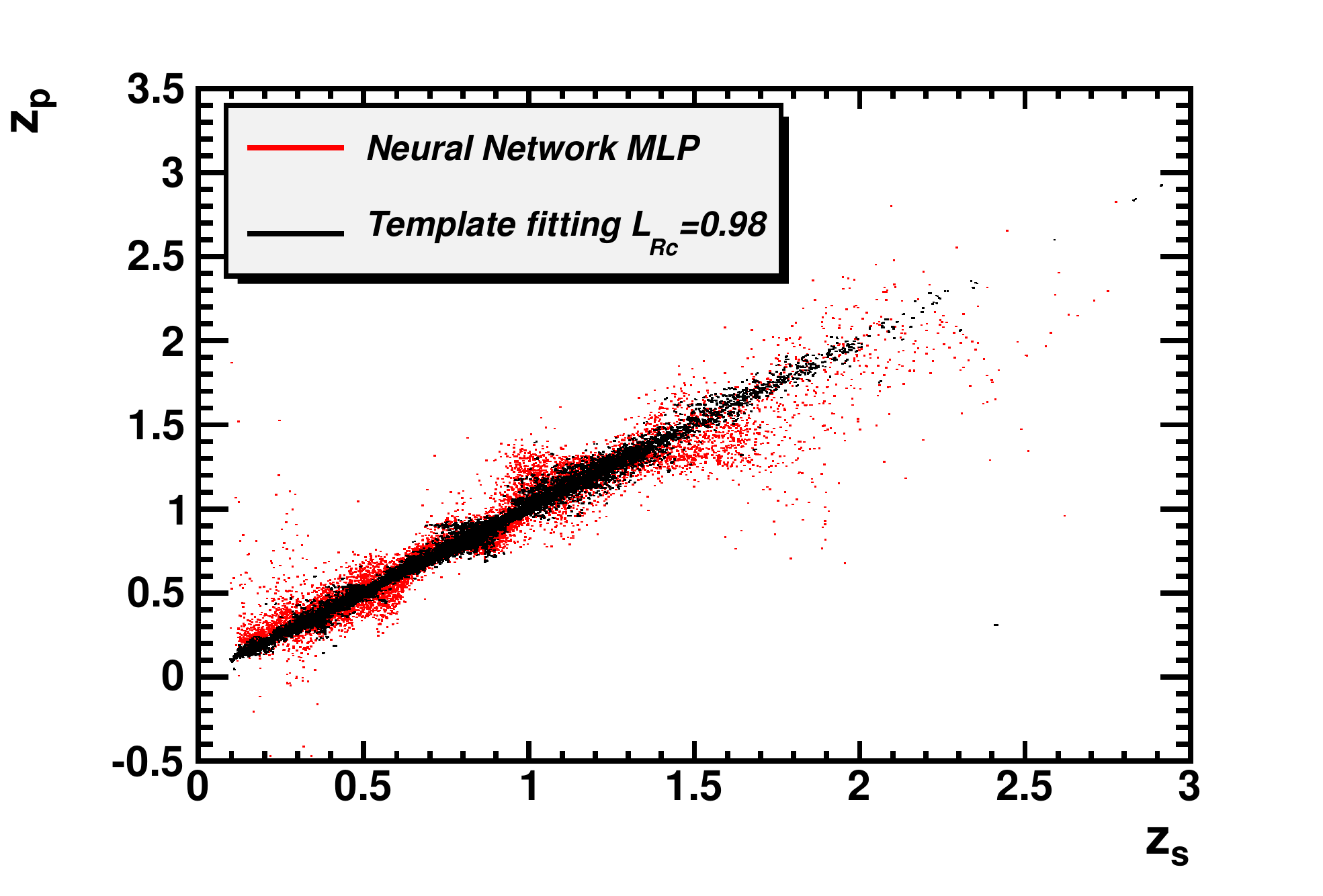}
	\includegraphics[width=9cm]{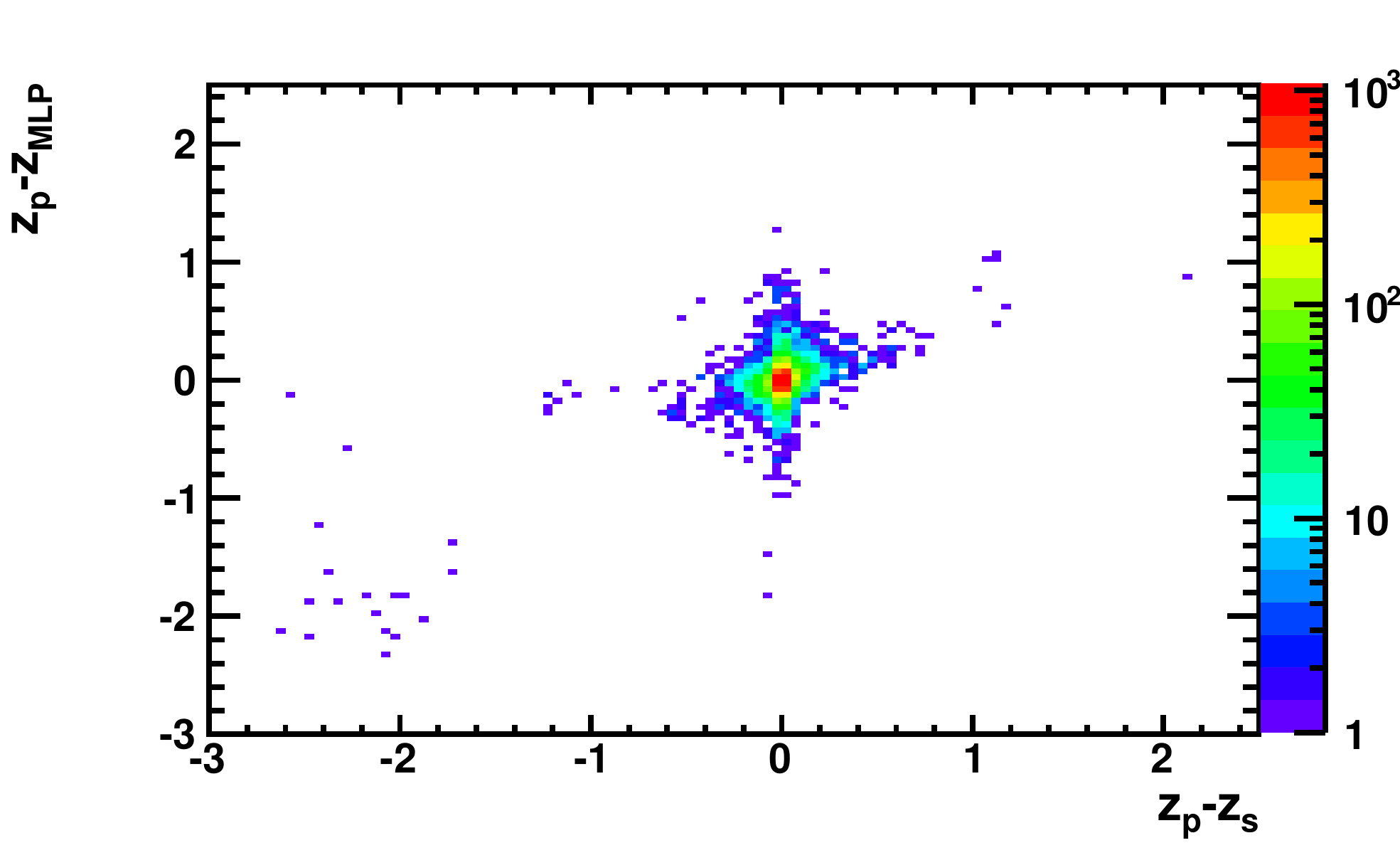}
      \caption{Top panel: $z_p$ \textit{vs.} $z_s$ with the template-fitting
      method in black (selection with $L_R>0.98$ ) and with the neural network in red for an LSST
      simulation of ten years of observations.
      Bottom panel: 2D histogram of $z_{MLP} - z_p$ as a function of $z_p - z_s$.}
         \label{Fig:CorrelationMLPTemplateFittingLSST}
\end{figure}

A scatter plot of photo-z versus spectroscopic redshift is shown on
the top panel of Fig.~\ref{Fig:CorrelationMLPTemplateFittingLSST}.
The black points show the results from the template-fitting method,
where a selection of $L_R>0.98$ was applied, and the red points show
the results from the neural network as described above.  
The plot compares the photo-z performance of the neural network method and the template-fitting method on the simulated LSST data.  Similar to \cite{Singal:2011}, we find that the neural network results in fewer outliers, although it has a larger rms for well-measured galaxies than the template-fitting method.

In the bottom panel of Fig.~\ref{Fig:CorrelationMLPTemplateFittingLSST}, the correlation between $z_p - z_{MLP}$ and $z_p-z_s$ is shown. Here, the correlation between both estimators is less useful for identifying outliers than it was for CFHTLS.  This is presumably due to both the simulation and the fit being performed with the same set of galaxy template SEDs. This should significantly reduce the fraction of outliers compared to a case where the templates used to estimate $z_p$ do not correctly represent the real galaxies.  For example, removing some of the templates from the $z_p$ fit reduces the photo-z quality, as demonstrated in \cite{Benitez:1998br}.  Therefore, the existence of a strong correlation between $z_p-z_{MLP}$ and $z_{p} - z_s$ may be useful in diagnosing and mitigating problems with the SED template set.

It is difficult to obtain a spectroscopic sample of galaxies that is truly representative of the photometric sample in terms of redshifts and galaxy types \citep{Cunha:2012}. For example, in the case of the LSST, the survey will be so deep that spectroscopic redshifts will be very hard to measure for the majority of faint galaxies or those within the ``redshift desert".   Here, we briefly investigate the effect of having the spectroscopic redshift distribution of the training sample biased with respect to the full photometric sample. 

\begin{figure}[!h]
   \centering
   	\includegraphics[width=9cm]{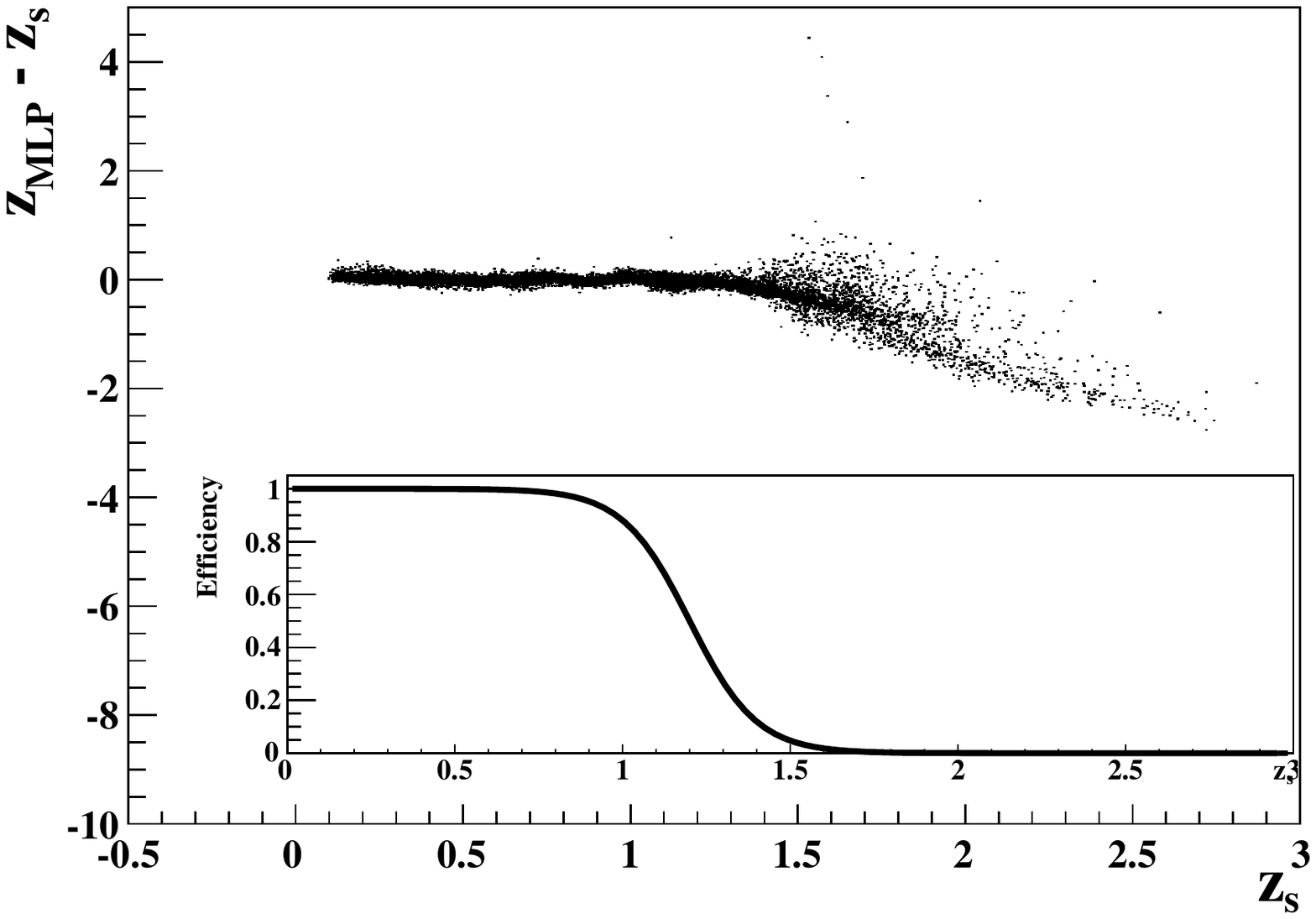}
   	\caption{$z_{MLP}-z_s$ as a function $z_s$ for the ten years of observations
   	of the LSST. The curve in the inset shows the efficiency function
   	$\epsilon$ as a function of the redshift, as it is used on the training
 	 sample to force a bias in the redshift selection.}
         \label{Fig:MLPBias}
\end{figure}

The fact that the distribution of redshifts in the spectroscopic sample is different from the underlying distribution is often (confusingly) termed redshift bias.
 The consequence of this bias can be seen by modifying the efficiency of detection as a function of the redshift. The efficiency
 function is chosen to be
 \begin{equation}
 \epsilon(z) = 1-1/\left(1+e^{-\left(\frac{z-1.2}{0.1}\right)}\right),
 \end{equation}
 and it is plotted in Fig.~\ref{Fig:MLPBias} (inset).
This efficiency function is then used to bias the training sample and the test
sample to compute new network weight coefficients. The photometric redshifts
for another unbiased sample are then computed using these weights. 

The scatter plot of $z_{MLP} - z_{s}$ as a function of $z_{s}$ is shown in Fig.~\ref{Fig:MLPBias}. We find that the photometric redshifts are well estimated as long as $\epsilon \geq 0.2$. 
This figure shows qualitatively that a bias in the training sample has a major impact on the photo-z reconstruction performance by the neural network, at least with the training
method used here.

\section{Discussion and future work}\label{Sec:Discussion}

In regard to simulations undertaken here, there are a number of simplifications that will be reconsidered in future work.  We discuss briefly some of these here.

\begin{itemize}
 \item \textit{Point source photometric errors}: We have assumed
   photometric errors based on estimates valid for point sources, and
   since galaxies are extended sources, we expect the errors to be
   larger in practice.  We made an independent estimate of the
   photometric errors as expected for the LSST, which includes the
   error degradation due to extended sources.  For the median expected
   seeing, we found that the photometric error scales as $\sigma_F/F =
   \theta/0.7$, where $\sigma_F$ is the error on the flux $F$, and
   $\theta$ is the size of the galaxy in arcseconds.  The next round
   of simulations will therefore include a prescription for simulating
   galaxy sizes to improve our simulation of photometric errors.  We will also compare our simple prescription to results obtained from the LSST image simulator (ImSim).
   \item \textit{Galactic extinction (Milky Way)}: Our current simulations effectively assume that i) Galactic extinction has been exactly corrected for, and ii) our samples of galaxies are all drawn from a direction of extremely low and uniform Galactic extinction.  In practice, there will be a contribution to the photometric errors due to the imperfect correction of the Galactic extinction, and this error varies in a correlated way across the sky. More problematically, the extinction has the effect of decreasing the depth of the survey as a function of position on the sky.  To account for these effects, we will construct a mapping between the coordinate system of our simulation and Galactic coordinates to apply the Galactic extinction in the direction to every galaxy in our simulation.  We can use the errors in those Galactic extinction values to propagate an error to the simulated photometry.
  \item \textit{Star contamination}: M-stars have extremely similar
    colors to early type galaxies and can easily slip into photometric
    galaxy samples.  Taking an estimate for the expected LSST
    star-galaxy separation quality, we plan to contaminate our catalog
    with stars.  This could have an important effect by biasing the
    clustering signal of galaxies, since the contamination increases
    when the line of sight is close to the Galactic plane. It should also be possible to use the photo-z algorithm (either a template fitting or neural network type) to identify stars within the catalog.
  \item \textit{Enhanced SEDs}: Our current simulations are probably more prone to problems with degeneracies in color space because we use a uniform interpolation between the main type SEDs. This may lead to poorer photometric redshifts than would be expected in reality, since galaxies might not exhibit such a continuous variation in SED type. In the future, we plan to implement a more
realistic interpolation scheme and the use of more complete template libraries, such as synthetic spectral libraries.
  \item \textit{Improved parameter estimation}: A better characterization of the set of locally optimal parameters when determining the photo-z through the template fitting method might help us in rejecting outliers.  We plan to further investigate this aspect in our future work.
\end{itemize}

Further improvements to the photo-z determination can be made by the use of angular cross-correlation
between objects in the photo-z sample and objects with spectroscopic
redshifts which are located in the
same area of the sky  (see \cite{2010ApJ...721..456M} and \cite{2011arXiv1109.2121M}).
This cross-correlation will help to characterize the redshift distribution
of the photometric sample, even though the spectroscopic sample may be incomplete or otherwise
not closely resemble the photometric sample.


\section{Conclusion}\label{Sec:Conclusion}

We have developed a set of software tools to generate mock galaxy catalogs
(as observed by the LSST or other photometric surveys), to compute photometric redshifts,
and study the corresponding redshift reconstruction performance.

The validity of these mock galaxy catalogs was carefully investigated (see Sect. \ref{Sec:MockCatValidation}).
We have shown that our simulation reproduces the photometric properties of the GOODS and CFHTLS
observations well, especially in regard to the number count, magnitude and color distributions.
We developed an enhanced template-fitting method for estimating the
photometric redshifts, which involved applying a new selection method, the likelihood ratio statistical test, that uses the posterior probability functions of the fitted photo-z parameters (z, galaxy type, extinction \ldots) and the galaxy colors to reject galaxies with outlier redshifts.

This method was applied to both the CFHTLS data and the LSST simulation to derive
photo-z performance, which was compared to the photo-z reconstruction
by using
a multilayer perceptron (MLP) neural network. We have shown how results from our template
fitting method and the neural network might be combined to provide a galaxy sample of
enriched objects with reliable photo-z measurements.

We find our enhanced template method produces photometric redshifts that are both realistic
and meet LSST science requirements, when the galaxy sample is selected using
the likelihood ratio statistical test. We have shown that a selection based on
the likelihood ratio test performs better than a simple selection based on apparent magnitude,
as it retains a significantly larger number of galaxies, especially at large redshifts
($z \gtrsim 1$), for a comparable photo-z quality.

We confirm that LSST requirements for
photo-z determination, which consists of a $(2-5)$\% dispersion on the photo-z estimate, with less
than $\sim$ 10\% outliers can be met, up to redshift $z\lesssim$2.5.
A number of enhancements for the mock galaxy catalog
generation and photo-z reconstruction have been identified and were
discussed in Sect. 8.

The photo-z computation presented here is designed for a full BAO
simulation that aims to
forecast the precision on the reconstruction of the dark energy equation-of-state parameter. This will be presented in a companion paper (Abate et al.,~\textit{in prep.}).

\begin{acknowledgements}
    Thanks to Pierre Astier and Delphine Hardin for their advice on CFHTLS data, to 
    Tomas Dahlen for his help on the simulation of the GOODS mock galaxy
    catalog and Eric Gawiser for his help with the IGM calculation.  We also thank the LSST internal reviewers: Patricia Burchat, Andy Connolly and Sam Schmidt for their constructive criticism of the original paper draft.
\end{acknowledgements}

\bibliographystyle{aa} 
\bibliography{biblio}     

\end{document}